\documentclass[11pt,a4paper]{article}
\pdfoutput=1

\usepackage{bm}
\usepackage{amsmath,amssymb}
\usepackage{cite}

\usepackage{graphicx}
\usepackage{color}
\usepackage{upgreek}

\usepackage{hyperref} %%put this package declaration last to make work since it redefines things!!%%
\hypersetup{colorlinks=true, citecolor=blue, filecolor=black, linkcolor=blue, urlcolor=blue, pdfpagemode=UseNone}
\numberwithin{figure}{section}
\numberwithin{equation}{section}

\newcommand{\be}{\begin{equation}}
\newcommand{\ee}{\end{equation}}
\newcommand{\bea}{\begin{eqnarray}}
\newcommand{\eea}{\end{eqnarray}}
\def\beal#1\eeal{\begin{align}#1\end{align}}   %one & only for aligning
\def\besp#1\eesp{\begin{multline}#1\end{multline}} %split an equation with first line left aligned & later right aligned

\newcommand{\TRM}[1]{#1} %{{\color{blue}\bf #1}}
\newcommand{\notes}[1]{}

\newcommand{\ph}{\varphi}
\newcommand{\vp}{\varphi}

\newcommand{\vpi}{\uppi}

\newcommand{\tg}{\tilde{g}}

\newcommand{\Lp}{a\Lambda_\p}

\newcommand{\p}{\mathrm{p}} %\newcommand{\p}{\sigma}
\newcommand{\dd}[2]{\delta_{\!\phantom{(} #1}^{\!(#2)}\!(\ph)}

\newcommand{\Lm}[1]{\mathfrak{L}_{#1}}
\newcommand{\Ll}{\mathfrak{L}}
\newcommand{\Lmm}{\Lm-}

\newcommand{\cc}[2]{\mathring{c}^{#2}_{#1}}

\newcommand{\ff}{\mathfrak{f}}
\newcommand{\fc}{\mathfrak{f}^{\sigma_0}_r}

\newcommand{\prop}{\triangle}

\newcommand{\rg}[1]{\mathring{#1}}

\newcommand{\hs}{\hat{s}}

\newcommand\ie{\textit{i.e.}\ }
\newcommand\eg{\textit{e.g.}\ }
\newcommand\cf{\textit{cf.}\ }

\newcommand{\aka}{{a.k.a.}\ }

\newcommand{\viz}{{\it viz.}\ }
\newcommand{\half}{\tfrac{1}{2}}

\newcommand{\eps}{\varepsilon}

\newcommand{\morri}{Morris:2018mhd}
\newcommand{\morrii}{Morris:2018axr}
\newcommand{\yuji}{Igarashi:2019gkm}
\newcommand{\nn}{\nonumber}
\newcommand{\propH}{\prop_\Lambda}
\newcommand{\proph}[1]{\prop_{#1}}

\newcommand{\cu}[1]{\!#1\!}

\newcommand{\cG}{\check{\Gamma}}
\newcommand{\nf}{N}
\newcommand{\Po}{\mathcal{P}}

\usepackage{datetime2}

\begin{document}

\begin{titlepage}
%\begin{flushright}
%%{\tt hep-ph/yymmnn}
%{\tt SHEP xx-xx}
%\end{flushright}

\begin{center}
{\huge \bf The continuum limit of the conformal sector at second order in perturbation theory}

%Renormalizable quantum gravity and diffeomorphism invariance}

%of the conformal sector in quantum gravity}
%Relevant directions for the conformal factor in perturbative quantum gravity 
%\emph{or: Through the conformal factor to(wards) perturbatively renormalizable quantum gravity }} 
%\vskip.3cm
%{\huge \bf  and etc} 
\end{center}
\vskip1cm

%\title{xxx}
%\author{Tim R. Morris}

\begin{center}
{\bf Tim R. Morris}
\end{center}

%\affiliation{
\begin{center}
{\it STAG Research Centre \& Department of Physics and Astronomy,\\  University of Southampton,
Highfield, Southampton, SO17 1BJ, U.K.}\\
\vspace*{0.3cm}
{\tt  T.R.Morris@soton.ac.uk}
\end{center}

\abstract{Recently a novel perturbative continuum limit for quantum gravity has been proposed and demonstrated to work at first order. Every interaction monomial $\sigma$ is dressed with a coefficient function $f^\sigma_\Lambda(\vp)$ of the conformal factor field, $\vp$. Each coefficient function is parametrised by an infinite number of underlying couplings, and decays at large $\vp$ with a characteristic amplitude suppression scale which can be chosen to be at a common value, $\Lambda_\p$. Although the theory is perturbative in couplings it is non-perturbative in $\hbar$.  At second order in perturbation theory, one must sum over all melonic Feynman diagrams to obtain the particular integral. We show that it leads to a well defined renormalized trajectory and thus continuum limit, provided it is solved by starting at an arbitrary cutoff scale $\Lambda\cu=\mu$ which lies in the range $0\cu<\mu\cu<a\Lambda_\p$ ($a$ some non-universal number). If $\mu$ lies above this range the resulting coefficient functions become singular, and the flow ceases to exist, before the physical limit is reached. To this one must add a well-behaved complementary solution, containing irrelevant couplings determined uniquely by the first-order interactions, and renormalized relevant couplings. 
Even though some irrelevant couplings diverge in the limit $\Lambda_\p\cu\to\infty$, domains for the underlying relevant couplings can be chosen such that diffeomorphism invariance will be recovered  in this limit, and where the underlying couplings disappear to be replaced by effective diffeomorphism invariant couplings.}

\end{titlepage}

\tableofcontents

%\newpage

\section{Introduction}
\label{sec:intro}

In refs. \cite{\morri,Kellett:2018loq,Morris:2018upm,\morrii,first} we discovered a new quantisation for quantum gravity, resulting in a \emph{perturbative} continuum limit. We established that this works to first order. In this paper we establish the existence of an appropriate continuum limit also to second order in perturbation theory. 

\TRM{To understand the continuum limit in depth, we need to use the Wilsonian RG (Renormalization Group) \cite{Wilson:1973,Wegner:1972my}. Then an essential ingredient is the concept of Kadanoff blocking \cite{Kadanoff:1966wm}, where one integrates out degrees of freedom at short distances to obtain effective short range interactions. Thus we must work in Euclidean signature, so that short distance really does imply short range. Similarly for RG fixed points to exist, the manifold itself must look the same at any scale. That tells us to work with fluctuations on flat $\mathbb{R}^4$. Thus we construct the theory in flat Euclidean space.\footnote{\TRM{Although we stay with this case in this paper we note that, having constructed the theory in flat space, one can then study the construction on other manifolds and the analytic continuation to Lorentzian signature \cite{\morrii} where the Wilsonian RG is strictly speaking  inapplicable. See however \cite{Manrique:2011jc}.}} This means the metric is given by $\delta_{\mu\nu}$. The interactions we start with are constrained not by diffeomorphism invariance but only by Lorentz invariance (actually $SO(4)$ invariance). }

%We thus work with perturbative fluctuations on a flat spacetime. 
In Euclidean signature, the partition function is ill defined due to the conformal factor instability \cite{Gibbons:1978ac}, but the Wilsonian exact RG flow equation continues to make makes sense \cite{Reuter:1996,\morri}. We therefore do not analytically continue the conformal factor as proposed in ref. \cite{Gibbons:1978ac}, but use the Wilsonian exact RG, which is anyway a more powerful route to define the continuum limit. Everything in the new quantisation follows from this observation.

\begin{figure}[ht]
\centering
\includegraphics[scale=0.3]{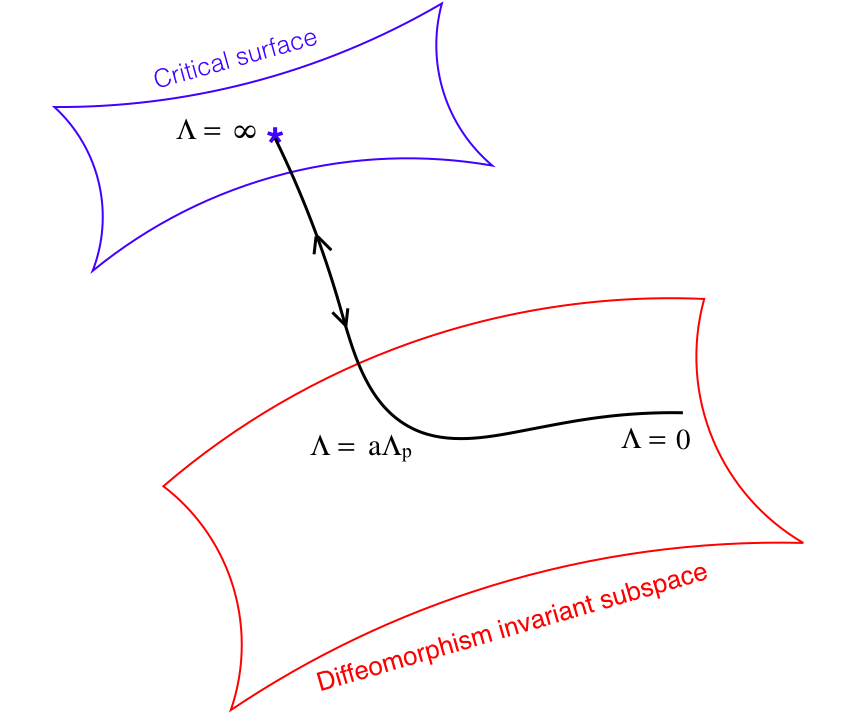}
%\vskip-30pt
\caption{The continuum limit is described by a renormalized trajectory that shoots out of the Gaussian fixed point (free gravitons) along relevant directions that cannot respect diffeomorphism invariance for $\Lambda>a\Lambda_\p$, where $\Lambda_\p$ is a characteristic of the renormalized trajectory and is called the \emph{amplitude suppression scale} (or \emph{amplitude decay scale}), and $a$ is a non-universal number. By appropriate choice of the \emph{underlying} couplings $g^\sigma_n$, diffeomorphism invariance is then recovered at scales $\Lambda,\vp\ll\Lambda_\p$ where also we recover an expansion in the \emph{effective} coupling  $\kappa\sim\sqrt{G}$.}
\label{fig:flow}
\end{figure}

With the effective cutoff $\Lambda$ in the far UV (ultraviolet) region, a perturbative continuum limit is constructed by expanding around the Gaussian fixed point (the action for free gravitons). We have shown that perturbations that are otherwise arbitrary functions of the conformal factor amplitude, $\vp$, can be expanded as a convergent sum over eigenoperators (and such convergence is a necessary condition for the Wilsonian RG to make sense) only if we construct them using a novel tower of operators $\dd\Lambda{n}$ ($n=0,1,\cdots$) \cite{\morri}. These operators have negative dimension $-1\cu-n$, and are therefore increasingly relevant as $n$ increases. Any interaction monomial $\sigma$ of the fields and their spacetime derivatives, thus ends up being dressed with a \emph{coefficient function} $f^\sigma_\Lambda(\vp)$, containing an infinite number of relevant couplings $g^\sigma_n$. 

In the UV regime these vertices cannot respect diffeomorphism invariance \cite{\morrii,first} or rather, precisely formulated, they cannot respect the quantum equivalent, which are the Slavnov-Taylor identities modified by the cutoff $\Lambda$ \cite{Ellwanger:1994iz,\yuji}. Succinctly stated, the interactions necessarily lie outside the \emph{diffeomorphism invariant subspace} defined by these identities. However the coefficient functions come endowed with an amplitude suppression scale $\Lambda_\p$, which characterises how fast they exponentially decay in the large $\vp$ limit \cite{\morri}. We have shown that to first order, provided that the underlying couplings $g^\sigma_n$ occupy appropriate domains, at scales much less than $\Lambda_\p$ the coefficient functions \emph{trivialise}. This means they become polynomials in $\vp$ times an overall constant which (for pure quantum gravity at vanishing cosmological constant) gets identified with 
\be 
\label{kappa}
\kappa=\sqrt{32\pi G}\,,
\ee
$G$ being Newton's constant. This property is sufficient to allow the modified Slavnov-Taylor identities to be recovered \cite{\morrii,first}. 
The renormalized trajectory thus takes the form sketched in fig. \ref{fig:flow}. 

%%PRD addition response referee place holder%%
\TRM{
In the end, UV completion is achieved because this part of the renormalized trajectory lies outside the diffeomorphism invariant subspace. Lorentz invariance is respected everywhere however. Nevertheless, in that certain conditions are relaxed in the UV to allow a continuum limit, it is similar to Ho\v rava-Lifshitz gravity \cite{Horava:2009uw}. There it is achieved instead by keeping diffeomorphisms but breaking Lorentz invariance (which then allows the time direction to have a different scaling dimension). 

Since in our case there an infinite number of relevant underlying couplings $g^\sigma_n$ for each interaction monomial $\sigma$, we would appear to have an infinite number of parameters for every interaction that would be required to be fixed experimentally. In this sense the theory would not be considered perturbatively renormalizable, even if it is UV complete. Actually as we indicate above, diffeomorphism invariance is recovered only at scales much less than $\Lp$. Equivalently, we must take the limit $\Lp\cu\to\infty$, holding everything else fixed.\footnote{\TRM{Note that the physical theory is produced only in this limit, since diffeomorphism invariance is non-negotiable. Without it the fluctuation field $H_{\mu\nu}$ would have propagating negative norm non-physical polarisations and thus unitarity would be destroyed. This is a significant difference with Ho\v rava-Lifshitz gravity where some breaking of Lorentz invariance can remain, although in practice it is hard to achieve whilst staying self-consistent and within stringent observational constraints \cite{Wang:2017brl}.} }  %Now, in this large $\Lambda_\p$ limit, 
Then, provided the couplings $g^\sigma_n$ lie within the above (infinite) domains, the theory `forgets' about them and only a single diffeomorphism invariant coupling remains for the given monomial $\sigma$, such as $\kappa$ above. This is a kind of universality property and is discussed in detail later in sec. \ref{sec:prelim}, see in particular equation \eqref{largeg}. It thus results in an infinite reduction in the number of couplings. 

However since at higher orders in perturbation theory there are an infinite number of monomials $\sigma$, there is still the opportunity for an infinite number of couplings to be left behind, only now as effective diffeomorphism invariant couplings associated to diffeomorphism invariant combinations of these monomials. This would then be essentially what one finds within the standard perturbative approach. Whether this is the outcome, or the effective couplings are themselves fixed by another mechanism, is debated in ref. \cite{second}. In the diffeomorphism invariant subspace, one must be left with an RG flow that is non-singular at all scales $\Lambda$, once  the limit $\Lp\cu\to\infty$ is taken. 
One such well-studied possibility is a non-perturbative (asymptotically safe) UV fixed point \cite{Weinberg:1980,Reuter:1996,Bonanno:2020bil}. We also highlight a novel mechanism for fixing the parameters that
could follow from the same mathematical properties of the partial differential flow equations that lead to the current formulation \cite{second}.
}
%%%end%%%

Now we can be precise about the steps we establish in this paper. We will show that at second order in perturbation theory, the renormalized trajectory is well defined and thus the continuum limit exists. We will show moreover that by choosing appropriate domains for the underlying relevant couplings, we can again ensure that all coefficient functions trivialise in an appropriate way to allow the modified Slavnov-Taylor identities (mST) to be recovered. Effectively, we therefore establish the existence of the renormalized trajectory down to the point where it can enter the diffeomorphism invariant subspace. This last IR (infrared) part of the renormalized trajectory will be treated in ref. \cite{second} where also we will recover the physical amplitudes. 

\begin{figure}[ht]
\centering
\includegraphics[scale=0.4]{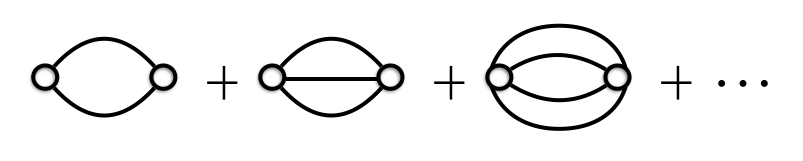}
%\vskip-30pt
\caption{The second-order part, $\Gamma_2$, of the effective action, resums an expansion over melonic Feynman diagrams, and appears in different guises. In \eqref{flowmelons}, the re-summation yields $\beta$ functions, with open circles given by the physical first-order vertices $\Gamma_{1\,\text{phys}}$. The $\beta$ functions integrate exactly to the $\rg\Gamma_2$ expression \eqref{ringGtwo}, which can be recast as our final expression for $\Gamma_2$  \eqref{Gammatwosol}. In this last version, the open circles are copies of $\Gamma_1$ whose solution \cite{\morrii,first} is illustrated in fig. \ref{fig:tadpoles}.}
\label{fig:melons}
\end{figure}

\begin{figure}[ht]
\centering
\includegraphics[scale=0.35]{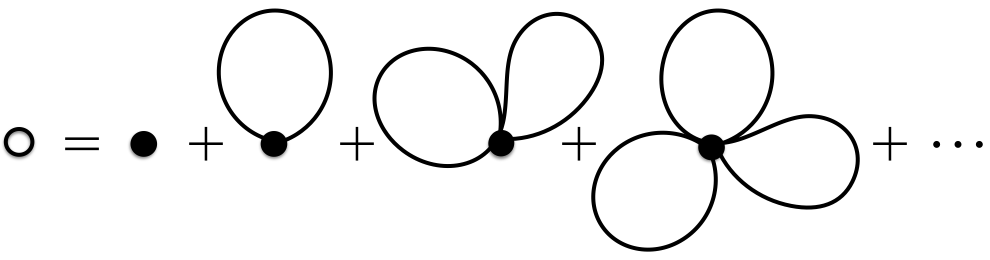}
%\vskip-30pt
\caption{The first-order effective action $\Gamma_1$ is an infinite sum over (marginally) relevant eigenoperators and their conjugate couplings $g^\sigma_n$. Each eigenoperator is equal to its physical limit
$\sigma\, \delta^{(n)}(\vp)$, plus all possible tadpole corrections. Those corrections generated by attaching to $\sigma$, terminate eventually (since the monomial runs out of fields), while $\vp$-tadpole corrections to $\delta^{(n)}(\vp)$ go on forever but resum to $\delta_{\, \Lambda}^{\!(n)}\!(\ph)$. We do not draw external legs, an infinite number of which attach to $\delta_{\, \Lambda}^{\!(n)}\!(\ph)$.} 
%\phantom{(} caused weird error!
\label{fig:tadpoles}
\end{figure}

What makes the steps in this paper particularly challenging, 
%are not only that there are now an infinite number of monomials $\sigma$, these being generated by derivative expansion of the effective action and each being associated to an infinite number of underlying couplings through their coefficient functions $f^\sigma_\Lambda$, but especially 
are that the operators $\dd\Lambda{n}$ are non-perturbative in $\hbar$. Thus while we can work perturbatively in the interactions, we must work non-perturbatively in the number of loops. Second order computations therefore require resumming to all loops the so-called melonic Feynman diagrams illustrated in fig. \ref{fig:melons}. 
To establish the above steps,  we need to show that this sum converges and leads to well-behaved coefficient functions possessing the right properties. The underlying couplings $g^\sigma_n(\Lambda)$ in the coefficient functions, now include irrelevant ones, and
run with $\Lambda$. In general their dimensionless versions must vanish in the UV limit, $\tg^\sigma_n(\Lambda)\cu\to0$ as $\Lambda\cu\to\infty$,  so that the renormalized trajectory indeed emanates from Gaussian fixed point.\footnote{In fact as we will see, to second order, one coupling behaves as exactly marginal, thus parametrising an `accidental' line of fixed points, which we compute. As we will see, it is not expected to remain exactly marginal at third order.}
We also need to show that the IR limit exists, since this corresponds to computing the physical Legendre effective action 
\be 
\label{physical}
\Gamma_\text{phys} = \lim_{\Lambda\to0} \Gamma\,.
\ee
This step in particular is non-trivial. Unless one is careful with the choice of underlying couplings,  coefficient functions become singular and the flow ceases to exist before the IR limit is reached,  even at the linearised level \cite{\morri}. However what allows us to make progress with all this is that at this stage we are only interested in establishing the existence of these various limits, rather than computing their precise values. Then it turns out we can work at a largely schematic level.

We now sketch our approach, and at the same time provide a guide to the reader for what is contained in each section of the paper. At second order, the Wilsonian effective interactions are no longer local, but quasi-local, \ie have a derivative expansion which continues indefinitely. This can be seen as originating from a Taylor expansion of the Feynman diagrams in their external momenta (this Taylor expansion converges for sufficiently small external momenta because the diagrams are IR regulated by $\Lambda$). It means however that at second order,  we now have infinitely many monomials $\sigma$ each with their own coefficient function $f^\sigma_\Lambda(\vp)$. At the beginning of sec. \ref{sec:second}, and in detail in sec. \ref{sec:open}, we show that the second-order flow equations then imply an open set of flow equations for these coefficient functions such that the flow of any one, depends not only on terms bilinear in the first-order coefficient functions, but also on tadpole corrections from higher-derivative second-order vertices and their coefficient functions. 

In sec. \ref{sec:model}, we gain a great deal of insight by temporarily truncating this ``tadpole cross-talk'', so that we get a closed model flow equation for a second-order coefficient function $f^{\sigma_0}_\Lambda(\vp)$, however still depending non-linearly on the first-order coefficient functions. Here we can analyse the required limits and verify the conclusions with closed form solutions. We first extract from the flow equations the infinite set of $\beta$ functions for the underlying second-order couplings. These $\beta$ functions are themselves an infinite sum over products of the first-order couplings.  These sums are guaranteed to converge for sufficiently high $\Lambda$, thanks to the required convergence conditions on the first-order coefficient functions \cite{\morrii,first}. We show that the requirement that the renormalized trajectory behaves correctly in the far UV, can be satisfied, and that as expected this fixes the irrelevant couplings in $f^{\sigma_0}_\Lambda(\vp)$ uniquely in terms of the (relevant and marginal) first-order couplings. At first sight the $\beta$ functions are badly divergent in the IR \cite{\morri}, but we see that the sums do not converge in this regime.  We get a sensible result instead by solving directly for the flow of the Fourier transform, $\dot\ff^{\sigma_0}(\vpi,\Lambda)$, \ie by working in conjugate momentum ($\vpi$) space. Then we see that the flow equation can be integrated, however the \emph{particular integral}  contributes a coefficient function $f^{\sigma_0}_\Lambda(\vp)$ that develops singularities, of the form highlighted above, unless we integrate from a starting point $\Lambda\cu=\mu\cu<a\Lambda_\p$. Furthermore since the derivative expansion breaks down in the limit $\Lambda\cu\to0$, we must choose the starting point to satisfy $\mu\cu>0$. To this particular integral we must add a \emph{complementary solution} $f^{\sigma_0}_\Lambda(\vp,\mu)$, a solution to just the homogeneous part of the flow equation which contains $g^{\sigma_0}_n(\mu)$: the irrelevant couplings and the renormalized relevant couplings evaluated at $\Lambda\cu=\mu$.

Returning to the true system of equations in sec. \ref{sec:open}, we also find a representation $\rg\Gamma_2$ where the infinite series of tadpole corrections are traded for new explicit contributions to the flow equation, these being the infinite sum over the melonic Feynman diagrams, while the $f^\sigma_\Lambda(\vp)$ get mapped to  \emph{stripped} coefficient functions $\rg{f}^\sigma_\Lambda(\vp)$, whose dependence on $\Lambda$ is only through the running couplings $g^\sigma_n(\Lambda)$ contained in $\ff^{\sigma}(\vpi,\Lambda)$. Although we show that the stripped coefficient functions $\rg{f}^\sigma_\Lambda(\vp)$ are singular in the limit $\Lambda\cu\to\Lambda_\p^-$,  the conjugate momentum expressions $\ff^{\sigma}(\vpi,\Lambda)$ continue to make sense for all $\Lambda\cu>0$. 

This is the starting point for the analysis of the full renormalized trajectory in sec. \ref{sec:fullrt}. We show that for each generated monomial $\sigma$, each melonic contribution is well defined, being fully regulated in both the IR and UV by $\Lambda$. We show that the sum over all the melonic contributions converges and yields a formula for $\dot\ff^{\sigma}(\vpi,\Lambda)$ whose asymptotic properties are the same as the one we derived for the model in sec. \ref{sec:model}. We derive these asymptotic properties first with a na\"\i ve estimate, for a general cutoff function. Then we verify the estimate exactly using a specific cutoff function of exponential form, by explicitly computing the large loop order behaviour of the integrals (with the help of app. \ref{app:largen}). What makes this possible is the fact that  the melonic Feynman diagrams of fig. \ref{fig:melons} are just pointwise products of propagators when written in position space, leaving only one space-time integral to be done to extract coefficients of the derivative expansion.

We again derive the form of the $\beta$ functions for the underlying couplings, this time for the full theory however, and thus demonstrate that  the renormalized trajectory behaves correctly in the far UV provided that the irrelevant couplings are set as determined by the first-order couplings. Along the way, we derive the dimension of the monomials $\sigma$ as a function of key properties, and similarly the parity of their coefficient functions and the dimension of the underlying couplings they contain. This establishes that the second-order couplings have only odd dimensions, and thus as a corollary that there are no new marginal couplings at this order and also that the first-order couplings do not run (since they are only even dimensional). We also demonstrate that this `accident' is not repeated at third order, so at third order we can expect the first-order couplings also to run.

Now, unlike in the model, the full flow equations for the stripped $\rg\Gamma_2$ are exactly integrable. However if we cast the integrals order by order in the loop expansion directly in terms of renormalized contributions that depend on only the one effective cutoff $\Lambda$, we find (with the help of app. \ref{app:fails}) that the resummation leads to a contribution to the coefficient functions that becomes singular for $\Lambda\cu\ge a\Lambda_\p/\sqrt{\textrm{e}-1}$ (after which the flow would cease to exist). Instead we must integrate from a finite starting point $\Lambda\cu=\mu$. This must satisfy $\mu\cu>0$ in order to make manifest the derivative expansion property. We then establish that the sum over melonic diagrams is convergent and leads to sensible coefficient functions although, just as happened in the model, this is only manifest if $\mu\cu<a\Lambda_\p$. Otherwise the particular integral creates coefficient functions that become singular at some critical  cutoff scale before reaching the IR limit. Inverting the map to the stripped representation $\rg\Gamma_2$ we arrive at our final form \eqref{Gammatwosol}, a well-defined renormalized trajectory for the full second-order contribution $\Gamma_2$. In app. \ref{app:alt}, we give a streamlined derivation of this key equation.

In the last part of this section we also characterise how the derivative expansion coefficients diverge as $\mu\cu\to0$.  Although these divergences are an artefact of the breakdown of the derivative expansion, they play an important r\^ole in characterising the large amplitude suppression scale limit, which we turn to in sec. \ref{sec:largeASS}. 
Recall that this limit is a necessary condition for recovering diffeomorphism invariance through the mST \cite{\morrii}. We show that in this limit the melonic expansion of the particular integral collapses to the difference of two one-loop diagrams in standard quantisation, while the second-order mST also collapses to something closely related to standard quantisation. 

We are left however to see if the relevant couplings can be constrained so that the  complementary solutions $f^\sigma_\Lambda(\vp,\mu)$ trivialise in this limit, the final condition that will be needed before the mST can be satisfied. Despite the fact that these coefficient functions are solutions of the linear flow equation, there is an apparent obstruction since their irrelevant couplings $g^\sigma_{2l+\eps}(\mu)$ are already determined non-linearly in terms of the first-order interactions.  Furthermore, some of these irrelevant couplings even diverge in the limit $\Lambda_\p\cu\to\infty$. At this point the observations we make in sec. \ref{sec:fixcouplings} become crucial. There we show that we can fix any finite set of couplings, $g^\sigma_\eps, g^\sigma_{2+\eps},\cdots,g^\sigma_{2N+\eps}$, to desired functions of $\Lambda_\p$, and yet still get linearised coefficient functions that trivialise in the limit $\Lambda_\p\cu\to\infty$, provided however that the \emph{reduced} form of these couplings $\bar{g}^\sigma_{2l+\eps}$ diverges slower than $\Lambda_\p^2$. These reduced couplings are certain dimensionless ratios \eqref{barg} and this requirement gives us the necessary convergence conditions.

In sec. \ref{sec:largeASSderiv0}, we gain further insight by returning to the model of sec. \ref{sec:model}. Apart from the factor of $\kappa^2$, the irrelevant couplings $g^\sigma_{2l+\eps}(\mu)$ depend on only two scales namely $\mu$ and $\Lambda_\p$. Thus the large $\Lambda_\p$ limit can be determined from the small $\mu$ behaviour which we already deduced in sec. \ref{sec:fullrt}. We confirm this by computing the limit and comparing to the exact expression we already derived in sec. \ref{sec:model}. We then show that the amplitude suppression scale for $f^{\sigma_0}_\Lambda(\vp)$ can be identified with $\Lambda_\p$ and that the convergence conditions \eqref{barg} can be satisfied so that it trivialises appropriately. 

Finally in sec. \ref{sec:lassgeneral}, we return to the true system of equations and derive the large $\Lambda_\p$ behaviour for all the irrelevant couplings in the same way. Then we show that all second-order amplitude suppression scales can be set to $\Lambda_\p$ and, by analysing various special cases, show that the convergence conditions can be met and relevant second-order couplings chosen to occupy domains, such that all the second-order coefficient functions trivialise appropriately in the large $\Lambda_\p$ limit. 

We start the paper in sec. \ref{sec:prelim} by collecting together the results we need from previous papers, while in sec. \ref{sec:conclusions} we summarise our key findings.

\section{Preliminaries}
\label{sec:prelim}

We recall material that we will need from the previous papers \cite{\morri,Kellett:2018loq,Morris:2018upm,\morrii,\yuji,first}. We are interested in using the Wilsonian RG to establish a perturbative continuum limit for quantum gravity. 
%We thus work with fluctuations on a flat Euclidean signature spacetime. The partition function is ill defined due to the conformal factor instability \cite{Gibbons:1978ac}, but the Wilsonian exact RG flow equation  still makes sense \cite{Reuter:1996,\morri}. We therefore do not analytically continue the conformal factor as proposed in ref. \cite{Gibbons:1978ac}, but use the Wilsonian exact RG, which is anyway a more powerful route to define the continuum limit. Everything in the new quantisation just follows from this one observation.
In terms of the interacting part of the infrared cutoff Legendre effective action,
the flow equation takes the form \cite{Nicoll1977, Wetterich:1992, Morris:1993} (see also  \cite{Weinberg:1976xy,Morris:2015oca,Bonini:1992vh,Ellwanger1994a,Morgan1991}):
\be %[18.9,10] = [6:8] is derived in p851 QG notes
\label{flow}
\dot{\Gamma}_I = -\half\, \text{Str}\left( \dot{\prop}_\Lambda\propH^{-1} \left[1+\propH \Gamma^{(2)}_I \right]^{-1}\right) 
%= -\half (-)^A \left( \dot{\prop}_\Lambda\propH^{-1}\left[1+\propH \Gamma^{(2)}_I \right]^{-1}\right)^{\!A}_{\ \ A} 
\,,
\ee
where the over-dot is $\partial_t =-\Lambda \partial_\Lambda$. The BRST invariance is  expressed through the mST (modified Slavnov-Taylor identity)  \cite{Ellwanger:1994iz,\yuji}:
\be 
\label{mST}
\Sigma := \half (\Gamma,\Gamma) - \text{Tr}\left( \!C^\Lambda\,  \Gamma^{(2)}_{I*} \left[1+\propH\Gamma^{(2)}_I\right]^{-1}\right) = 0\,,
\ee
where $\Gamma=\Gamma_0+\Gamma_I$, $\Gamma_0$  being the action for free gravitons and their BRST transformations \cite{\morrii,first} (we do not actually need its explicit form in this paper).
These equations are both ultraviolet (UV) and infrared (IR) finite thanks to the presence of the UV cutoff function $C^\Lambda(p)\equiv C(p^2/\Lambda^2)$ which, since it is multiplicative, satisfies $C(0)=1$, and its associated IR cutoff
%\footnote{These cutoffs were written as $C^\Lambda\equiv C$ and $C_\Lambda \equiv \bar{C}$ in refs. \cite{\morri,\morrii}.} 
$C_\Lambda= 1-C^\Lambda$, which appears in the IR regulated propagators as $\propH^{AB} = C_\Lambda\prop^{AB}$. The cutoff function is chosen so that $C(p^2/\Lambda^2)\cu\to0$ sufficiently fast as $p^2/\Lambda^2\cu\to\infty$ to ensure that all momentum integrals are indeed UV regulated (faster than power fall off is necessary and sufficient). It is also required to be smooth (differentiable to all orders), corresponding to a local Kadanoff blocking. It thus permits for $\Lambda\cu>0$, a quasi-local solution for $\Gamma_I$, namely one that has a space-time derivative expansion to all orders. We need this since it is equivalent to imposing locality on a bare action.

In the above equations we have introduced Str$\,\mathcal{M} = (-)^A\, \mathcal{M}^{A}_{\ \,A}$ and Tr$\,\mathcal{M} = \mathcal{M}^{A}_{\ \,A}$, and set
\be 
\label{Hessian}
\Gamma^{(2)}_{I\ AB} = \frac{\partial_l}{\partial\Phi^A}\frac{\partial_r}{\partial\Phi^B}\Gamma_I\,,\qquad \
\left(\Gamma^{(2)}_{I*}\right)^{A}_{\ \ B} 
\,=\,  \frac{\partial_l}{\partial\Phi^*_A}\frac{\partial_r}{\partial\Phi^B}\Gamma_I\,,
\ee 
%=  \frac{\dell{}}{\partial\Phi^A}\Gamma_I\frac{\delr{}}{\partial\Phi^B}\,.
Here $\Phi$ and $\Phi^*$ are the collective notation for the classical fields (the graviton $H_{\mu\nu}$ and ghost $c_\mu$) and antifields (sources $H^*_{\mu\nu}$ and $c^*_\mu$ of the corresponding BRST transformations) respectively.  Splitting 
\be 
H_{\mu\nu} = h_{\mu\nu}+\half \vp\delta_{\mu\nu}
\ee 
into its traceless and traceful (\aka conformal factor) parts, the propagators we need are
\beal
%\langle H_{\mu\nu}(p)\,H_{\alpha\beta}(-p)\rangle &= \frac{\delta_{\mu(\alpha}\delta_{\beta)\nu}}{p^2}
%-\frac12\frac{\delta_{\mu\nu}\delta_{\alpha\beta}}{p^2}\,,\nn \\
\label{hh}
\langle h_{\mu\nu}(p)\,h_{\alpha\beta}(-p)\rangle &= \frac{\delta_{\mu(\alpha}\delta_{\beta)\nu}-\frac14\delta_{\mu\nu}\delta_{\alpha\beta}}{p^2} 
\,,\\
%\langle h_{\mu\nu}(p)\, \ph(-p)\rangle &= \langle \ph(p)\,h_{\mu\nu}(-p)\rangle = 0\,,\nn\\
\label{pp}
\langle \ph(p)\,\ph(-p)\rangle &=  -  \frac1{p^2}\,,\\
\label{cc}
\langle c_\mu(p)\, \bar{c}_\nu(-p)\rangle &= -\langle \bar{c}_\mu(p)\, c_\nu(-p) \rangle =  \delta_{\mu\nu}/{p^2}\,,
\eeal
where we have written \notes{477.7}
\be 
\label{defs}
\prop^{AB}  = \langle\Phi^A\,\Phi^B\rangle\,,\qquad
%\quad\text{and}\quad
\Phi^A(x) = \int_p  \text{e}^{-i p\cdot x}\, \Phi^A(p)\,,\qquad
\int_p \equiv \int\!\! \frac{d^4p}{(2\pi)^4}\,.
\ee
Note that $h_{\mu\nu}$ propagates with the right sign, and that the numerator  is just the projector onto traceless tensors, while the conformal factor $\vp$ propagates with wrong sign (a consequence of the conformal factor instability). 

In the limit $\Lambda\cu\to0$, the IR cutoff is removed and we get back the standard Legendre effective action, 
$\Gamma_\text{phys} = \lim_{\Lambda\to0} \Gamma$.
On the other hand the flow equation \eqref{flow} and the mST \eqref{mST}
are compatible: %so that 
if $\Sigma=0$ at some generic scale $\Lambda$, it remains so on further evolution, in particular as $\Lambda\to0$. The second term in the mST
is a quantum modification due to the cutoff $\Lambda\cu>0$. At non-exceptional momenta (\ie such that no internal particle in a vertex can go on shell) it remains IR finite, and thus vanishes as $\Lambda\to0$, thanks to the UV regularisation. We are then left with just the first term which is the Batalin-Vilkovisky antibracket \cite{Batalin:1981jr,Batalin:1984jr}, \ie we are left with the
 Zinn-Justin equation $\half(\Gamma,\Gamma)=0$ \cite{ZinnJustin:1974mc,ZinnJustin:1975wb}. Thus in the limit $\Lambda\to0$ we recover both the Legendre effective action and the standard realisation of quantum BRST invariance through the Slavnov-Taylor identities for the corresponding vertices.

%while $\Gamma$ is the ``effective average action'' \cite{Wetterich:1992} part of the infrared cutoff Legendre effective action \cite{\yuji,Morris:1993}:
%\be 
%\label{LegendreEffAct}
%\Gamma^{tot} = \Gamma + \half \Phi^A \mathcal{R}_{AB} \Phi^B\,,  \qquad \prop^{-1}_{\Lambda\, AB} = \prop^{-1}_{AB} + \mathcal{R}_{AB}\,,
%\ee
%where $\mathcal{R}_{AB}$ is the infrared cutoff expressed in additive form.
%$\Gamma$ is expressed in terms of a free part, $\Gamma_0$, which includes the free BRST transformations, plus the interaction part $\Gamma_I[\Phi,\Phi^*]$:
%\be 
%\label{Gamma}
%\Gamma = \Gamma_0 + \Gamma_I\,,\qquad \Gamma_0 =
%\half\, \Phi^A \prop^{-1}_{AB}\Phi^B -  (Q_0\Phi^A) \Phi^*_A \,.  %R^A_{\ B}\Phi^B\,
%\ee
%The non-vanishing free BRST transformation is in fact

We expand $\Gamma_I$ perturbatively in its interactions, assuming the existence of an appropriate small parameter $\epsilon$:
\be 
\label{expansion}
\Gamma_I = \sum_{n=1}^\infty\Gamma_n\,{\epsilon^n}/{n!} \,.
\ee
\TRM{In this paper we show that to second order we have a well defined renormalized trajectory that shoots out of the Gaussian fixed point as in fig. \ref{fig:flow}. In particular this means that  we establish that the interactions can indeed be constructed so as to vanish (when written in dimensionless terms) as $\Lambda\cu\to\infty$. In this way the above assumption is justified in the UV. On the other hand at scales $\Lambda\cu<\Lp$, $\epsilon$ eventually becomes identified with $\kappa$, as we will see. In this regime, by dimensions, higher orders are accompanied by increasing numbers of space-time derivatives, just as is found in the standard approach. However, since we are now dealing with a theory with a genuine continuum limit, the fact that perturbation theory breaks down in the regime\footnote{Here $\partial$ stands for the typical magnitude of space-time derivatives.} $\kappa\partial>1$, just indicates that the theory becomes non-perturbative in this regime and not, as usually interpreted in the standard approach, a signal of breakdown of an effective quantum field theory description. 
 }
 
At first order the flow equation \eqref{flow} and mST \eqref{mST} become
\beal 
\label{flowone}
\dot{\Gamma}_1 &=  \half\, \text{Str}\, \dot{\prop}_\Lambda \Gamma^{(2)}_1 \,, \\
\label{mSTone}
0 &=  (\Gamma_0,\Gamma_1) - \text{Tr}\left( \!C^\Lambda\,  \Gamma^{(2)}_{1*} \right) 
%= (Q_0+Q^-_0-\Delta^--\Delta^=) \Gamma_1 
=: \hs_0\, \Gamma_1\,,
\eeal
where the first equation is the flow equation satisfied by eigenoperators: their RG time derivative is given by the action of the tadpole operator \cite{\morrii}, while the second equation defines the total free quantum BRST operator \cite{\morrii,\yuji,first}. We will mostly not need its explicit form in this paper. 

%We will use only the property that the action can be graded by antighost number so that an operator $\Op = \sum_{m=0}^n \Op^m$ with some maximum antighost number $n$, that is annihilated by $\hat{s}_0$, must satisfy the descent equations:
%\be 
%\label{descendants}
%Q_0\, \Op^n = 0\,, \quad Q_0\, \Op^{n-1} = (\Delta^--Q^-_0)\, \Op^n\,,\quad Q_0\,\Op^{n-2} =  (\Delta^--Q^-_0) \,\Op^{n-1} + \Delta^=\,\Op^n\,, \cdots\,.
%\ee

The linearised flow equation \eqref{flowone} was used to derive the first order interactions in refs. \cite{\morrii,first}. It continues to play a very important r\^ole at higher order, as we will see.
Its general solution is a sum over eigenoperators with constant coefficients. These latter are nothing but the associated couplings, which at the linearised level do not run with cutoff scale, $\Lambda$. The eigenoperator equation follows from separation of variables, the RG eigenvalue being the scaling dimension of the coupling. Since we are working perturbatively, thus constructing the eigenoperators  around the free action (Gaussian fixed point), the scaling dimension of the coupling is just its (engineering) mass dimension. Since the eigenoperator equations are of Sturm-Liouville type, any perturbation can be expanded over eigenoperators as a convergent sum (in the square integrable sense) provided that the amplitude dependence is square integrable under the Sturm-Liouville measure. This measure turns out to be:
\be 
\label{measureAll}
\exp \frac{1}{2\Omega_\Lambda}\left(\vp^2- h^2_{\mu\nu} -2\, \bar{c}_{\mu} c_{\mu}\right),
\ee
as determined by the UV regularised tadpole integral: 
\be 
\label{Omega}
\Omega_\Lambda= |\langle \ph(x) \ph(x) \rangle | =  \int_q \frac{{C}(q^2/\Lambda^2)}{q^2} = \frac{\Lambda^2}{2a^2}\,,
\ee
$a\cu>0$ being a dimensionless non-universal constant. Since we need the sum over eigenoperators to converge in order for the Wilsonian RG to make sense \cite{Dietz:2016gzg}  we insist that at sufficiently high scales $\Lambda$, perturbations must lie inside the Hilbert space, $\Lm{}$, defined by the measure \eqref{measureAll}. This can be interpreted as a `quantisation condition' that is thus both natural and necessary for the exact RG. 

The wrong-sign propagator \eqref{pp} leads to the exponentially growing $\vp$ amplitude dependence in \eqref{measureAll} and will thus force all perturbations in $\Ll$ to decay exponentially in $\vp$. This has profound effects on RG properties. While for the graviton and ghosts the eigenoperators are built from Hermite polynomials, justifying the usual expansion in powers of these fields, the eigenoperators for the conformal factor take the form
\be
\label{physical-dnL}
\dd{\Lambda}{n} := \frac{\partial^n}{\partial\vp^n}\, \dd{\Lambda}{0}\,, \qquad{\rm where}\qquad \dd{\Lambda}0 := \frac{1}{\sqrt{2\pi\Omega_\Lambda}}\,\exp\left(-\frac{\vp^2}{2\Omega_\Lambda}\right)
\ee
(integer $n\ge0$). They span the Hilbert space $\Lmm$ defined by the $\vp$ part of the measure \eqref{measureAll}, under which they are also orthonormal.
%Their dimension is $[\dd{\Lambda}{n}]=-1-n$. 
%and thus these relevant. are all relevant, their scaling dimensions being equal to their engineering dimensions in mass units, namely $-1\!-\!n$. 
Since $\Omega_\Lambda\propto \hbar$, the $\dd{\Lambda}{n}$ are non-perturbative in $\hbar$. For this reason we must develop the theory whilst remaining non-perturbative in $\hbar$. Note that the physical operators, gained by sending $\Lambda\to0$, are $\dd{}n$, the $n^\text{th}$-derivatives of the Dirac delta function. 

Writing the linearised flow equation \eqref{flowone} as
\be 
\label{flowoneexpanded}
\dot{\Gamma}_1 = -\frac12\, \dot{\prop}^{\Lambda\,AB} \frac{\partial^2_l}{\partial\Phi^B\partial\Phi^A}\,\Gamma_1\,,
\ee
where $\prop^{\Lambda\,AB} = C^\Lambda \prop^{AB}$ is the UV regulated propagator, the general eigenoperator solution can be seen to be expressed via the appropriate integrating factor, in terms of its physical ($\Lambda\cu\to0$) limit as
\be 
\label{eigenoperatorsol}
 \exp\left(-\frac12 {\prop}^{\Lambda\,AB} \frac{\partial^2_l}{\partial\Phi^B\partial\Phi^A}\right)\, \Gamma_{1\,\text{phys}}\,, \qquad\text{where}\quad \Gamma_{1\,\text{phys}} =\sigma(\partial,\partial\vp,h,c,\Phi^*) \ \delta^{(n)}(\vp)\,.
\ee
Here $\sigma$ is a Lorentz invariant monomial in gauge invariant minimal basis, involving some or all of the components indicated, in particular the arguments $\partial\vp,h,c,\Phi^*$  can appear as they are, or differentiated any number of times, but $\sigma$ cannot depend on the undifferentiated amplitude $\vp$ itself, this being taken care of by the last term. If $d_\sigma=[\sigma]$ is the mass dimension of $\sigma$,  then the dimension of the corresponding eigenoperator is just the sum of the  dimensions, namely $d_\sigma\cu-1\cu-n$. 

%\begin{figure}[ht]
%\centering
%\includegraphics[scale=0.35]{tadpoles.png}
%%\vskip-30pt
%\caption{The eigenoperator is equal to its physical limit
%$\sigma\, \delta^{(n)}(\vp)$, plus all possible tadpole corrections. Those corrections generated by attaching to $\sigma$, terminate eventually (since the monomial will run out of fields), while $\vp$-tadpole corrections to $\delta^{(n)}(\vp)$ go on forever but resum to $\delta_{\, \Lambda}^{\!(n)}\!(\ph)$. We do not draw the external legs, an infinite number of which attach to $\delta_{\, \Lambda}^{\!(n)}\!(\ph)$.} 
%%\phantom{(} caused weird error!
%\label{fig:tadpoles}
%\end{figure}

After mapping to gauge fixed basis \cite{\morrii,first},
\be
\label{gaugeFixed}
%\bar{c}^*_\mu \,|_\text{gf}  &= \bar{c}^*_\mu \,|_\text{gi} + F_\mu\,,\\ \nonumber
H^*_{\mu\nu} \mapsto  H^*_{\mu\nu} +\partial_{(\mu} \bar{c}_{\nu)} -\half\,\delta_{\mu\nu}\, \partial\cdot \bar{c}\,,
%\partial_\alpha\bar{c}_\alpha\,.
\ee
the exponential operator in the eigenoperator solution \eqref{eigenoperatorsol} can be evaluated. It just generates all the Wick contractions for the propagator, as illustrated in fig. \ref{fig:tadpoles}. For each functional derivative in the exponential operator we can write by the Leibniz rule
\be 
\frac{\partial_l}{\partial\Phi^A} = \frac{\partial^L_l}{\partial\Phi^A} + \frac{\partial^R_l}{\partial\Phi^A}\,
\ee
where $\partial^{L}$ acts only on the left-hand factor, here $\sigma$, and $\partial^R$ acts only the right-hand factor, here $\delta^{(n)}(\vp)$. Factoring out $-C^\Lambda$ for later convenience, we see that the exponential factors into three:
\be 
\label{WickTwoFactorId}
\frac12{\prop}^{AB} \frac{\partial_l^2}{\partial\Phi^B\partial\Phi^A} = 
\frac12{\prop}^{AB} \frac{{\partial^{L}_l}^2}{\partial\Phi^B\partial\Phi^A} 
+{\prop}^{AB} \frac{\partial^L_l}{\partial\Phi^B}\frac{\partial^R_l}{\partial\Phi^A} 
+\frac12 {\prop}^{AB} \frac{{\partial^{R}_l}^2}{\partial\Phi^B\partial\Phi^A}\,.
\ee
Since $\delta^{(n)}(\vp)$ only depends on $\vp$, the third exponential collapses to
\be
\label{sumtadpoles}
\exp\!\left(\!-\frac12 {\prop}^{\Lambda\,AB} \frac{{\partial^{R}_l}^2}{\partial\Phi^B\partial\Phi^A}\right) \delta^{(n)}(\vp) = \mathrm{e}^{\frac12\Omega_\Lambda\partial^2_\vp}\, \delta^{(n)}(\vp) %\nn\\
= \partial^n_\vp\int^\infty_{-\infty}\!\! \frac{d\vpi}{2\pi}\, \, \mathrm{e}^{-\frac12\vpi^2\Omega_\Lambda+i\vpi\vp} = \dd\Lambda{n}\,,
\ee
where we used the expression for the $\vp$ propagator \eqref{pp}, giving the tadpole integral \eqref{Omega} and derivatives $\partial_\vp$ with respect to the amplitude (\ie no longer functional), and then expressed the result in conjugate momentum ($\vpi$) space, after which the integral evaluates to the expressions we already gave for the pure-$\vp$ eigenoperators  \eqref{physical-dnL}. Thus the entire eigenoperator can be written as
\be 
\label{eigenoperator}
\exp\left(-{\prop}^{\Lambda\,\vp\vp} \frac{\partial^L}{\partial\vp}\frac{\partial^R}{\partial\vp}\right) \left\{ \exp\left(-\frac12 {\prop}^{\Lambda\,AB} \frac{\partial^2_l}{\partial\Phi^B\partial\Phi^A}\right)\sigma\right\} \dd\Lambda{n}\,,
\ee
where the term in braces expresses all the tadpole corrections acting purely on $\sigma$, in particular for each component of ghost and graviton amplitudes these build the corresponding Hermite polynomials, and the left-most term generates $\vp$-propagator \eqref{pp} tadpole corrections that attach to both $\sigma$ and $\dd\Lambda{n}$ (from the above we see that each such attachment will increase $n\cu\mapsto n\cu+1$). 

Since the operator is relevant as soon as $n\cu>d_\sigma\cu-5$, it follows from \eqref{eigenoperator} that every monomial $\sigma$ is associated to an infinite tower of operators, which can be subsumed into 
%a \emph{coefficient function} $f^{\sigma}_\Lambda(\vp)$ \cite{\morrii} such that 
\be 
\label{firstOrder}
f^{\sigma}_\Lambda(\vp)\, \sigma(\partial,\partial\vp,h,c,\Phi^*)+\cdots = \exp\left(-{\prop}^{\Lambda\,\vp\vp} \frac{\partial^L}{\partial\vp}\frac{\partial^R}{\partial\vp}\right) \left\{ \exp\left(-\frac12 {\prop}^{\Lambda\,AB} \frac{\partial^2_l}{\partial\Phi^B\partial\Phi^A}\right)\sigma\right\} f^{\sigma}_\Lambda(\vp)\,,
\ee
where the ellipses stand for the finite number of tadpoles generated by the exponential operators on the RHS, and the \emph{coefficient function} of the top term is given by
\be 
\label{coefff}
f^\sigma_\Lambda(\vp) = \sum^\infty_{l} g^\sigma_{2l+\varepsilon}\, \dd{\ \Lambda}{2l+\varepsilon} \,.
\ee
Here we have also taken into account that we can specialise to coefficient functions of definite parity \cite{first}, with $\varepsilon=0$ or $1$ according to whether the coefficient function is even or odd. The sum converges for sufficiently high $\Lambda$ such that $f^\sigma_\Lambda\in\Lmm$. At the linearised level, the \emph{underlying couplings} $g^\sigma_{2l+\varepsilon}$ are constant, and the expansion is only over the marginal and relevant eigenoperators, thus the dimensions
\be
\label{gdimension}
[g^\sigma_{2l+\varepsilon}] = 4-(d_\sigma\cu-1\cu-2l-\varepsilon) = 5+2l+\varepsilon-d_\sigma\,,
\ee
must all be non-negative, with those low-$l$ couplings that do not satisfy this, set to zero.

From the first order flow equation \eqref{flowone}, the coefficient function satisfies the linearised flow equation
\be
\label{flowf}
\dot{f}^\sigma_\Lambda(\vp) = \half\,\dot{\Omega}_\Lambda\,  f^{\sigma\prime\prime}_\Lambda(\vp)\,,
\ee
where prime is $\partial_\vp$. 
We define the \emph{amplitude suppression scale} $\Lambda_\sigma\ge0$ to be the smallest scale such that for all $\Lambda\cu>a\Lambda_\sigma$, the coefficient function is inside $\Lmm$. The coefficient function exits $\Lmm$ as $\Lambda$ falls below $a\Lambda_\sigma$, either because it develops singularities after which the flow to the IR ceases to exist, or because  it  decays too slowly at large $\vp$. We need to choose the underlying couplings so that the flow all the way to $\Lambda\to0$ does exist, so that all modes can be integrated over and the physical Legendre effective action can thus be defined. Since the coefficient function thus exits $\Lmm$ by decaying too slowly, we deduce from the Liouville measure \eqref{measureAll} its asymptotic exponential dependence at large $\vp$, as it exits (up to subleading terms):
\be
%\label{asympfLp}
f^\sigma_{a\Lambda_\sigma}(\vp)\sim \mathrm{e}^{-\vp^2/4\Omega_{a\Lambda_\sigma}} = \mathrm{e}^{-\vp^2/2\Lambda_\sigma^2}\,.
\ee
This provides us with a boundary condition for the linearised flow equation  \eqref{flowf}, which then fixes the asymptotic exponential dependence for all $\Lambda$:
\be 
\label{largephi}
f^\sigma_\Lambda(\vp) \sim   \exp\left(-\frac{a^2\vp^2}{\Lambda^2+a^2\Lambda^2_\sigma}\right)\,.
\ee
Setting $\Lambda=0$ shows that the physical coefficient function $f^\sigma_\text{phys}(\vp)$, which following \cite{\morrii} we write simply as $f^\sigma\!(\vp)$, is characterised by the decay:
\be 
\label{largephys}
f^\sigma\!(\vp) \sim \mathrm{e}^{-\vp^2/\Lambda_\sigma^2}\,.
\ee 
This physical behaviour is the reason for calling $\Lambda_\sigma$ an amplitude suppression scale. 

The general solution to the linearised flow equation \eqref{flowf} for the coefficient function, can be given by working in conjugate momentum space:
\be 
\label{fourier-sol}
f_\Lambda^\sigma(\vp) = \int^\infty_{-\infty}\!\frac{d\vpi}{2\pi}\, \ff^\sigma\!(\vpi)\, {\rm e}^{%\exp\left(
-\frac{\vpi^2}{2}\Omega_\Lambda+i\vpi\vp} \,, %\right)\,,
\ee
where $\ff^\sigma$ is $\Lambda$-independent and is thus actually the Fourier transform of the physical $f^\sigma(\vp)$. Remarkably, from the expansion over eigenoperators \eqref{coefff} and the last equality in the sum over tadpoles identity \eqref{sumtadpoles}, we see that the couplings are coefficients of powers of $\vpi$ (rather than powers of $\vp$ as would be the case for a theory with right-sign propagator):
\be 
\label{fourier-expansion}
\ff^\sigma\!(\vpi) = i^\varepsilon \sum_l (-)^l g_{2l+\varepsilon}^\sigma \vpi^{2l+\varepsilon}\,.
\ee
In field-amplitude-space, the couplings are given by moments of the physical coefficient function:\footnote{Notice that this is consistent with the fact that couplings of the wrong parity actually vanish.}
\be 
\label{gnfphys}
g^\sigma_n = \frac{(-)^n}{n!}\int^\infty_{-\infty}\!\!\!\!\!d\vp\,\vp^n\, f^\sigma\!(\vp)\,,
\ee
as can be derived by substituting the Fourier transform and converting $\vp$ to $-i\partial_\vpi$ (see \cite{\morri} for alternative derivations). 

In fact since the linearised flow equation \eqref{flowf} is parabolic in the IR $\to$ UV direction, the solution $f^\sigma_\Lambda(\vp)$ exists for all $\Lambda\cu\ge0$ and is unique, once the physical coefficient function is specified. This latter is subject only to the asymptotic constraint \eqref{largephys} and that its lowest $l$ couplings vanish if their dimensions \eqref{gdimension} are negative. In particular, the asymptotic exponential decay  \eqref{largephys} of the physical coefficient function implies the asymptotic exponential decay \eqref{largephi} at all higher $\Lambda$, and thus as required that $f^\sigma_\Lambda\in\Lmm$ once $\Lambda\cu>a\Lambda_\sigma$. 

The most general linearised solutions for such coefficient functions involve a spectrum of amplitude suppression scales \cite{\morri,first} so that asymptotically the function has subleading parts that decay exponentially at a faster rate than \eqref{largephys}, \ie contain amplitude suppression scales that are smaller than $\Lambda_\sigma$. Rather than working with the most general such coefficient functions, we simplify the analysis by working with linearised solutions that contain only one amplitude suppression scale \cite{first}. Then this asymptotic behaviour in $\vp$-space, \eqref{largephys}, fixes the asymptotic behaviour in $\vpi$-space. For later purposes we write this latter asymptotic relation in terms of a reduced Fourier transform, $\ff^\sigma(\vpi)\propto \bar\ff^\sigma(\vpi^2\Lambda^2_\sigma)$, where $\bar{\ff}^{\sigma}$ which is any dimensionless entire function of a dimensionless argument that asymptotically satisfies at large $\bar{\vpi}\cu=\vpi\Lambda_\sigma$,
\be 
\label{largevpibar}
\bar{\ff}^{\sigma}(\bar{\vpi}^2) \sim {\rm e}^{-\bar{\vpi}^2/4}\,.
\ee

Now notice that as $\Lambda_\sigma\cu\to\infty$ the exponential decay (\ref{largephi},\ref{largephys})  becomes instead a statement that, up to sub-exponential factors, the coefficient function tends to a constant. In refs. \cite{\morrii,first} it was shown that this limit of large amplitude suppression scale (holding everything else fixed) is required to recover BRST invariance. Equivalently this corresponds to taking the limit where $\vp,\Lambda\cu\ll \Lambda_\sigma$, holding $\Lambda_\sigma$ fixed.
In general to recover BRST invariance, we require the physical coefficient function \emph{trivialises} in this limit \cite{\morrii,first}, \ie
\be 
\label{flatphys}
f^\sigma(\vp) \to A_\sigma\,\vp^\alpha\qquad\text{as}\quad\Lambda_\sigma\to\infty\,,
\ee
for some non-negative integer $\alpha$. Note that this determines the parity of the coefficient function. $A_\sigma$ is a constant. From \eqref{flatphys} we read off its dimensions
\be
\label{dimA}
[A_\sigma]=4-d_\sigma-\alpha\,.
\ee 
In the great majority of cases, $\alpha\cu=0$, however if BRST invariance requires appearance of undifferentiated $\vp$, then $\alpha\cu>0$. The trivialisation limit of the physical coefficient function \eqref{flatphys} implies that its Fourier transform must satisfy
\be 
\label{fflatp}
\ff^{\sigma}(\vpi) \to 2\pi A_\sigma\, i^\alpha \delta^{(\alpha)}(\vpi)\qquad\text{as}\quad\Lambda_{\sigma}\to\infty\,,
\ee
understood in the usual distributional sense.
This constraint is satisfied (on finite smooth functions) provided that (for $n\cu\ge0$)
\be 
\label{ffconditions}
\int^\infty_{-\infty}\!\frac{d\vpi}{2\pi}\, \frac{(i\vpi)^n}{n!} \,\ff^{\sigma}(\vpi) \to  A_\sigma\,\delta_{n\alpha} \qquad\text{as}\quad\Lambda_\sigma\to\infty\,.
\ee
Either from \eqref{fflatp} or directly from the limit of the physical coefficient function \eqref{flatphys} and the parabolic property discussed above, we see that the limit at $\Lambda\cu>0$ is uniquely determined to be
\be 
\label{flatp}
f^{\sigma}_\Lambda(\vp) \to A_{\sigma} \left({\Lambda}/{2ia}\right)^\alpha H_\alpha\!\left({ai\vp}/{\Lambda}\right)
\qquad\text{as}\quad\Lambda_{\sigma}\to\infty\,,
\ee
where $H_\alpha$ is the $\alpha^\text{th}$ Hermite polynomial:
\be 
\label{Hermite}
\left({\Lambda}/{2ia}\right)^\alpha H_\alpha\!\left({ai\vp}/{\Lambda}\right) = \vp^\alpha + \alpha(\alpha-1)\,\Omega_\Lambda\vp^{\alpha-2}/2+\cdots\,.
\ee
Now one can see that the general solution for the Fourier transform takes the form \cite{first}
\be 
\label{ffforma}
\ff^{\sigma}(\vpi) = 2\pi\, i^\alpha A_\sigma\,\Lambda^{\alpha+1}_{\sigma}\, \partial^\alpha_{\bar{\vpi}}\left[\bar{\vpi}^{2\bar{n}_{\sigma}}\, \bar{\ff}^{\sigma}(\bar{\vpi}^2)\right]\,,
\ee
where again $\bar\ff^\sigma$ is any entire function satisfying the asymptotic condition \eqref{largevpibar} however now the extra conditions \eqref{ffconditions}, imply that additionally it must satisfy the normalisation constraint:
\be 
\label{normalised}
\int^\infty_{-\infty}\!\!\!\!\!\!d\bar{\vpi}\ \bar{\vpi}^{2\bar{n}_{\sigma}}\, \bar{\ff}^{\sigma_\alpha}(\bar{\vpi}^2)\  =\ 1\,, 
\ee
and the vanishing limits
\be 
\label{vanishes}
\frac1{\Lambda^{2p}_{\sigma}}\int^\infty_{-\infty}\!\!\!\!\!\!d\bar{\vpi}\ \bar{\vpi}^{2(\bar{n}_{\sigma}+p)}\, \bar{\ff}^{\sigma_\alpha}(\bar{\vpi}^2)\ \to \ 0\,, \qquad\text{as}\quad\Lambda_{\sigma}\to\infty\,.
\ee
for any integer $p>0$. Note that these integrals converge for large $\bar{\vpi}$ by virtue of  \eqref{largevpibar}. The constraints \eqref{vanishes} are trivially satisfied if $\bar{\ff}^{\sigma}$ is a finite function independent of $\Lambda_\sigma$, which it is at first order. At second order in perturbation theory, we will find that we need linearised coefficient functions for which $\bar{\ff}^{\sigma}$ depends on $\Lambda_{\sigma}$. In the majority of cases we can choose it to tend to a finite function as $\Lambda_{\sigma}\cu\to\infty$, but exceptionally it will prove useful to allow it to contain terms with coefficients that diverge logarithmically with $\Lambda_\sigma$. Clearly this mild divergence is well within the bounds implied by the vanishing limits \eqref{vanishes}. Finally, $\bar{n}_\sigma$ is just there to ensure that the Taylor expansion \eqref{fourier-expansion} starts at a high enough power such that the low-$l$ irrelevant underlying couplings are missing (see \cite{first} for the precise formula), as they should be at the linearised level. 

Since (for fixed $\Lambda_\sigma$) the reduced Fourier transform $\bar\ff^\sigma$ is any normalised \eqref{normalised} entire function satisfying the asymptotic condition \eqref{largevpibar} we still have an infinite dimensional function space of solutions. The underlying couplings are thus very weakly constrained. Indeed, the asymptotic condition \eqref{largevpibar} translates, via the Taylor expansion formula \eqref{fourier-expansion},  into only an asymptotic constraint on the large-$n$ behaviour of the couplings \cite{first}:
\be 
\label{largeg}
g^\sigma_n \sim A_\sigma \left(\frac{\mathrm{e}}{2n}\right)^{\frac{n}{2}}\! \Lambda^{n+1}_\sigma\,\qquad\text{as}\quad n\to\infty\,.
\ee
In particular note that the trivialisation property \eqref{flatp} does not require specific values for any of the underlying couplings, but is rather a universal result that follows in the large amplitude suppression scale limit for infinitely many sets of couplings that satisfy \eqref{largeg} asymptotically.\footnote{Note that \eqref{largeg} implies that large-$n$ couplings diverge in this limit, even though the coefficient function remains finite. Alternatively one can scale $A_\sigma$ in this limit to keep the couplings finite in this asymptotic expansion \cite{\morrii,first}.}

Substituting the general solution for the Fourier transform \eqref{ffforma} into the Fourier transform formula for the linearised coefficient function, \eqref{fourier-sol}, one can derive more refined trivialisation limits than \eqref{flatp} \cite{first}. In particular the approach to trivialisation is characterised by Taylor series corrections in $\Lambda^2/\Lambda^2_\sigma$ and $\vp^2/\Lambda^2_\sigma$, except for those cases at second order where these corrections will also include a single factor of $\ln(\Lambda_{\sigma})$. Thus for large $\Lambda_{\sigma}$,
\beal 
\label{regularity}
\partial^p_\vp \left[ f^{\sigma}_\Lambda(\vp) - A_{\sigma} \left({\Lambda}/{2ia}\right)^\alpha H_\alpha\!\left({ai\vp}/{\Lambda}\right) \right]\ &=\ O(1/\Lambda^2_{\sigma}) \qquad &\text{for}\quad p\le\alpha\,,\nn\\
 \partial^p_\vp f^{\sigma}_\Lambda(\vp)\ &=\ O(1/\Lambda_{\sigma}^{2\lceil\tfrac{p-\alpha}2\rceil})\qquad &\text{for}\quad p>\alpha\,, %\qquad\text{as}\quad\to\infty\,.
\eeal
\eqref{flatp} being the $p=0$ case, where the RHS is corrected by a factor of $\ln(\Lambda_{\sigma})$ in some cases at second order. 

At first order \cite{first}, we further specialised to keeping just two coefficient functions, $f^1_\Lambda$ and $f^{1_1}_\Lambda$, of positive and negative parity respectively, with their amplitude suppression scales set equal to a common scale, $\Lambda_\sigma\cu=\Lambda_\p$. Further restricting the parametrisation in this way still leaves us with infinite dimensional function spaces, each parametrised by an infinite number of freely variable underlying couplings, so represents a mild restriction on testing universality \cite{first}. We will find that at second order we can continue to ensure that the amplitude suppression scales are all identified with the one scale, $\Lambda_\p$. Since the amplitude suppression scale is sent to infinity, this amounts to a simplification of the limiting process where otherwise parts are sent to this limit independently.

In terms of these coefficient functions, the first order vertices are given by the sum of three contributions with definite antighost number. At antighost level two, we have
\be 
\label{Gammaonetwo}
\Gamma^2_1 = -c_\nu\,\partial_\nu c_\mu\, c^*_\mu\, f^1_\Lambda(\vp)\,,
\ee
at antighost level one:
\be 
\label{Gammaoneone}
\Gamma^1_1 = - \left(c_\alpha \partial_\alpha H_{\mu\nu} + 2\, \partial_\mu c_\alpha h_{\alpha\nu}\right) H^*_{\mu\nu}\, f^1_\Lambda(\vp) -\partial_\mu c_\nu H^*_{\mu\nu}\, f^{1_1}_\Lambda (\vp)\,,
\ee
and at antighost level zero:
\besp 
\label{Gammaonezero}
\Gamma^0_1 = \Big( 
\frac14h_{\alpha\beta}\partial_\alpha\vp\partial_\beta\vp
-h_{\alpha\beta}\partial_\gamma h_{\gamma\alpha}\partial_\beta\vp
-\frac12 h_{\gamma\delta}\partial_\gamma h_{\alpha\beta}\partial_\delta h_{\alpha\beta}
-h_{\beta\mu}\partial_\gamma h_{\alpha\beta}\partial_\gamma h_{\alpha\mu} \\
+2h_{\mu\alpha}\partial_\gamma h_{\alpha\beta}\partial_\mu h_{\beta\gamma} 
+h_{\beta\mu}\partial_\gamma h_{\alpha\beta}\partial_\alpha h_{\gamma\mu}
-h_{\alpha\beta}\partial_\gamma h_{\alpha\beta} \partial_\mu h_{\mu\gamma}
+\frac12h_{\alpha\beta}\partial_\gamma h_{\alpha\beta} \partial_\gamma \vp \Big) f^1_\Lambda \\
+\left( \frac38(\partial_\alpha\vp)^2
-\frac12\partial_\beta h_{\beta\alpha}\partial_\alpha\vp
-\frac14(\partial_\gamma h_{\alpha\beta})^2 
+\frac12 \partial_\gamma h_{\alpha\beta}\partial_\alpha h_{\gamma\beta} \right) f^{1_1}_\Lambda
+\frac72 b \Lambda^4 f^{1_1}_\Lambda \,.
\eesp
Expanding the coefficients over the $\delta$-operators as in \eqref{coefff}, gives
\be 
\label{coeffone}
f^1_\Lambda(\vp) = \sum^\infty_{l=0} \,g^1_{2l}\, \dd\Lambda{2l}\,, \qquad f^{1_1}_\Lambda(\vp) = \sum^\infty_{l=0}\, g^{1_1}_{2l+1}\, \dd\Lambda{2l+1}
\ee
these sums converging (in the square integrable sense) for $\Lambda>a\Lambda_\p$, as a consequence of the asymptotic condition \eqref{largeg} on the underlying couplings. Here the sums are unrestricted since, by the dimension formula \eqref{gdimension},  the couplings have dimension:
\be 
\label{gonedimensions}
[g^1_{2l}] = 2l\,,\qquad [g^{1_1}_{2l+1}] = 2l+2\,.
\ee
In particular, all are relevant except $g^1_0$, which is marginal. We will see in sec. \ref{sec:fullrt} that at second order in perturbation theory, these couplings remain independent of $\Lambda$, \ie do not run, although they will run for the first time at third order. In particular this means that to second order as we work in this paper, $g^1_0$ continues to behave as though it is exactly marginal \cite{first}, parametrising a line of fixed points that includes the Gaussian ($g^1_0\cu=0$) one.

The coefficient functions have the trivialisation limits of the  form \eqref{flatp} with $\alpha\cu=0,1$:
\be 
\label{fonelimit}
f^1_\Lambda(\vp) \to \kappa\,,\qquad f^{1_1}_\Lambda(\vp)\to\kappa\,\vp\,,
\qquad\text{as}\quad\Lambda_\p\to\infty\,,
\ee
where the refined regularity properties \eqref{regularity} also apply, in particular in these cases the limits are reached at least as fast as $1/\Lambda_\p^2$. For the first time, Newton's constant $G$ makes its appearance. It does so through the proportionality constant $A_\sigma\cu=\kappa=\sqrt{32\pi G}$, as a collective effect of the underlying couplings, as encoded by the common proportionality constant in their asymptotic behaviour \eqref{largeg}. Together with the monomials $\sigma$ specified in (\ref{Gammaonetwo},\ref{Gammaoneone},\ref{Gammaonezero}), the first order vertices have the property that
\be 
\label{Gammaonelimits}
\Gamma^n_1\to \kappa\,\cG^n_1\,,\qquad \text{as}\quad \Lambda_\p\to\infty\,,
%\Gamma^2_1 \to \kappa\,\cG^2_1\,, \qquad \Gamma^1_1\to\kappa\,\cG^1_1\,,\qquad\Gamma^0_1\to \kappa\,\cG^0_1+\half\kappa\, b\Lambda^4\vp\,,\qquad \text{as}\quad \Lambda_\p\to\infty\,.
\ee
where $\cG^n_1$ are the antighost level parts of the non-trivial quantum BRST cohomology representative \cite{\morrii,first}. They correspond to expressions for the first order vertices in standard (polynomial) quantisation together with a one-loop tadpole correction (the last term in \eqref{Gammaonezero}), as required to solve the first order flow equation \eqref{flowone} and mST \eqref{mSTone} in standard quantisation. 

In terms of the general solution \eqref{ffforma} for their Fourier transform,  
$\ff^1$ takes the $\alpha\cu=\bar{n}\cu=0$ form:
\be 
\label{ffsigmaEg}
\ff^1(\vpi) = 2\pi A_\sigma\,\Lambda_\sigma\, \bar{\ff}^1(\bar{\vpi}^2)\,,
\ee
where $A_\sigma\cu=\kappa$ and $\Lambda_\sigma\cu=\Lambda_\p$, and $\bar{\ff}^1$ is the reduced Fourier transform with limiting behaviour \eqref{largevpibar} at large $\bar\vpi$, and
satisfying the normalisation condition \eqref{normalised} with $\bar{n}_\sigma=0$. Similarly $\ff^{1_1}$ is expressed through its own reduced Fourier transform as $\ff^{1_1}(\vpi)= 2\pi i\kappa\Lambda_\p^2\,\partial_{\bar{\vpi}}\bar{\ff}^{1_1}(\bar{\vpi}^2)$. 

Although in the following, we will deal with the most general coefficient functions satisfying these properties, it is helpful for interpretation to refer to some simple examples \cite{first}. For a coefficient function $f^\sigma_\Lambda$ having the same properties as $f^1_\Lambda$ (\ie $\alpha\cu=0$,  and all couplings switched on so $\bar{n}_\sigma\cu=0$), if we set its reduced Fourier transform for simplicity to be equal to the RHS of \eqref{largevpibar} for all $\bar\vpi$, then 
the normalisation condition  \eqref{normalised} implies
\be 
\label{ffoneEg}
\bar{\ff}^\sigma(\bar{\vpi}^2) =  \frac{{\rm e}^{-\bar{\vpi}^2/4}}{2\sqrt{\pi}}\,,
\ee
which using \eqref{ffsigmaEg} gives us the simplest example used previously \cite{\morri,\morrii}:
\be 
\label{foneEg}
f^\sigma_\Lambda(\vp) = \frac{a A_\sigma\Lambda_\sigma}{\sqrt{\Lambda^2+a^2\Lambda^2_\sigma}}\, {\rm e}^{-\frac{a^2\vp^2}{\Lambda^2+a^2\Lambda^2_\sigma}}\,,\quad f^{\sigma}\!(\vp) = A_\sigma\, {\rm e}^{-{\vp^2}/{\Lambda_\sigma^2}}\,, \quad g^\sigma_{2l} =\frac{\sqrt{\pi}}{l!4^l}\, A_\sigma\,\Lambda_\sigma^{2l+1}
\ee
($l=0,1,\cdots$). Here the first expression follows from performing the Fourier integral in \eqref{fourier-sol}, the second is its $\Lambda\cu\to0$ limit, and the couplings follow from the Taylor expansion relation \eqref{fourier-expansion}. To switch off the first coupling $g^\sigma_0=0$ we set instead  $\bar{n}_\sigma=1$ in the general formula \eqref{ffforma}. If we take its reduced Fourier transform $\check{\bar{\ff}}^\sigma$ to again be equal to its asymptotic limit, the normalisation condition \eqref{normalised} now implies $\check{\bar{\ff}}^\sigma = \bar{\ff}^\sigma/2$, where $\bar{\ff}^\sigma$ is our previous example \eqref{ffoneEg}, so that the Fourier transform $\check{\ff}^\sigma(\vpi)$ and couplings $\check{g}^\sigma_{2l}$ now take the form
\be 
\label{ffcheckEg}
\check{\ff}^\sigma(\vpi) = \pi A_\sigma\,\Lambda_\sigma\, \bar{\vpi}^2\,\bar{\ff}^\sigma(\bar{\vpi}^2)\,,\qquad
\check{g}^\sigma_{2l} =-\frac{2\sqrt{\pi}}{(l\cu-1)!\,4^l}\, A_\sigma\,\Lambda_\sigma^{2l+1} %\quad (n=1,2,\cdots)\,,
\ee
($\check{g}^\sigma_0\cu=0$, the non-vanishing couplings taking $l=1,2,\cdots$).
Performing the Fourier integral gives
\be
\label{fhi2}
\check{f}^\sigma_\Lambda(\vp) = \frac{a^3\Lambda_{\sigma}^3 A_\sigma}{\left(\Lambda^2+a^2\Lambda_{\sigma}^2\right)^{3/2}}\left(1-\frac{2a^2\vp^2}{\vphantom{\tilde\Lambda}\Lambda^2+a^2\Lambda_{\sigma}^2}\right) {\rm e}^{-\frac{a^2\vp^2}{\vphantom{\tilde\Lambda}\Lambda^2+a^2\Lambda_{\sigma}^2}}\,,\quad 
\check{f}^{\sigma}\!(\vp) = A_\sigma\, \left(1-\frac{2\vp^2}{\Lambda^2_\sigma}\right) {\rm e}^{-{\vp^2}/{\Lambda_\sigma^2}}\,,
\ee
which one sees explicitly still satisfies the same $\alpha\cu=0$ trivialisation limit (\ref{flatphys},\ref{flatp}) as before.

\section{Pointwise versus uniform convergence}
\label{sec:fixcouplings}

We have seen that it is possible to zero any number of couplings, namely the irrelevant low-$l$ $g^{\sigma}_{2l+\varepsilon}$, and still satisfy the desired pointwise trivialisation limits (\ref{flatp}) for the coefficient function \cite{\morrii}. In fact we also have the flexibility to choose at will any finite number of the remaining constituent couplings. This property will prove crucial above first order in perturbation theory. The reason why this is possible is because the couplings are given by an integral \eqref{gnfphys} over the physical coefficient function.\footnote{The same comments apply to the corresponding formula at finite $\Lambda$ which can be found in ref. \cite{\morri}.} The key then is to recognise the difference between point-wise and uniform convergence. There are again infinitely many solutions. 
Suppose we want to fix the first $\nf\cu+1$ couplings,
\be
\label{fixedcouplings}
 g^{\sigma}_{\varepsilon}\,,\ g^{\sigma}_{2+\varepsilon}\,,\ \cdots\,,\ g^{\sigma}_{2\nf+\varepsilon}
\ee  
(recall below \eqref{coefff} that $\varepsilon$ is fixed by parity) to some desired values, or in general to some desired functions of the amplitude suppression scale.  Clearly this subsumes the previous case where we required the low-order couplings just to vanish if they are irrelevant.  As we will justify shortly, all we need is to include $\nf\cu+\lfloor\tfrac{\alpha}2\rfloor\cu+2$ parameters in some sensible way into $\ff^{\sigma}$.
For example, pulling out the required dimensions we can set
\be 
\label{ffPoalpha}
\ff^{\sigma}(\vpi) = 2\pi\,  A_\sigma\,\Lambda^{\alpha+1}_{\sigma} (i\bar{\vpi})^{\varepsilon}\, \Po(\bar{\vpi}^2)\, \bar{\ff}^{\sigma}(\bar{\vpi}^2)\,,
\ee
where 
\be 
\Po(\bar{\vpi}^2) = \sum_{r=0}^{\nf\cu+\lfloor\tfrac{\alpha}2\rfloor\cu+1}\!\! p_r \,\bar{\vpi}^{2r}
\ee
is a polynomial containing the required number of parameters $p_r$,
and the new reduced Fourier transform $\bar{\ff}^{\sigma}$ is some fixed dimensionless entire function satisfying the asymptotic constraint \eqref{largevpibar}. Then the fixed couplings \eqref{fixedcouplings} via the Taylor series \eqref{fourier-expansion}, provide $\nf\cu+1$ constraints, while the trivialisation conditions \eqref{ffconditions} provide the other $\lfloor\tfrac{\alpha}2\rfloor\cu+1$, together with convergence conditions analogous to the vanishing limit conditions \eqref{vanishes}:\footnote{Note that $2\lceil\tfrac{\alpha}2\rceil = \alpha+\varepsilon$.}
\beal 
\int^\infty_{-\infty}\!\!\!\!\!\!d\bar{\vpi}\ \bar{\vpi}^{2(n+\varepsilon)}\, \Po(\bar{\vpi}^2)\, \bar{\ff}^{\sigma}(\bar{\vpi}^2)\  &=\ 0\,, \qquad n=0,1,\cdots,\lfloor\tfrac{\alpha}2\rfloor\cu-1\,,\label{constraintPoalpha}\\ 
\int^\infty_{-\infty}\!\!\!\!\!\!d\bar{\vpi}\ \bar{\vpi}^{2\lceil\tfrac{\alpha}2\rceil}\, \Po(\bar{\vpi}^2)\, \bar{\ff}^{\sigma}(\bar{\vpi}^2)\  &=\ (-)^{\lceil\tfrac{\alpha}2\rceil} \alpha!\,,\label{normalisedPoalpha}\\
\frac1{\Lambda^{2p}_{\sigma}}\int^\infty_{-\infty}\!\!\!\!\!\!d\bar{\vpi}\ \bar{\vpi}^{2\lceil\tfrac{\alpha}2\rceil+2p}\, \Po(\bar{\vpi}^2)\, \bar{\ff}^{\sigma}(\bar{\vpi}^2)\  &\to\ 0\,, \qquad\text{as}\quad\Lambda_{\sigma}\to\infty\,, \label{vanishesPoalpha}
\eeal
(integer $p\cu>0$).
Note that if we do choose the fixed couplings \eqref{fixedcouplings} to vanish when they are irrelevant, this will fix the polynomial $\Po$'s first non-vanishing power, while the $\lfloor\tfrac{\alpha}2\rfloor$ constraints \eqref{constraintPoalpha} guarantee that the polynomial parametrisation \eqref{ffPoalpha} can then be recast into the earlier general form \eqref{ffforma}.
From the polynomial parametrisation \eqref{ffPoalpha} and the Taylor expansion formula \eqref{fourier-expansion}, we have that the $p_r$ actually depend linearly on the dimensionless ratios:
\be 
\label{barg}
\bar{g}^{\sigma}_{2l+\varepsilon} =
\frac{g^{\sigma}_{2l+\varepsilon}}{A_\sigma\,\Lambda^{2l+\varepsilon+\alpha+1}_\sigma}\,.
\ee
($l\cu=0,1,\cu\cdots,\nf$. Recall dimensions are set for $A_\sigma$ by \eqref{dimA} and for the couplings by \eqref{gdimension}.) We see that the convergence conditions \eqref{vanishesPoalpha} are met provided only that these ratios diverge slower than $\Lambda^2_\sigma$. It is these ratios that will %be required to 
%in fact 
tend to a finite limit as $\Lambda_{\sigma}\cu\to\infty$ in the majority of cases, or exceptionally diverge as $\ln(\Lambda_{\sigma})$.
%$\Lambda_\sigma$ or
%at worst as $\Lambda_\sigma\ln(\Lambda_{\sigma})$.  
The properties we recalled at the end of sec. \ref{sec:prelim} then apply, in particular the refined limits \eqref{regularity}.

As an example we set $\alpha\cu=0$,  and use again the simplest choice for the reduced Fourier transform \eqref{ffoneEg}. Dividing through by $\bar{\ff}^\sigma$ one readily derives from the polynomial parametrisation \eqref{ffPoalpha} and the Taylor expansion formula \eqref{fourier-expansion} that
\be 
\label{pr}
p_r = \frac1{\sqrt{\pi}}\sum_{s=0}^r\frac{(-)^s}{(r-s)!\,4^{r-s}}\,\bar{g}^\sigma_{2s}\,,\qquad r=0,1,\cdots,\nf\,,
\ee
and from the new normalisation condition \eqref{normalisedPoalpha},
\be 
p_{\nf+1} = \frac1{(2\nf+1)!!\,2^{\nf+1}}\left(1-\sum_{r=0}^\nf  (2r-1)!!\, 2^rp_r\right)\,.
\ee
If we fix just $g^\sigma_0$ then the polynomial is
\be 
\label{Pofix0}
\Po(\bar{\vpi}^2) = \frac{\bar{\vpi}^2}{2}+\frac{\bar{g}^\sigma_0}{\sqrt{\pi}} \left(1-\frac{\bar{\vpi}^2}{2}\right)\,,
\ee
and if we fix also $g^\sigma_2$ then
\be 
\Po(\bar{\vpi}^2) = \frac{\bar{\vpi}^2}{12}+\frac{\bar{g}^\sigma_0}{\sqrt{\pi}} \left(1+\frac{\bar{\vpi}^2}{4}-\frac{\bar{\vpi}^4}{8}\right)-\frac{\bar{g}^\sigma_2}{\sqrt{\pi}}\left(\bar{\vpi}^2-\frac{\bar{\vpi}^4}{6}\right) \,.
\ee
Combining with the rest of the polynomial parametrisation \eqref{ffPoalpha}, the reduced couplings \eqref{barg}, reduced Fourier transform \eqref{ffoneEg} and the Taylor expansion formula \eqref{fourier-expansion}, one easily verifies by inspection that the respective couplings are fixed as desired. Fixing just $g^\sigma_0$,
so using \eqref{Pofix0} in the rest of the polynomial parametrisation \eqref{ffPoalpha}, and comparing to the general form \eqref{ffsigmaEg} and \eqref{ffcheckEg} for the examples given at the end of sec. \ref{sec:prelim}, we see that the corresponding coefficient function $\mathring{f}^{\sigma}_\Lambda$ is just a linear combination of the previous example solutions \eqref{foneEg} and \eqref{fhi2}:
\be 
\label{pointvuniformEg}
\mathring{f}^{\sigma}_\Lambda(\vp)  = \check{f}^{{\sigma}}_\Lambda(\vp) + \frac{\bar{g}^{\sigma}_0}{\sqrt{\pi}} \left[ f^\sigma_\Lambda(\vp) -\check{f}^{{\sigma}}_\Lambda(\vp)\right]\,,\quad
 \mathring{g}^\sigma_{2n} = -\frac{2\sqrt{\pi}}{(n\cu-1)!4^n}\, A_\sigma\,\Lambda_\sigma^{2n+1} + \frac{2n+1}{n!4^n}\,g^\sigma_0\Lambda^{2n}_\sigma\,.
\ee
Its properties are readily visible from the first formula. It has coupling $\mathring{g}^\sigma_0\cu=g^\sigma_0$ since $\check{f}^{{\sigma}}_\Lambda(\vp)$ has vanishing integral over $\vp$, while $f^\sigma_\Lambda(\vp)$ has integral $\sqrt{\pi}A_\sigma\Lambda_\sigma$ as follows from the moment formula \eqref{gnfphys} and its couplings \eqref{foneEg}. Since $\check{f}^{{\sigma}}_\Lambda(\vp)$ and $f^\sigma_\Lambda(\vp)$ have the same point-wise trivialisation limit \eqref{flatp}, so does $\mathring{f}^{\sigma}_\Lambda(\vp)$. In particular comparing their explicit formulae \eqref{foneEg} and \eqref{fhi2}, we see that the part of the approach to the limit  that is proportional to $\bar{g}^\sigma_0$, indeed goes as $O(1/\Lambda^2_\sigma)$ in agreement with the new vanishing conditions \eqref{vanishesPoalpha}.

%\section{Second order}
%\label{sec:second}

\section{Second order}
\label{sec:second}

At second order in the perturbative expansion \eqref{expansion}, the flow equation \eqref{flow},  and mST \eqref{mST}, become
\beal 
\label{flowtwo}
\dot{\Gamma}_2 - \half\, \text{Str}\, \dot{\prop}_\Lambda \Gamma^{(2)}_2 &= 
- \half\, \text{Str}\, \dot{\prop}_\Lambda \Gamma^{(2)}_1 \propH \Gamma^{(2)}_1 \,, \\
\label{mSTtwo}
 \hs_0\,\Gamma_2 &= -  \half\,(\Gamma_1,\Gamma_1) -\text{Tr}\,C^\Lambda\, \Gamma^{(2)}_{1*}\propH \Gamma^{(2)}_1 \,.
%\left( \!C_\Lambda\,  \Gamma^{(2)}_{1*}\propH \Gamma^{(2)}_1  \right) \,.
\eeal
Again the strategy is to first construct the continuum limit, \ie  solutions to \eqref{flowtwo} that realise the full renormalized trajectory $\Lambda\cu\ge0$, and then by appropriate choice of the solutions for the corresponding coefficient functions, arrange to satisfy \eqref{mSTtwo} in the limit of large amplitude suppression scale $\Lambda_\p$, or what is the same, in limit that  $\Lambda$  and $\vp$ are much less than this scale.

Although these equations 
are second order in perturbative expansion \eqref{expansion} they are, as required, non-perturbative in $\hbar$. Initial explorations of such second order computations were made in the $\vp$-sector  in ref. \cite{\morri} both in terms of a standard treatment involving resumming `melonic' \cite{Groote:1998ic} Feynman diagrams  to all loops,  as illustrated in fig. \ref{fig:melons}, and through direct solution of the flow equation (the diagrams of fig.\ref{fig:melons} can also be derived by iterating \eqref{flowtwo} perturbatively in $\hbar$).
%however in both cases without regard to BRST invariance. 
%Here we will gain further insight into the difficulties encountered in trying to pursue a more standard treatment. In ref. \cite{\morri} we found instead that solving \eqref{flowtwo} directly for the continuum solution provides the more powerful route. We will see that it is this approach that can naturally be adapted and developed into a complete solution of the above equations. 
%Furthermore this avoids the problems that arise from the impossibility of imposing \eqref{mSTtwo} on a local bare action (\ie one with finitely many space-time derivatives) at some initial UV scale $\Lambda\cu=\Lambda_0$ \cite{Becchi:1996an,Igarashi:2019gkm}. %Ouch- Rubbish! mST never satisfied there in this quantisation!

As reviewed in sec. \ref{sec:prelim}, we need solutions that have a derivative expansion.
As at first order \cite{\morrii,first}, we construct such solutions by starting at the largest antighost level and then working downwards. The non-linear terms in the second order equations (\ref{flowtwo},\ref{mSTtwo}) contribute a maximum antighost number as determined by the solution $\Gamma_1$ recalled in sec. \ref{sec:prelim}. At first sight we could consider solutions for $\Gamma_2$ that have greater antighost number than this, but the parts with greater antighost number would have to satisfy just the linearised equations. Solutions to these latter are just solutions to the first order equations  (\ref{flowone},\ref{mSTone}), which we have fixed already via our choice of non-trivial quantum BRST cohomology representative $\cG_1$. 
Therefore we can restrict the solution to have the maximum antighost number generated by the RHS of the above equations. 

By inspection, this is antighost number four. At this antighost level only the flow equation contributes, by attaching $\vp$ propagators \eqref{pp} from one coefficient function in the antighost level two first-order vertex \eqref{Gammaonetwo} to another copy. Thus the RHS of the flow equation \eqref{flowtwo} reads:
\be 
\label{twofourRHS}
- \half\, \text{Str}\, \dot{\prop}_\Lambda \Gamma^{(2)}_1 \propH \Gamma^{(2)}_1\, \Big|^{4} =
\left(c^*_\mu c_\alpha\partial_\alpha c_\mu f^{1\prime\prime}_\Lambda\right) 
D(-\Box/\Lambda^2) \left(c^*_\nu c_\beta\partial_\beta c_\nu f^{1\prime\prime}_\Lambda\right)\,,
\ee
where $|^4$ means the antighost-four part, and
\be 
\label{Dp}
D(p^2/\Lambda^2) = \frac12 \int_q \frac{\dot{C}^\Lambda(q)\,C_\Lambda(q\cu+p)}{q^2\,(q\cu+p)^2} %\,,
\ee
is well defined, dimensionless, and has a derivative expansion:
\be 
\label{D}
D(-\Box/\Lambda^2) = \sum^\infty_{m=0} \frac{D_m}{m!} \left(-\frac{\Box}{\,\Lambda^2}\right)^m\,.
\ee
The $D_m$ are thus non-universal numbers, apart from the
lowest term that happens to be universal:
\be 
\label{Dzero}
D_0 = -\frac{1}{32\pi^2} \int^\infty_0\!\!\!\!\!\! dq\,\, \frac{\partial}{\partial q} C_\Lambda^2(q) = -\frac{1}{32\pi^2} \,.
\ee
Taking this lowest order term as an example, it implies that the second order antighost level-four contribution  $\Gamma^4_2$ must contain a vertex
\be 
\label{twofour}
\sigma_0\, f^{\sigma_0}_\Lambda(\vp) = \left(c^*_\mu c_\alpha\partial_\alpha c_\mu\right)^2 f^{\sigma_0}_\Lambda(\vp)\,.
%\qquad\text{where}\qquad \sigma_0 = \left(c^*_\mu c_\alpha\partial_\alpha c_\mu\right)^2 \,,
\ee
Since there is no possibility of attaching tadpoles to $\sigma_0$, there are no such terms generated by the LHS of the second order flow equation \eqref{flowtwo}. %(\cf the ellipses in \eqref{firstOrder}). 
However, expanding out the action of the rest of the $D$ operator \eqref{D} on the RHS of the level-four expression \eqref{twofourRHS}, generates a sum over infinitely many other higher-derivative monomials $\sigma'$ which thus also correspond to vertices in $\Gamma^4_2$. They include cases where the derivatives hit the second $f^{1\prime\prime}_\Lambda$ generating space-time differentiated conformal factors $\sim\partial^s\vp$. 
Infinitely many of these $\sigma' f^{\sigma'}_\Lambda$ do have ($\vp$-)tadpole corrections generated by the LHS of the second order flow equation \eqref{flowtwo}, and out of these, infinitely many result in the remaining monomial being $\sigma_0$ again. This would not matter if the general form of the linearised solution \eqref{firstOrder} still correctly subsumed these tadpole corrections, but since the second order flow equation \eqref{flowtwo} is non-linear, even if we package them up using \eqref{firstOrder} we are left with a remainder. Thus the LHS of \eqref{flowtwo} leads to an open equation such that $f^{\sigma_0}_\Lambda$ picks up an infinite series of tadpole corrections from the higher-derivative $\sigma'$ and their associated coefficient functions $f^{\sigma'}_\Lambda$. Together with the flow of the coefficient functions themselves, these corrections reconstruct the melonic diagrams in fig. \ref{fig:melons}, as we will see explicitly when we set out precisely the form of these equations in \eqref{open} and show how to eliminate this ``tadpole cross-talk'' in sec. \ref{sec:fullrt}.

\subsection{A model for the renormalized trajectory}
\label{sec:model}

In this subsection,  and later also in sec. \ref{sec:largeASSderiv0}, we study a simple model. To get this model we just discard
%now we just drop 
all these higher-derivative tadpole corrections and thus take
\be 
\label{flowftwofour}
\dot{f}^{\sigma_0}_\Lambda(\vp) = \half\,\dot{\Omega}_\Lambda\,  f^{\sigma_0\prime\prime}_\Lambda + D_0 \left(f^{1\prime\prime}_\Lambda\right)^2\,,
\ee
as the flow equation for $f^{\sigma_0}_\Lambda$. Despite the severity of the truncation, and despite the fact that ultimately this antighost level will anyway not survive imposing the second-order mST \eqref{mSTtwo} in the ensuing limit of large amplitude suppression scale, we will gain powerful intuition from studying this model. We thus analyse the continuum limit solution to this equation in some detail.

%We will now analyse the continuum limit solution to this equation in some detail.

%We concentrate first on the unique term with the lowest number of derivatives. We see that it implies that $\Gamma^4_2$ must have a vertex
%\be 
%%\label{twofour}
%\sigma_0\, f^{\sigma_0}_\Lambda(\vp) = \left(c^*_\mu c_\alpha\partial_\alpha c_\mu\right)^2 f^{\sigma_0}_\Lambda(\vp)\,,
%%\qquad\text{where}\qquad \sigma_0 = \left(c^*_\mu c_\alpha\partial_\alpha c_\mu\right)^2 \,,
%\ee
%%where the monomial $\sigma_0$ is defined in \eqref{sigma0} 
%where the coefficient function satisfies
%\be 
%\label{flowftwofour}
%\dot{f}^{\sigma_0}_\Lambda(\vp) = \half\,\dot{\Omega}_\Lambda\,  f^{\sigma_0\prime\prime}_\Lambda + D_0 \left(f^{1\prime\prime}_\Lambda\right)^2\,.
%\ee
%We will now analyse the continuum limit solution to this equation in some detail, and then generalise this to the rest of the derivative expansion.

Given the symmetry of $f^1_\Lambda$ we see that $f^{\sigma_0}_\Lambda$ must also be symmetric, since we insist that the coefficient functions have definite parity. By the quantisation condition, see below \eqref{measureAll}, $f^{\sigma_0}_\Lambda$ must have an amplitude suppression scale $\Lambda_{\sigma_0}$.
According to its definition, for $\Lambda\cu>a\Lambda_{\sigma_0}$ we thus have that $f^{\sigma_0}_\Lambda$ is an expansion over the operators $\dd\Lambda{2l}$ with corresponding couplings $g^{\sigma_0}_{2l}$. This already corresponds to expanding the level-four vertex \eqref{twofour}  over eigenoperators since there is no other opportunity to attach tadpoles. Since the homogeneous part of the above flow equation \eqref{flowftwofour} coincides with the linearised flow equation \eqref{flowf}, if we only had this part the couplings would be constant. The  inhomogeneous term however induces these couplings to run. Furthermore irrelevant operators are generated, whose couplings should not be freely variable but fixed by  exactly marginal and (marginally) relevant couplings in the continuum limit. Therefore we have
\be 
\label{coefftwofour}
f^{\sigma_0}_\Lambda(\vp) = \sum^\infty_{l=0} g^{\sigma_0}_{2l}\!(\Lambda)\,\dd\Lambda{2l}\,,
\ee
where by definition this sum converges for $\Lambda\cu>a\Lambda_{\sigma_0}$. 
%which will be true for $\Lambda>a\Lambda_{\sigma_0}$, where $\Lambda_{\sigma_0}$ is an amplitude suppression scale to be determined shortly. 
By dimensions \eqref{gdimension} since $d_{\sigma_0}\cu=10$, we have $[g^{\sigma_0}_{2l}] = 2l\cu-5$, and thus
$g^{\sigma_0}_{0}$, $g^{\sigma_0}_{2}$ and  $g^{\sigma_0}_{4}$ are irrelevant while all the rest are relevant. 

Using the asymptotic behaviour \eqref{largephi}, together with the square-integrability constraint under the measure \eqref{measureAll},
a little algebra establishes that $\left(f^{1\prime\prime}_\Lambda\right)^2\in\Lmm$ for $\Lambda > a\Lambda_\p/\sqrt{3}$. From our model flow equation \eqref{flowftwofour} we see therefore that the new amplitude suppression scale must satisfy
\be
\label{Lambdatwofour}
\Lambda_{\sigma_0}\ge \Lambda_\p/\sqrt{3}\,, 
\ee
since only then can $f^{\sigma_0}_\Lambda(\vp)\in\Lmm$ for $\Lambda\cu>a\Lambda_{\sigma_0}$.
In this regime we just have that $\dot{g}^{\sigma_0}_{2l}$ is given by the coefficient of $\dd\Lambda{2l}$ in the inhomogeneous term. Using the fact that $\delta_{\!\phantom{(} \Lambda}^{\!(m)\, \prime\prime}\!(\ph) = \dd\Lambda{m+2}$, as is evident from their definition \eqref{physical-dnL}, and the `operator product' rule:
%\footnote{This is a Hermite polynomial identity in disguise \cite{\morri}.} 
\be 
\label{prod}
\dd\Lambda{m}\,\dd\Lambda{n} = \Lambda^{-1-m-n}\sum_{j=0}^\infty \Lambda^j\cc{mn}j\, \dd\Lambda{j}\,, \notes{[794.3]}
\ee
a Hermite polynomial identity where the expansion coefficients are the numbers \cite{\morri,Gradshteyn1980}:
\be 
\label{cnumbers}
\cc{mn}j=\frac{2^{s-j}a^{2s-2j}}{2\pi^2j!} \Gamma(s-j)\Gamma(s-m)\Gamma(s-n)\, \delta_{j+m+n\,=\,{\rm even}} \,,\quad {\rm with}\quad 2s = j+m+n+1\,,
\ee
we thus find the $\beta$-function equations
\be 
\label{beta-physical}
\dot{g}^{\sigma_0}_{2l} = D_0 \!\!\sum^\infty_{m,n=0}\!\cc{2m+2,2n+2}{2l}\, g^1_{2m}\, g^1_{2n}\,\Lambda^{2(l-n-m)-5}\,.
\ee
Since the couplings $g^1_{2m}$ do not run at this order (as we show in sec. \ref{sec:fullrt}) and $D_0$ and $\cc{2m+2,2n+2}{2l}$ are (known) numbers, we can integrate this immediately to give
\be 
\label{gtwofours}
g^{\sigma_0}_{2l}(\Lambda) = D_0 \!\!\sum^\infty_{m,n=0} \frac{\cc{2m+2,2n+2}{2l}}{5+2(m\cu+n\cu-l)}\, g^1_{2m}\, g^1_{2n}\, \Lambda^{2(l-n-m)-5} +\mathring{g}^{\sigma_0}_{2l}\,,
\ee
where $\mathring{g}^{\sigma_0}_{2l}$ are finite $\Lambda$-integration constants (of dimension $2l\cu-5$), which we will shortly confirm vanish for $l\le2$. Since the expansion of $f^1_\Lambda$ over eigenoperators \eqref{coeffone}, converges absolutely (in the square integrable sense) for $\Lambda\cu>a\Lambda_\p$, the sum above converges absolutely in this regime. (Note that this is different from the expansion of $f^{\sigma_0}_\Lambda$ over eigenoperators \eqref{coefftwofour} which converges for $\Lambda>a\Lambda_\p/\sqrt{3}$ as we have already remarked.)

%In \eqref{beta-physical} the sum continued below this boundary, if plugged directly into
%\eqref{coefftwofour}, gives an expansion over eigenoperators that converges for $\Lambda>a\Lambda_\p/\sqrt{2}$ as we have already noted. Therefore it seems reasonable to assume that for suitable choices of $\mathring{g}^{\sigma_0}_{2l}$, substituting a resummed \eqref{gtwofours} into \eqref{coefftwofour} also leads to an expansion that converges for $\Lambda>a\Lambda_\p/\sqrt{2}$. 
%By the quantisation condition, the integration constants regarded as couplings themselves, \viz $g^{\sigma_0}_{2l}=\mathring{g}^{\sigma_0}_{2l}$, separately give a solution \eqref{coefftwofour}  to the homogeneous part of \eqref{flowftwofour}, and must themselves have an associated amplitude suppression scale $\mathring{\Lambda}_{\sigma_0}>0$. The combined result in \eqref{gtwofours} therefore has an amplitude suppression scale $\Lambda_{\sigma_0}$ which is the larger of $\mathring{\Lambda}_{\sigma_0}$ and $\Lambda_\p/\sqrt{2}$, verifying \eqref{Lambdatwofour}. Shortly, we will use a more powerful method to solve for $f^{\sigma_0}_\Lambda(\vp)$ where we can prove convergence for all $\Lambda\cu>0$.

Since the sum above \eqref{gtwofours} converges for large $\Lambda$, we can read off some useful properties in this UV limit. Firstly note that the relevant couplings $g^{\sigma_0}_{2l}(\Lambda)$, those with $l\cu\ge3$, diverge in this limit. These correspond, more or less \cite{Morris:1993,Morris:2015oca}, to the bare couplings. We avoid constructing explicitly such a bare action by solving directly for the continuum limit solution to the second-order flow equation \eqref{flowtwo} (and indeed the fact that flows to the IR generically fail makes it much harder to begin by constructing the bare action \cite{\morri} as we will also see below).
However for the above solution \eqref{gtwofours} to be genuinely such a renormalized trajectory, we better have that the dimensionless couplings tend to a finite limit:
\be 
\label{twofourfpline}
\tg^{\sigma_0}_{2l}(\Lambda) = \Lambda^{5-2l} g^{\sigma_0}_{2l}(\Lambda)\, \to \tg^{\sigma_0}_{2l*}\,\,,\qquad\text{as}\quad\Lambda\to\infty\,, 
\ee
where $\tg^{\sigma_0}_{2l*}(g^1_0)$ parametrise the line of fixed points that exist if $g^1_0\ne0$ (recalling the remark below %\eqref{dimensionless} and also
 \eqref{gonedimensions} \cite{first}).  Multiplying the solution \eqref{gtwofours} through by $\Lambda^{5-2l}$, we see that this is the case 
%This follows trivially by dimensions, however  
if and only if 
%\be 
%\label{irrelevanttwofour}
$\mathring{g}^{\sigma_0}_{2l}=0$ %\qquad\text{for}\quad l=0,1,2\,. 
%\ee
for $l\cu\le2$.
We thus confirm that the irrelevant couplings are indeed determined by the marginal and marginally relevant couplings, namely all the $g^1_{2n}$. The $\mathring{g}^{\sigma_0}_{2l}$ for $l\cu>2$ so far remain allowed, and are the freely variable finite parts of the corresponding relevant couplings $g^{\sigma_0}_{2l}$. We can also read off in this model approximation, an explicit expression for the line of fixed points at second order:
\be 
\label{twofourfplineExplicit}
\tg^{\sigma_0}_{2l*} = D_0\, \frac{\cc{2,2}{2l}}{5\cu-2l}\, \left(g^1_{0}\right)^2 = (-1)^l \frac{(2l\cu-1)(2l\cu-3)}{(2l\cu-5)l! (8a^2)^l}\frac{a^5}{32\sqrt{2\pi^5}}\, \left(g^1_{0}\right)^2\,,\qquad l\ge0\,,
\ee
where in the second equality we used the formula for $D_0$ \eqref{Dzero} and the $\cc{}{}$ numbers \eqref{cnumbers}.

The UV limit thus behaves as desired. On the other hand, the continuum limit solution \eqref{gtwofours} appears to be badly IR divergent (\ie as $\Lambda\to0$) \cite{\morri}, with power law divergences of arbitrarily high order forced by dimensions, as expected of a theory with infinitely many super-renormalizable couplings (\ie ones with positive mass dimension).\footnote{In statistical models these divergences can signal the existence of an IR fixed point, see \eg \cite{ZinnJustin:2002ru}.}
 However the sum does not converge in this regime. As before we get a sensible result by utilising conjugate momentum space: 
\be 
\label{fourier-sol-twofour}
f^{\sigma_0}_\Lambda(\vp) = \int^\infty_{-\infty}\frac{d\vpi}{2\pi}\, \ff^{\sigma_0}(\vpi,\Lambda)\, {\rm e}^{-\frac{\vpi^2}{2}\Omega_\Lambda+i\vpi\vp} \,, 
\ee
%\cf \eqref{fourier-sol}, 
where in contrast to the solution at first order (\ref{fourier-sol},\ref{fourier-expansion},\ref{coeffone}), we now have an $\ff$ that runs with $\Lambda$ and whose Taylor expansion starts at $l=0$:
\be 
\label{fourier-expansion-twofour}
\ff^{\sigma_0}(\vpi,\Lambda) = \sum_{l=0}^\infty (-)^l g^{\sigma_0}_{2l}\!(\Lambda)\, \vpi^{\,2l}\,.
\ee
%\cf \eqref{fourier-expansion}. 
%\TRM{\old{With an eye to later purposes we define}
%\be 
%%\label{Fn}
%\mathfrak{F}^{i_1,i_2}_{n_{1},n_{2}}(\vpi,\Lambda) = {\rm e}^{\frac{\vpi^2}{2}\Omega_\Lambda} \int^\infty_{-\infty}\!\!\!\!\!d\vp\, {\rm e}^{-i\vpi\vp} f^{i_1\, (n_{1})}_\Lambda(\vp)\, f^{i_2\, (n_{2})}_\Lambda(\vp)\,,
%\ee
%where $i_j$ is $1$ or $1_1$,  and $n_{j}$ is the number of times the corresponding coefficient function is differentiated with respect to $\vp$.}
%\old{Then \eqref{flowftwofour} gives $\dot{\ff}^{\sigma_0}(\vpi,\Lambda) = D_0\, \mathfrak{F}^{1,1}_{2,2}(\vpi,\Lambda)$.} 
Since $f^1_\Lambda$ is expressed in terms of a Fourier transform as \eqref{fourier-sol}, the model second order flow equation \eqref{flowftwofour} gets expressed in terms of the convolution
\beal 
\label{convolutionFirst}
\dot{\ff}^{\sigma_0}(\vpi,\Lambda) &= D_0\, {\rm e}^{\frac{\vpi^2}{2}\Omega_\Lambda} \int^\infty_{-\infty}\frac{d\vpi_1}{2\pi}\,\, \vpi^2_1 (\vpi-\vpi_1)^2\, \ff^1(\vpi_1) \ff^1(\vpi-\vpi_1)\, {\rm e}^{-\Omega_\Lambda\left[{\vpi^2_1}+(\vpi-\vpi_1)^2\right]/2}\,,\\
\label{convolution}
&= D_0\, {\rm e}^{\frac{\vpi^2}{4}\Omega_\Lambda} \int^\infty_{-\infty}\frac{d\vpi_1}{2\pi}\,\, \left(\vpi_1^2-\frac{\vpi^2}{4}\right)^2\! \ff^1\!\left(\vpi_1+\frac\vpi2\right) \ff^1\!\left(\vpi_1-\frac\vpi2\right)\, {\rm e}^{-\Omega_\Lambda{\vpi^2_1}}\,.
\eeal
%where the RHS of the top line is, apart from the initial exponential, just the Fourier transform of $D_0 \left(f^{1\prime\prime}_\Lambda\right)^2$, and 
In the second line we have shifted the integration variable to make the symmetry under $\vpi\mapsto-\vpi$ manifest. Since $\ff^1(\vpi)$ is entire and decays exponentially for large $\vpi$, it is clear that the RHS (of either alternative) converges for all $\Lambda\ge0$. Indeed recall that $\ff^1$ takes the general form \eqref{ffsigmaEg}, %with $A_\sigma\cu=\kappa$, $\Lambda_\sigma\cu=\Lambda_\p$, and 
with the reduced Fourier transform $\bar{\ff}^\sigma=\bar{\ff}^1(\bar{\vpi}^2)$ having limiting behaviour \eqref{largevpibar}.

%Since $\ff^1$ takes the form \eqref{ffsigmaEg}, with $A_\sigma\cu=\kappa$, $\Lambda_\sigma\cu=\Lambda_\p$, 

%from \eqref{ffform} we have that
%\be 
%\label{ffone}
%\ff^1(\vpi) = 2\pi\kappa\Lambda_\p\bar{\ff}^1(\Lambda^2_\p\vpi^2)
%\ee
%(since by \eqref{kappa},  $A_1=\kappa$, and by \eqref{fonelimit}, $m=\bar{n}_{\sigma}=n_\sigma=0$), where $\bar{\ff}^1$ has limiting behaviour  \eqref{largevpibar}. 

We need to split this equation for $\ff^{\sigma_0}$ \eqref{convolution} into its relevant and irrelevant parts, \ie splitting off the $\vpi$-Taylor expansion up to $\vpi^4$. Thus we write
\beal 
\label{frelevant}
\ff^{\sigma_0}(\vpi,\Lambda) &=  \ff^{\sigma_0}_r(\vpi,\Lambda) + \ff^{\sigma_0}_{ir}(\vpi,\Lambda)\,,\\
\label{firrelevant}
\ff^{\sigma_0}_{ir}(\vpi,\Lambda)  &= g^{\sigma_0}_{0}(\Lambda) - g^{\sigma_0}_{2}(\Lambda)\,\vpi^2 + g^{\sigma_0}_{4}(\Lambda)\,\vpi^4\,.
\eeal
%We already know from \eqref{twofourfpline} the irrelevant couplings behave as $\propto1/\Lambda^{2l-5}$ for large $\Lambda$, and thus we get well-defined solutions for them by integrating down from the UV:
Taylor expanding the convolution \eqref{convolution} with respect to $\vpi$ yields resummed expressions for the $\dot{g}^{\sigma_0}_{2l}$. 
Since the irrelevant couplings  have no $\Lambda$-integration constants, and from the asymptotic behaviour \eqref{twofourfpline} they decay for large $\Lambda$ (as $1/\Lambda^{5-2l}$), we then get their values uniquely, and as well-defined expressions, by 
integrating down from the UV:
\beal 
\label{ftwofourirrelevant}
g^{\sigma_0}_{0}(\Lambda) &=D_0 \int^\infty_\Lambda\! \frac{d\Lambda'}{\Lambda'}\! \int^\infty_{-\infty}\!\frac{d\vpi_1}{2\pi}\,\, \vpi_1^4\, (\ff^1)^2 \, {\rm e}^{-\Omega_{\Lambda'}{\vpi^2_1}} \,,\\
g^{\sigma_0}_{2}(\Lambda) &= -\frac{D_0}{4}\int^\infty_\Lambda\! \frac{d\Lambda'}{\Lambda'}\! \int^\infty_{-\infty}\!\frac{d\vpi_1}{2\pi}\,\, \vpi_1^2\left\{ \vpi_1^2 \ff^1 \ff^{1\prime\prime} -\vpi_1^2 (\ff^{1\prime})^2 +\left(\Omega_{\Lambda'}\vpi_1^2-2\right)(\ff^1)^2 \right\} \, {\rm e}^{-\Omega_{\Lambda'}{\vpi^2_1}}\,,\nn
\eeal
where $\ff^1\equiv\ff^1(\vpi_1)$ in the above, and we omit a similar but longer expression for $g^{\sigma_0}_{4}$.

A large part of the value of the expressions we are deriving lies in their generality: that they hold whatever choice we make for the coefficient functions subject to the general form \eqref{ffforma}, in this case the first-order trivialisation limits \eqref{fonelimit}.  Our final results for continuum physics better be universal, and we will get confirmation of that when they become independent of these choices. However for this model system, we pause the development to give an explicit example. Setting $A_\sigma\cu=\kappa$ and $\Lambda_\sigma\cu=\Lambda_\p$ in \eqref{foneEg} gives the simplest example for $f^1_\Lambda$ as we saw \cite{\morri,\morrii,first}. 
Substituting its Fourier representation (\ref{ffsigmaEg},\ref{ffoneEg})  into the first expression above \eqref{ftwofourirrelevant},  gives a well-defined closed-form expression for this second order coupling:
\be 
\label{gtwofourzeroEg}
g^{\sigma_0}_{0}(\Lambda) = \frac{\kappa^2}{32\sqrt{2\pi^3}}\left\{ \frac{3}{\Lambda_\p^3}\ln\left(\frac{\Lambda}{a\Lambda_\p+\sqrt{\Lambda^2+a^2\Lambda_\p^2}}\right)+\frac{3a}{\Lambda^2_\p\sqrt{\Lambda^2+a^2\Lambda^2_\p}}+\frac{a^3}{\left(\Lambda^2+a^2\Lambda^2_\p\right)^{3/2}}\right\}\,,
\ee
where we used the explicit form for $D_0$ \eqref{Dzero}.
This has the desired and expected properties, for example we see from the singularity structure in the complex plane that expanding in $1/\Lambda$ will give a series that converges $\Lambda\cu>a\Lambda_\p$, and furthermore for large $\Lambda$ we recover 
\be 
g^{\sigma_0}_{0}(\Lambda) \sim -\frac{3a^5}{160\sqrt{2\pi^3}}\frac{\kappa^2\Lambda^2_\p}{\Lambda^5}\qquad\text{as}\qquad\Lambda\to\infty
\ee
(using $\sim$ in the strict asymptotic sense \ie that the ratio of left and right hand sides tends to one),
verifying the line of fixed points behaviour at second order (\ref{twofourfpline},\ref{twofourfplineExplicit}), since in this example $g^1_0=\sqrt{\pi}\kappa\Lambda_\p$.

A standard procedure at this stage would be to find the relevant couplings  also by integrating downwards, this time starting at some UV scale $\Lambda=\Lambda_0$:
\be 
\label{fftwofourfrombare}
\ff^{\sigma_0}(\vpi,\Lambda) =\ff^{\sigma_0}(\vpi,\Lambda_0)+ D_0\! \int^{\Lambda_0}_\Lambda\! \frac{d\Lambda'}{\Lambda'}\!\left\{ 
{\rm e}^{\frac{\vpi^2}{4}\Omega_{\Lambda'}}\!\! \int^\infty_{-\infty}\frac{d\vpi_1}{2\pi}\,\, \left(\vpi_1^2-\frac{\vpi^2}{4}\right)^2\! \ff^1\!\left(\vpi_1+\frac\vpi2\right) \ff^1\!\left(\vpi_1-\frac\vpi2\right)\, {\rm e}^{-\Omega_{\Lambda'}{\vpi^2_1}}\right\}
\ee
The integration constants in the bare $\ff^{\sigma_0}(\vpi,\Lambda_0)$, play the r\^ole of bare couplings. In particular the relevant ones would need to be chosen to diverge in such a way that in the limit $\Lambda_0\to\infty$, we are left with a finite solution at finite scales.\footnote{From the solution \eqref{gtwofours} and UV limit (\ref{twofourfpline},\ref{twofourfplineExplicit}) we know how this starts: $g^{\sigma_0}_{2l}(\Lambda_0)= \tg^{\sigma_0}_{2l*}(g^1_0)\,\Lambda^{2l-5}_0 + O(\Lambda^{2l-7}_0)$.}
However this route works against the natural direction of the flow and thus almost certainly ends in a singular coefficient function before reaching the physical limit $\Lambda\to0$ \cite{\morri}. The problem here comes from the first exponential  in the $\Lambda'$-integrand which grows quadratically with $\Lambda'$. It cannot be compensated by the $\vpi$ dependence in the decaying exponentials in the $\ff^1$ terms, since their decay (\ref{largevpibar}) is set by the amplitude suppression scale $\Lambda_\p$ that must be held finite until we have formed the renormalized trajectory. 
%finite\footnote{At this stage of forming the renormalized trajectory, $\Lambda_\p$ is finite. Only after taking the continuum limit $\Lambda_0\to\infty$, do we take $\Lambda_\p\to\infty$ to recover diffeomorphism invariance.} 
Written in the above form \eqref{fftwofourfrombare},  using the symmetrised form of the convolution \eqref{convolution}, ensures that the leading behaviour of the explicit exponentials and those from $\ff^1$  \eqref{largevpibar}, depend on $\vpi$ only through $\vpi^2$, eliminating the $\vpi\vpi_1$ mixed terms that appear in the first form of the convolution \eqref{convolutionFirst}.
Collecting these $\vpi^2$ exponents, we see that integrating down in this way means that we are thus including exponentials of 
\be 
\label{piexponentials}
{\vpi^2}\Omega_{\Lambda'}/4-\Lambda_\p^2\,\vpi^2/8 = \frac{\vpi^2}{8a^2}\left\{ (\Lambda')^2 -(a\Lambda_\p)^2\right\}\,,
\ee
 for $\Lambda\cu< \Lambda' \cu<\Lambda_0$, where we substituted  \eqref{Omega} for $\Omega_\Lambda$. But the exponentials at the  UV limit $\Lambda'\cu=\Lambda_0$, overwhelm the damping factor at large $\vpi$ in the Fourier integral for $f^{\sigma_0}_\Lambda$ \eqref{fourier-sol-twofour}, as soon as 
\be 
\label{downlimit}
2\Lambda^2 < \Lambda^2_0-a^2\Lambda_\p^2\,.
\ee
In the limit of large $\Lambda_0$ as needed to form the complete renormalized trajectory and thus the continuum limit, the solution therefore ends in a singularity already at $\Lambda= \Lambda_0/\sqrt{2}$. The same conclusion was reached in ref. \cite{\morri} using a standard treatment of summing over the melonic Feynman diagrams fig. \ref{fig:melons}, with vertices formed from one $\dd\Lambda{n}$ operator at a time. To make further progress along these lines, the large $\vpi$ behaviour in the integral above \eqref{fftwofourfrombare}, has to be ameliorated by a careful cancellation against the large $\vpi$ behaviour of the chosen bare $\ff^{\sigma_0}(\vpi,\Lambda_0)$, so that the Fourier integral  \eqref{fourier-sol-twofour} converges not only at $\Lambda= \Lambda_0/\sqrt{2}$  but also at all lower scales where the constraints actually get more severe.

However the same arguments show us that this issue is solved by instead integrating up from some arbitrary finite scale $\Lambda\cu=\mu\cu>0$. From the above inequality \eqref{downlimit} we can even integrate down from $\mu$, provided that we do not violate the inequality 
$
2\Lambda^2\cu > \mu^2 \cu- a^2\Lambda_\p^2
$.
Thus if we choose 
\be 
\label{murange}
0< \mu < a\Lambda_\p\,,
\ee
we can now form the complete renormalized trajectory: 
%Provided \eqref{murange}, for or all finite $\Lambda>0$, we have
\be
\label{ftwofourrelevant}
\ff^{\sigma_0}_r(\vpi,\Lambda) = \fc(\vpi,\mu)
- D_0 \!\int^\Lambda_\mu\! \frac{d\Lambda'}{\Lambda'}\!\left\{ 
{\rm e}^{\frac{\vpi^2}{4}\Omega_{\Lambda'}}\! \int^\infty_{-\infty}\frac{d\vpi_1}{2\pi}\,\, \left(\vpi_1^2-\frac{\vpi^2}{4}\right)^2\! \ff^1\!\left(\vpi_1+\frac\vpi2\right) \ff^1\!\left(\vpi_1-\frac\vpi2\right)\, {\rm e}^{-\Omega_{\Lambda'}{\vpi^2_1}}\right\}\Bigg|_r\! ,
\ee
where $|_r$ on the RHS means that we take the relevant part, \ie in this case that the Taylor expansion in $\vpi$ up to $\vpi^4$ is subtracted, the irrelevant part having already been constructed through integrating down from $\Lambda\cu=\infty$ \eqref{ftwofourirrelevant}, and where $\fc(\vpi,\mu)$ now provides the integration constants: 
\be 
\label{fcdef}
\fc(\vpi,\mu) = \sum_{l=3}^\infty (-)^n g^{\sigma_0}_{2l}(\mu)\, \vpi^{\,2l}\,.
\ee
We recognise that the integration constants $g^{\sigma_0}_{2l}(\mu)$ are nothing but the finite renormalized relevant couplings, which at the interacting level are $\mu$ dependent. Substituting $\ff^{\sigma_0}(\vpi,\Lambda) = \fc(\vpi,\mu)$  into the Fourier integral \eqref{fourier-sol-twofour}, provides a renormalized trajectory solution to the homogeneous part of our model second-order flow equation \eqref{flowftwofour}. The $g^{\sigma_0}_{2l}(\mu)$ are freely variable except for the fact that asymptotically at large-$l$ they obey \eqref{largeg}, leading to a physical coefficient function with its own finite amplitude suppression scale $\Lambda_{\sigma_0}$ \eqref{largephys}. 

Note that we have already established that our solution for the relevant part \eqref{ftwofourrelevant} has the correct UV properties since expanding the integrand for large $\Lambda$ gives back the RHS of the $\beta$-function equations \eqref{beta-physical}, which integrated thus gives the explicit solution in terms of underlying couplings \eqref{gtwofours}. However we also note that the integration constants in this former solution \eqref{gtwofours} are not the same as the $g^{\sigma_0}_{2l}(\mu)$.  Formally they are related through \eqref{gtwofours}
by
\be
g^{\sigma_0}_{2l}(\mu) =  D_0 \!\!\sum^\infty_{m,n=0} \frac{\cc{2m+2,2n+2}{2l}}{5+2(m\cu+n\cu-l)}\, g^1_{2m}\, g^1_{2n}\, \mu^{2(l-n-m)-5} +\mathring{g}^{\sigma_0}_{2l}\,,
\ee
however as noted below \eqref{gtwofours} this sum converges only for $\mu>a\Lambda_\p$, which is the regime excluded by the required range for $\mu$ \eqref{murange}. Therefore the above can only be used after resummation. This resummation is provided by our solution for the relevant part \eqref{ftwofourrelevant}. Thus the explicit values of the $\mu$-independent constants $\mathring{g}^{\sigma_0}_{2l}$ can be extracted by subtracting the divergent $\Lambda$ dependence in the  explicit solution in terms of underlying couplings \eqref{gtwofours} from our solution for the relevant part \eqref{ftwofourrelevant}, and then taking the (now finite) limit as $\Lambda\cu\to\infty$.  

Since \eqref{ftwofourrelevant} provides the most general well-defined solution for the relevant part of the renormalized trajectory, it also solves the problem of finding the correct form for the bare couplings in the more standard procedure \eqref{fftwofourfrombare}. Indeed putting $\Lambda\cu=\Lambda_0$ in \eqref{ftwofourrelevant} provides the \emph{most general} expression for the relevant bare couplings $\ff^{\sigma_0}_r(\vpi,\Lambda_0)$ such that the resulting coefficient function $f^{\sigma_0}_\Lambda(\vp)$ survives evolution down to any positive $\Lambda$. 

Although $f^{\sigma_0}_\Lambda(\vp)$ is finite for all finite $\Lambda\cu>0$, it is
still subject to a logarithmic divergence as $\Lambda\to0$, as a result of the $1/\Lambda'$ measure factor in both the irrelevant \eqref{ftwofourirrelevant} and relevant \eqref{ftwofourrelevant} parts. This is why we choose to define the relevant part \eqref{ftwofourrelevant} from $\mu\cu>0$ rather than attempting to integrate up from $\mu\cu=0$. Just as in normal quantum field theories, such as Yang-Mills \cite{\yuji}, this is related to the fact that the derivative expansion diverges there, in particular in the expansion of the Feynman integral \eqref{D} and likewise is cured by using the exact expression for the Feynman integral \eqref{Dp} instead, as we will see explicitly in sec. \ref{sec:fullrt}.

In preparation, we split the integral for the irrelevant couplings \eqref{ftwofourirrelevant} about $\Lambda'\cu=\mu$,
writing:
\be 
g^{\sigma_0}_{0}(\Lambda) =D_0 \int^\infty_\mu\! \frac{d\Lambda'}{\Lambda'}\! \int^\infty_{-\infty}\!\frac{d\vpi_1}{2\pi}\,\, \vpi_1^4\, (\ff^1)^2 \, {\rm e}^{-\Omega_{\Lambda'}{\vpi^2_1}}\quad -D_0 \int^\Lambda_\mu\! \frac{d\Lambda'}{\Lambda'}\! \int^\infty_{-\infty}\!\frac{d\vpi_1}{2\pi}\,\, \vpi_1^4\, (\ff^1)^2 \, {\rm e}^{-\Omega_{\Lambda'}{\vpi^2_1}}\,,
\ee
and similarly for the other two. The first term is just the same definition for the irrelevant couplings but given at scale $\mu$ rather than $\Lambda$, while substituting into $f^{\sigma_0}_{ir}$ \eqref{firrelevant}, we see that the second term provides the missing irrelevant components for the integral in the solution \eqref{ftwofourrelevant}, so that it can be written without the $r$ subscript:
\be
\label{ftwofour}
\ff^{\sigma_0}(\vpi,\Lambda) = \ff^{\sigma_0}(\vpi,\mu)
- D_0 \!\int^\Lambda_\mu\! \frac{d\Lambda'}{\Lambda'}\!\left\{ 
{\rm e}^{\frac{\vpi^2}{4}\Omega_{\Lambda'}}\! \int^\infty_{-\infty}\frac{d\vpi_1}{2\pi}\,\, \left(\vpi_1^2-\frac{\vpi^2}{4}\right)^2\! \ff^1\!\left(\vpi_1+\frac\vpi2\right) \ff^1\!\left(\vpi_1-\frac\vpi2\right)\, {\rm e}^{-\Omega_{\Lambda'}{\vpi^2_1}}\right\} .
\ee
Altogether we can write the most general well-defined solution for the renormalized trajectory in a form that will apply more generally:
%for all the other vertices:
\be 
\label{gensolution}
f^{\sigma}_\Lambda(\vp) = f^{\sigma}_\Lambda(\vp,\mu) - \int^\infty_{-\infty}\!\frac{d\vpi}{2\pi}\,{\rm e}^{-\frac{\vpi^2}{2}\Omega_\Lambda+i\vpi\vp} \!\!\int^\Lambda_\mu\! \frac{d\Lambda'}{\Lambda'}\,  \dot{\ff}^{\sigma}(\vpi,\Lambda')\,,
\ee
where the integrals converge for $\Lambda\cu>0$ provided $\mu$ lies in the range \eqref{murange}, and the explicit form of $\dot{\ff}^{\sigma}(\vpi,\Lambda)$ is ${\rm e}^{\frac{\vpi^2}{2}\Omega_\Lambda}$ times the Fourier transform of the inhomogeneous part of the flow equation. (In the current case of $\sigma\cu=\sigma_0$ we have the model flow equation \eqref{flowftwofour} giving the convolutions \eqref{convolutionFirst} or \eqref{convolution}.) And $f^{\sigma}_\Lambda(\vp,\mu)$ is the following solution of the homogeneous part of the flow equation:
\beal 
\label{homogeneous}
f^{\sigma}_\Lambda(\vp,\mu) &= \int^\infty_{-\infty}\frac{d\vpi}{2\pi}\, \ff^{\sigma}(\vpi,\mu)\, {\rm e}^{-\frac{\vpi^2}{2}\Omega_\Lambda+i\vpi\vp} \,,\\
\ff^{\sigma}(\vpi,\mu) &= \ff^{\sigma}_r(\vpi,\mu) +\int^\infty_\mu\! \frac{d\Lambda'}{\Lambda'}\,  \dot{\ff}^{\sigma}_{ir}(\vpi,\Lambda')\,.
\label{homoirr}
\eeal
In the second line, $\ff^{\sigma}_r(\vpi,\mu)$ contains the new renormalized (marginally) relevant couplings evaluated at $\mu$, in the current case as expanded in \eqref{fcdef}, while the integral computes the irrelevant couplings at $\mu$, by  taking the irrelevant part $\dot{\ff}^{\sigma}_{ir}(\vpi,\Lambda)$ which is the first few terms in the Taylor expansion of $\dot{\ff}^{\sigma}(\vpi,\Lambda)$, namely those with negative dimension coefficients  (up to $\vpi^4$ in the current case). 

Note that even though the above \eqref{homogeneous} is a solution of the homogeneous part of the flow equation, it is not  a valid linearised renormalized trajectory because we now include irrelevant couplings through \eqref{homoirr}, and furthermore
our particular linearised solution depends on the inhomogeneous part through the $\Lambda'$-integral. Using standard terminology from the theory of differential equations, we will refer to this $f^\sigma_\Lambda(\vp,\mu)$ as the \emph{complementary solution} and to the second term in the general solution \eqref{gensolution} as the \emph{particular integral}.
% Nevertheless we will refer to the first term on the RHS of \eqref{gensolution} as the ``homogeneous part'' and the second term as the ``inhomogeneous part'' and trust that the reader will forgive us.\footnote{From the theory of differential equations, we could call the first term the ``general solution'' and the second term the ``special solution'' but these names seem no better.} 
Evidently from \eqref{gensolution}, this particular complementary solution has the property that it coincides with the full solution at $\Lambda\cu=\mu$:
\be 
\label{equalitymu}
f^{\sigma}_\mu(\vp) = f^{\sigma}_\mu(\vp,\mu)\,.
\ee

\subsection{Open system of flow equations}
\label{sec:open}

%We return to the real thing and solve \eqref{flowtwo} directly for the continuum limit. 
%\old{We will see that the solution involves the sum over `melonic' Feynman diagrams to all loop orders \cf fig. \ref{fig:melons}. }
%%as expected since it is second order in perturbation theory but non-perturbative in the loop expansion.
%First we derive the form of the open system of equations discussed above sec. \ref{sec:model}, for which \eqref{flowftwofour} is the truncation.

We return to the real thing and begin by deriving the form of the open system of equations discussed above sec. \ref{sec:model} (and for which the model flow equation \eqref{flowftwofour} is a truncation). Note that establishing that renormalized trajectories exist with the right properties, amounts to establishing the existence of a number of limits. In order to do this, we 
only need the structure of the flow equation and its solution mapped out in a rather schematic way. It is in this spirit that we begin by writing the complete RHS of the second-order flow equation \eqref{flowtwo} as 
\be 
\label{twoexpand}
- \half\, \text{Str}\, \dot{\prop}_\Lambda \Gamma^{(2)}_1 \propH \Gamma^{(2)}_1 = \sum_{rc}\sum_{\sigma^a\sigma^b} \sigma^a\! f^{a(n_a)}_\Lambda F_{rc}(i\partial)\,\sigma^b\!f^{b(n_b)}_\Lambda\,,
\ee
where we suppress all Lorentz indices, $F$ acts on everything to its right, 
$f^a_\Lambda$ is $f^1_\Lambda(\vp)$ or $f^{1_1}_\Lambda(\vp)$, with $n_{a}\cu\le2$ being the number of times it is differentiated with respect to $\vp$ in forming the two propagators, and similarly for $f^b_\Lambda$. As well as attaching $\vp$-propagators \eqref{pp} to coefficient functions, they can also be attached directly to some of the monomials in $\Gamma_1$ (\ref{Gammaonetwo},\ref{Gammaoneone},\ref{Gammaonezero}),  as can the $c$ and $h_{\mu\nu}$ propagators (\ref{cc},\ref{hh}), the former after mapping to gauge fixed basis \eqref{gaugeFixed}.
For each option, the net result is the coefficient functions as displayed, the remaining monomials $\sigma^a$ and $\sigma^b$, and (up to some coefficient of proportionality) the Feynman diagram 
\be 
\label{generalFeynman}
(-i)^d\int_q \frac{\dot{C}^\Lambda(q)\,C_\Lambda(q\cu+p)}{q^2\,(q\cu+p)^2} (q\hbox{ or }p)^d\,,
\ee
where $d\cu\le4$ factors of momentum appear in the numerator as a result of attaching propagators to differentiated fields in the monomials in $\Gamma_1$ (and thus the corresponding factors \eqref{defs}  of $-i$). Using Lorentz invariance we can recast the Feynman integrals as a sum over scalar integrals multiplying tensor expressions containing $r$ instances of $p_\mu$. The scalar integrals will be logarithmically, quadratically, or quartically UV divergent according to whether $c\cu=0$, $1$, or $2$ respectively, where %thus 
\be 
\label{rcd}
r+2c=d\,.
\ee 
Since these Feynman integrals are UV regulated by $\dot{C}^\Lambda$, they take the form
\be 
\label{Frc}
F_{rc}(p) = (-ip)^r \Lambda^{2c} F_{rc}(p^2/\Lambda^2)\,,
\ee
and since they are IR regulated by $C_\Lambda$ the dimensionless scalar factor has a Taylor expansion:
\be 
\label{FTaylor}
F_{rc}(p^2/\Lambda^2) = \sum_{m=0}^\infty \frac{F^m_{rc}}{m!} \left(\frac{p^2}{\Lambda^2}\right)^m\,.
\ee
We organise the resulting derivative expansion  \eqref{twoexpand}  on the RHS of the flow equation, by taking first the tensor factor and $k$ factors of $\Box$ and letting these act in all possible ways on the two terms to their right but such that at least one $\partial$ from each $\Box$ is involved in differentiating $f^b$:
\be 
\label{kraction}
\frac{(m-k)!}{m!}(-\Box)^k \partial^r \, \sigma^b\!f^{b(n_b)}_\Lambda = \sum^{2k+r}_{n=k}\sum_{\sigma^b_{kn}} \sigma^b_{kn}\, f^{b(n_b+n)}_\Lambda\,.
\ee
Thus the monomials $\sigma^b_{kn}$ gain $n$ factors of $\partial\vp$ which may or may not then be further differentiated. For each $m\cu\ge k$, we insist that the remaining $(-\Box)^{m-k}$ acts exclusively on these resulting monomials: 
\be 
\label{mkaction}
\frac{1}{(m-k)!}(-\Box)^{m-k} \sigma^b_{kn} = \sum_{\sigma^b_{kmn}\, |\, \sigma^b_{kn}}\!\!\!\! \sigma^b_{kmn}\,,
\ee
where the factorial factor cancels that in the previous equation \eqref{kraction} and is for later convenience,  and the sum is over all (linearly independent) monomials generated in this way given the initial (linearly independent) $\sigma^b_{kn}$. 
For given $\sigma^a$, $\sigma^b$, $r$ and $c$, we have now expanded over a full set of monomials $\sigma^a \sigma^b_{kmn}$ such that we can write for the RHS of the flow equation \eqref{twoexpand}:
%\be 
%\label{defsigma}
%\sigma = \sigma^a \sigma^b_{kmn}\,.
%\ee
%Thus finally  the RHS:
\be 
\label{flowtwoRHS}
- \half\, \text{Str}\, \dot{\prop}_\Lambda \Gamma^{(2)}_1 \propH \Gamma^{(2)}_1 =  \sum_{rc}\sum_{\sigma^a \sigma^b_{kmn}}\mkern-10mu \sigma^a \sigma^b_{kmn}\, F^m_{rc}\,\Lambda^{2(c-m)} f^{a(n_a)}_\Lambda f^{b(n_b+n)}_\Lambda\,.
\ee
Turning to the LHS of the second-order flow equation \eqref{flowtwo}, since it is of the same form as the first-order flow equation \eqref{flowoneexpanded},  we recognise that we can reuse the integrating factor \eqref{eigenoperatorsol}, by setting:
\be 
\label{ringG}
\rg{\Gamma}_2 = \exp\left(\frac12 {\prop}^{\Lambda\,AB} \frac{\partial^2_l}{\partial\Phi^B\partial\Phi^A}\right) \Gamma_{2}\,.
\ee
This would be independent of $\Lambda$ if it were not for the RHS of the flow equation, which now reads:
\be 
\label{integfactflow}
\frac{\partial}{\partial t}\, \mathring{\Gamma}_2 = -\frac12\exp\left(\frac12 {\prop}^{\Lambda\,AB} \frac{\partial^2_l}{\partial\Phi^B\partial\Phi^A}\right)  \text{Str}\, \dot{\prop}_\Lambda \Gamma^{(2)}_{1} \propH \Gamma^{(2)}_{1} \,.
\ee
Expanding $\rg{\Gamma}_2$ \eqref{ringG} over a complete set of monomials $\sigma$ (extending $\{\sigma^a \sigma^b_{kmn}\}$ to a set that span all of $\rg{\Gamma}_2$) we parametrise their coefficient functions via Fourier transform, as:
\be 
\label{ringGexp}
\rg{\Gamma}_2 = \sum_\sigma \sigma \mathring{f}^\sigma_\Lambda(\vp)\,,\qquad \rg{f}_\Lambda^\sigma(\vp) = \int^\infty_{-\infty}\!\frac{d\vpi}{2\pi}\, \ff^\sigma\!(\vpi,\Lambda)\, {\rm e}^{%\exp\left(
i\vpi\vp} \,.
\ee
Thanks to the exponential operator in the transformed flow equation \eqref{integfactflow}, the set $\{\sigma\}$ has to be larger than $\{\sigma^a \sigma^b_{kmn}\}$. However if $\sigma$ does not appear on the RHS of \eqref{integfactflow}, its  $\ff^\sigma(\vpi)$ is $\Lambda$ independent. By inverting the $\rg{\Gamma}_2$ definition \eqref{ringG} and comparing to the first-order solution \eqref{eigenoperatorsol} we see that it corresponds to adding a linearised solution. In principle such linearised solutions might need to be added in order to satisfy the second-order mST \eqref{mSTtwo} in the large amplitude suppression scale limit. They are straightforward to treat using the methods in ref. \cite{first} since they correspond to complementary solutions with no irrelevant couplings. However it turns out that all the monomials $\sigma$ needed, are already generated on the RHS of \eqref{integfactflow}  \cite{second}.
%By the same arguments as at the beginning of sec. \ref{sec:second}, these would just change our choice of first-order solution $\Gamma_1$ and are thus dealt with already and now excluded.
Therefore in the following we restrict $\{\sigma\}$ to a minimal set necessary to span the RHS of the transformed flow equation  \eqref{integfactflow} and thus all of these will have $\Lambda$-dependent $\ff^\sigma$. Inverting the definition of $\rg{\Gamma}_2$ \eqref{ringG} gives 
\be 
\label{ringinvert}
\Gamma_2\ =\ \exp\left(-\frac12 {\prop}^{\Lambda\,AB} \frac{\partial^2_l}{\partial\Phi^B\partial\Phi^A}\right) \rg{\Gamma}_2 
\ =\ \sum_\sigma \left( \sigma f^\sigma_\Lambda(\vp)+\cdots \right)\,,
\ee
where the tadpoles are generated by the same formula as in the general first-order solution \eqref{firstOrder} and the coefficient functions 
\be 
\label{coeffgen}
f^{\sigma}_\Lambda(\vp) = \int^\infty_{-\infty}\frac{d\vpi}{2\pi}\, \ff^{\sigma}(\vpi,\Lambda)\, {\rm e}^{-\frac{\vpi^2}{2}\Omega_\Lambda+i\vpi\vp} \,, 
\qquad
\ff^{\sigma}(\vpi,\Lambda) = i^{\,\eps}\sum_{l=0}^\infty (-)^l g^{\sigma}_{2l+\eps}\!(\Lambda)\, \vpi^{\,2l+\eps}\,,
\ee
$\eps$ according to parity and $\ff^\sigma$ as defined above in \eqref{ringGexp},
are the general case which (\ref{fourier-sol-twofour},\ref{fourier-expansion-twofour}) modelled. That is we recognise that they again take the same form as the linearised Fourier transform solution (\ref{fourier-sol},\ref{fourier-expansion}), except that $\ff^\sigma$ and couplings run, and the sum is also over the irrelevant couplings. 

We see that $\rg\Gamma_2$ therefore has an expansion \eqref{ringGexp} over just the top terms, being the first terms in the final bracket of \eqref{ringinvert}, however with  `stripped' coefficient functions $\rg{f}^\sigma_\Lambda$ that do not include the exponential damping factor present in the Fourier transform of the \textit{bona fide} coefficient functions \eqref{coeffgen} above. Actually that leads to a problem for this representation, which we can already see from the model expression for $\ff^{\sigma_0}(\vpi,\Lambda)$ \eqref{ftwofour},
and which we will confirm in full in sec. \ref{sec:fullrt}. Since there is no damping factor, the Fourier transform \eqref{ringGexp} for the stripped coefficient function, fails to converge for $\Lambda\cu>a\Lambda_\p$, and thus $\rg{f}^\sigma_\Lambda(\vp)$ becomes distributional as $\Lambda\cu\to a\Lambda_\p^-$ and if analytically continued above this, will be complex in general (compare the cases described in ref. \cite{\morri}). Indeed the model expression \eqref{ftwofour} is an integral over the $\vpi^2$ exponentials \eqref{piexponentials} which have positive exponents once $\Lambda\cu\ge\Lambda'\cu>a\Lambda_\p$. The differentiated version $\partial_t \rg{f}^\sigma_\Lambda$ in \eqref{integfactflow} suffers the same problem for $\Lambda\cu>a\Lambda_\p$, as is obvious from the same arguments applied to the $t$-differential of the model answer \eqref{ftwofour}, \ie the original convolution expression \eqref{convolution}. Since the problem is only in the $\vpi$-integral for the stripped coefficient function \eqref{ringGexp}, we can make sense of equations involving $\rg\Gamma_2$ even in the region $\Lambda\cu>a\Lambda_\p$ if we interpret them as defining the Fourier transforms $\ff^\sigma(\vpi,\Lambda)$, \ie understand that to get equations that make sense for all $\Lambda\cu>0$, we should work in Fourier transform space.

Using the map \eqref{ringinvert} from $\rg\Gamma_2$ to $\Gamma_2$,  we can however cast the $\rg\Gamma_2$ flow equation \eqref{integfactflow} in terms of the \textit{bona fide} coefficient functions and thus in a form that genuinely exists in $\vp$-space at all $\Lambda\cu>0$.
Differentiating the middle equation in \eqref{ringinvert} with respect to $t$, using the expansion over stripped coefficient functions \eqref{ringGexp}, and comparing again to \eqref{ringinvert} we see that
\be 
\label{flowtwoLHS}
\dot{\Gamma}_2 - \half\, \text{Str}\, \dot{\prop}_\Lambda \Gamma^{(2)}_2 \ =\ \exp\left(-\frac12 {\prop}^{\Lambda\,AB} \frac{\partial^2_l}{\partial\Phi^B\partial\Phi^A}\right) \partial_t\rg{\Gamma}_2
\ =\ \sum_\sigma \left( \sigma \stackrel{\odot}{f}\mkern-4.5mu{}^\sigma_\Lambda(\vp)+\cdots \right)\,,
\ee  
%% NB \! = \mkern-3mu
where the tadpoles are again generated in the same way as the first-order formula \eqref{firstOrder}, and 
\be 
\label{solar}
\stackrel{\odot}{f}\mkern-4.5mu{}^\sigma_\Lambda(\vp) = \dot{f}^{\sigma}_\Lambda(\vp) - \half\,\dot{\Omega}_\Lambda\,  f^{\sigma\prime\prime}_\Lambda(\vp) =
 \int^\infty_{-\infty}\frac{d\vpi}{2\pi}\, \dot{\ff}^{\sigma}(\vpi,\Lambda)\, {\rm e}^{-\frac{\vpi^2}{2}\Omega_\Lambda+i\vpi\vp}\,,
\ee
is the coefficient function with the necessary damping factor but obtained by applying the RG time derivative only to $\ff^\sigma$. 
%(Thus if $\sigma$ does not appear on the RHS, $f^\sigma_\Lambda$ is a linearised solution and $\stackrel{\odot}{f}\mkern-4.5mu{}^\sigma_\Lambda$ vanishes.) 
Finally we can combine the expansion over monomials of the RHS of the second-order flow equation \eqref{flowtwoRHS} and the above expression for its LHS \eqref{flowtwoLHS} to get the advertised open system of flow equations:
\be 
\label{open}
\stackrel{\odot}{f}\mkern-4.5mu{}^\sigma_\Lambda(\vp) + \sum_{\sigma',n'=0} a_{\sigma',n'}\, \Lambda^{d_{\sigma'}+\eps-d_\sigma-\eps'+n'} \stackrel{\odot}{f}\mkern-4.5mu{}^{\sigma'(n')}_\Lambda(\vp)
= %\sum_{rckmn\, |\, \sigma=\sigma^a\sigma^b_{kmn} }\mkern-25mu 
\sum F^m_{rc}\,\Lambda^{2(c-m)} f^{a(n_a)}_\Lambda f^{b(n_b+n)}_\Lambda\,,
\ee
where the RHS sums over those cases (if any) for which for given $r,c,k,m,n,a,b$, $n_a$ and $n_b$, we have $\sigma =\sigma^a\sigma^b_{kmn}$,
and on the LHS we recognise that $\sigma$ will appear in the tadpole corrections of some $d_{\sigma'}\cu>d_\sigma$ dimensional $\sigma'$s in \eqref{flowtwoLHS} either through tadpole corrections that act exclusively on $\sigma'$ or through attaching $n'$ $\vp$-tadpoles to both $\sigma'$ and $f^{\sigma'}$, $a_{\sigma',n'}$ being the resulting numerical coefficient. 

%Also as advertised, the truncation \eqref{flowftwofour} is a special case but which only follows from neglecting these corrections.

\subsection{The full renormalized trajectory}
\label{sec:fullrt}

However, writing the flow equation in its $\rg\Gamma_2$ form \eqref{integfactflow}, we factor out the tadpole corrections, producing instead an equation which relates the conjugate momentum coefficient functions to an infinite set of loop corrections as generated by the exponential on the RHS of \eqref{integfactflow}. We will shortly see that these loop corrections are the melonic Feynman diagrams of fig. \ref{fig:melons}. We use this form for very general sets of couplings, to show that a well-defined renormalized trajectory can be constructed, and thus that the continuum limit exists at second order. 

Notice that, since the open system of equations \eqref{open} are equivalent to this new form, we thus confirm  that the $\vp$-tadpoles already captured in \eqref{solar} and  the rest of the sum over tadpole corrections on the LHS of \eqref{open} reconstruct these melonic diagrams. 

%\begin{figure}[ht]
%\centering
%\includegraphics[scale=0.4]{melons.png}
%%\vskip-30pt
%\caption{$\Gamma_2$ resums an expansion over melonic Feynman diagrams, and appears in different guises. In \eqref{flowmelons}, the re-summation yields $\beta$ functions, with open circles given by $\Gamma_{1\,\text{phys}}$. They integrate exactly to the $\rg\Gamma_2$ expression \eqref{ringGtwo}, which can be recast as an expression for $\Gamma_2$ \eqref{Gammatwosol}. In this last version, the open circles are copies of $\Gamma_1$ which itself resums the expansion over tadpole diagrams in fig. \ref{fig:tadpoles}.}
%\label{fig:melons}
%\end{figure}

The exponential in the $\rg\Gamma_2$ flow \eqref{integfactflow}, attaches propagators in all possible ways to the two copies of $\Gamma_{1}^{(2)}$. We factor it into three pieces using Leibniz, in the same way as before \eqref{WickTwoFactorId}, and introduce the notation
\be 
\label{Posimple}
\Po^\Lambda = {\prop}^{\Lambda\, AB} \frac{\partial^L_l}{\partial\Phi^B}\frac{\partial^R_l}{\partial\Phi^A}
\ee
for the middle exponent. Introducing correspondingly $\Po_\Lambda$ for the IR regulated propagator, the one-loop expression that the exponential acts on, can be expressed as:
\be 
\label{oneloopPsimple}
-\tfrac12\,\text{Str}\, \dot{\prop}_\Lambda \Gamma^{(2)}_{1} \propH \Gamma^{(2)}_{1} = \half\,\dot{\Po}^\Lambda\,\Po_\Lambda\,\Gamma_{1}\,\Gamma_{1}\,.
\ee
The purely L and purely R pieces of the exponential operate exclusively on their own copy of $\Gamma_{1}$ and by the latters' solution  \eqref{eigenoperatorsol}, turn them into $\Gamma_{1\,\text{phys}}$. Therefore the second-order flow equation \eqref{flowtwo} can be written as
\be 
\label{flowclosed}
\partial_t\,\rg{\Gamma}_2 = \half\, \dot{\Po}^\Lambda \,\Po_\Lambda \,\mathrm{e}^{\Po^\Lambda}\, \Gamma_{1\,\text{phys}}\,\Gamma_{1\,\text{phys}}\,.
\ee
Expanding the exponential yields the sum over $\ell$-loop melonic Feynman diagrams in fig. \ref{fig:melons}:
\be 
\label{flowmelons}
\partial_t\,\rg{\Gamma}_2\ =\ \frac12\, \sum_{\ell=1}^\infty \frac{1}{(\ell-1)!}\,\dot{\Po}^\Lambda \,\Po_\Lambda (\Po^\Lambda)^{\ell-1}\ \Gamma_{1\,\text{phys}}\,\Gamma_{1\,\text{phys}}\,.
\ee
If we would attach propagators only to the monomials in $\Gamma_{1\,\text{phys}}$, the expansion would terminate at three propagators, \ie two loops, this contribution coming exclusively from the first bracket in the level-zero expression \eqref{Gammaonezero}. The expansion however continues forever by attaching propagators to the coefficient functions. Each diagram can be written in a similar way to the expansion \eqref{twoexpand} of the one-loop expression on the RHS of the original form for the second-order flow equation, \ie we can express this as
\be 
\label{allexpand}
\partial_t\,\rg{\Gamma}_2\ =\  \sum_{rc\ell} \sum_{\sigma^a\sigma^b} \sigma^a\! f^{a(n_{a\ell})} F_{rc\ell}(i\partial)\,\sigma^b\!f^{b(n_{b\ell})}\,,
\ee
where again $r$ is the power of external momentum, but $c$ has a modified meaning explained below, and now $n_{a\ell}, n_{b\ell}\cu\le\ell\cu+1$ from attaching propagators, while the vertices are supplied by the physical limit of the first-order solution (\ref{Gammaonetwo},\ref{Gammaoneone},\ref{Gammaonezero}) and thus expressed in terms of $f^1(\vp)$ and $f^{1_1}(\vp)$. As we will show shortly, at each loop order the contributions are finite, being both IR and UV regulated by $\Lambda$, and furthermore the sum over loops converges such as to ensure well defined conjugate momentum space expressions $\ff^\sigma(\vpi,\Lambda)$ for the stripped coefficient functions   \eqref{ringGexp},  for all $\Lambda\cu>0$. Therefore the conjugate momentum space version of the flow equation  in either form (\ref{flowmelons},\ref{allexpand}) is well defined.

Each Feynman diagram is the obvious generalisation of the previous one-loop expression  \eqref{generalFeynman}, in particular we again have $d\cu\le4$ factors of momentum in the numerator. Thus by dimensions, %and reality of \eqref{allexpand},
\be 
\label{Frcl}
F_{rc\ell}(p) = (-ip)^r \Lambda^{2(c+\ell-1)} F_{rc\ell}(p^2/\Lambda^2)\,,
\ee
where the dimensionless scalar factor has a Taylor expansion:
\be 
\label{FrclTaylor}
F_{rc\ell}(p^2/\Lambda^2) = \sum_{m=0}^\infty \frac{F^m_{rc\ell}}{m!} \left(\frac{p^2}{\Lambda^2}\right)^m.
\ee
Here we recognise that by Lorentz invariance, $r$ must differ from $d$ by an even integer, which we call $2c$ as we did previously \eqref{rcd}. However the diagram is now overall UV divergent with power index $2(c\cu+\ell\cu-1)$.

The contributions are well defined at each loop order, because $\ell$ propagators are UV regulated by $\Lambda$ and two are IR regulated by $\Lambda$. (Recall that $\dot{\prop}^{\!\Lambda} = -\dot{\prop}_\Lambda$.) In fact in general such melonic contributions are UV finite provided at least $\ell$ (of the total $\ell\cu+1$) propagators are UV regulated, as is clear since then at least one propagator is UV regularised around any loop. And in general such melonic contributions are IR finite provided at least one propagator is IR regulated, as is particularly clear in position space. Thus to see this last statement, recast the expansion \eqref{allexpand} as:
\be 
\label{allposition}
\partial_t\,\rg{\Gamma}_2\ =\  \sum_{rc\ell} \sum_{\sigma^a\sigma^b}\int\!\!d^4\!x\,d^4\!y\, \left[\mkern-1mu\sigma^a f^{a(n_{a\ell})}\right]\!\!(x)\,F_{rc\ell}(x\cu-y)\left[\mkern-1mu\sigma^b f^{b(n_{b\ell})}\right]\!\!(y)\,.
\ee
In this form the Feynman diagrams are just the product of the propagators written in position space.
For example if $n_{a\ell} \cu= n_{b\ell}\cu=\ell\cu+1$, all propagators attach to the coefficient functions, and thus
\be 
\label{Fprod}
F_{rc\ell}(x\cu-y) = F_{rc\ell}(i\partial_x)\, \delta(x\cu-y) = \frac{(-)^{\ell}}{2(\ell-1)!}\,\dot{\prop}_\Lambda\prop_\Lambda (\prop^{\!\Lambda})^{\ell-1}\,,
\ee
from the melonic expansion \eqref{flowmelons}, where here we mean the standard scalar propagator:
\be 
\label{scalarnoreg}
\prop \equiv \prop(x\cu-y) = \int_q \frac1{q^2}\, \text{e}^{-i q\cdot (x-y)}\,,
\ee
in position space,
the sign having been factored out of the $\vp$-propagator \eqref{pp}, and the cutoff function $C^\Lambda(p)$ or $C_\Lambda = 1\cu-C^\Lambda$ inserted as requested. The derivative expansion 
\be 
\label{Dexp}
F_{rc\ell}(i\partial) = \sum_{m=0}^\infty \frac{(-)^m}{m!} D^{m}_{rc\ell}\, \partial^{2m+r}\,,
\ee
can be computed by Taylor expanding the last square-bracketed term in the position space expression \eqref{allposition} about $x$, and thus 
\be 
\label{Fderiv}
D^{m}_{rc\ell} = \frac{(-)^{r+m}m!}{(2m\cu+r)!}\int\!\!d^4\!x\, x^{2m+r}\, F_{rc\ell}(x)
\ee
(continuing to ignore tensor contractions),
as is also clear directly from substituting the derivative expansion \eqref{Dexp} into the expression for $F_{rc\ell}(x)$  \eqref{Fprod} and then into the above \eqref{Fderiv}. The coefficients 
%In terms of the previous expansion (\ref{Frcl},\ref{FrclTaylor}) we see that 
\be 
\label{FderivLambda}
D^{m}_{rc\ell} = 
F^{m}_{rc\ell}\, \Lambda^{2(c+\ell-m-1)}\,
\ee
are just the dimensionful versions, as is clear by comparing the previous expansion (\ref{Frcl},\ref{FrclTaylor}).
%, and will prove useful later. 
Now, the UV regulated propagators are free of regularisation at large $x$, \ie become the standard expression $\prop(x) = 1/(4\pi x)^2$ in this regime, but the IR regulated propagator $\prop_\Lambda(x)$ (also $\dot{\prop}_\Lambda$) is quasi-local, decaying faster than a power outside $x\sim1/\Lambda$. (We will see an explicit example shortly.) In the cases where the propagators attach to space-time differentiated fields in the original monomials, we get space-time differentiated versions $\partial^s\prop(x)$ ($s\cu=1,2$) but this does not change the quasi-locality property. Thus the product of propagators also vanishes faster than a power at large $x$, provided at least one is IR regulated, which confirms the general statement and that integral expression for the derivative expansion coefficient  \eqref{Fderiv} in particular is IR finite for $\Lambda\cu>0$. For completeness we note that UV finiteness is almost as obvious in position space as it was in momentum space. Since $\prop^\Lambda(x)$ (also $\dot{\prop}^\Lambda$) is smooth at $x\cu=0$, while $\prop_\Lambda(x)\sim1/(4\pi x)^2$ as $x\cu\to0$, we get an integrable divergence in the integral expression \eqref{Fderiv}. At first sight, $\partial^2\prop_\Lambda(x)=O(x^{-4})$ could be problematic, but the full expression is still integrable, as follows by integrating by parts.

Organising the derivative expansion in the same way as in we did in the previous subsection, \cf \eqref{kraction} -- \eqref{flowtwoRHS}, we see that the expansion \eqref{allexpand} can be written as
\be 
\label{flowallRHS}
\partial_t\,\rg{\Gamma}_2\ =\  \sum_{rc\ell} \sum_{\sigma^a \sigma^b_{kmn}}\mkern-10mu \sigma^a \sigma^b_{kmn}\, %F^m_{rc\ell}\,\Lambda^{2(c+\ell-m-1)} 
D^{m}_{rc\ell}\,
f^{a(n_{a\ell})} f^{b(n_{b\ell}+n)}\,,
\ee
where however these monomials are now the full set necessary to span the RHS of the $\rg\Gamma_2$ flow equation \eqref{integfactflow}, and thus
\be 
\label{flowfo}
\partial_t\, \rg{f}^\sigma_\Lambda(\vp) = \sum %F^m_{rc\ell}\,\Lambda^{2(c+\ell-m-1)} 
D^{m}_{rc\ell}\,f^{a(n_{a\ell})}(\vp)\, f^{b(n_{b\ell}+n)}(\vp)\,,
\ee
where the sum is non-empty and is over all  $r,c,\ell,k,m,n,a,b$, $n_{a\ell}$ and $n_{b\ell}$ that give a match for
%where however these monomials are now the full set necessary to span \eqref{integfactflow}, \ie we can now identify in one-to-one fashion these with the expansion \eqref{ringGexp}:
\be 
\label{defsigma}
\sigma = \sigma^a \sigma^b_{kmn}\,.
\ee
Recalling the discussion in the previous subsection, to define this for all $\Lambda\cu>0$, summing over all loops, we transfer to conjugate momentum space. Substituting the conjugate momentum expression  \eqref{ringGexp} for the stripped coefficient function $\rg{f}^\sigma_\Lambda$,  into the above flow equation \eqref{flowfo} and for the first-order functions the physical limit of their conjugate momentum expressions \eqref{fourier-sol}, we get the analogous expression to the convolution formula \eqref{convolution} in the model approximation:
\be
\label{flowff}
\dot{\ff}^{\sigma}(\vpi,\Lambda) = (-)^{\eps_b}\sum D^m_{rc\ell}\, i^{n_{a\ell}-n_{b\ell}-n} 
%\,\Lambda^{2(c+\ell-m-1)} \\ 
\!\int^\infty_{-\infty}\!\frac{d\vpi_1}{2\pi}\, \left(\vpi_1+\frac\vpi2\right)^{n_{a\ell}}\! \left(\vpi_1-\frac\vpi2\right)^{n_{b\ell}+n}\! \ff^a\!\left(\vpi_1+\frac\vpi2\right) \ff^b\!\left(\vpi_1-\frac\vpi2\right)\,,
\ee
$\eps_b$ being the parity of $f^b$.
The exponentials over $\Omega_\Lambda\vpi^2$ and $\Omega_\Lambda\vpi^2_1$ that are evident in convolution formula \eqref{convolution} are still present, hidden in this sum. To see this note that $n_{a\ell}$ and $n_{b\ell}$ are, up to an additive constant, equal to $\ell$. Thus the sum over all loops involves
\be 
\label{sumD}
\sum_\ell D^m_{rc\ell} \left(\vpi_1^2-\frac{\vpi^2}{4}\right)^{\!\ell-1}\,.
\ee
Now the $\ell$-loop contribution is given by the integral expression \eqref{Fderiv}, with $F_{rc\ell}(x)$ given by the product over propagators \eqref{Fprod}, apart from an $\ell$-independent factor from tensor contractions and from propagators attaching to monomials in $\Gamma_{1\,\text{phys}}$. Since each propagator \eqref{scalarnoreg} is dimension two, a na\"\i ve estimate of the sum over $\ell$ is thus\footnote{For a tidy answer we match the proportionality to the one loop integral \eqref{Dp} contribution, which is dimensionless.} 
%from the $n_{a,b\ell}$ powers.}
\be 
\label{naivel}
\sum_\ell D^m_{rc\ell} \left(\vpi_1^2-\frac{\vpi^2}{4}\right)^{\!\ell-1}
%\propto F^m_{rc\ell}\Lambda^{2\ell} 
\!\!\propto\, \sum_\ell \frac{(-)^{\ell}}{(\ell-1)!}\, \Omega_\Lambda^{\,\ell-1} \left(\vpi_1^2-\frac{\vpi^2}{4}\right)^{\!\ell-1}\!\! =\, {\rm e}^{\left(\frac{\vpi^2}{4}-\vpi^2_1\right)\Omega_\Lambda}\,,
\ee
\ie that each new regularised loop integral contributes the same magnitude as the tadpole integral \eqref{Omega}. Then up to a sum of multiplicative power corrections (in $\vpi$, $\vpi_1$ and $\Omega_\Lambda$),  the flow equation for $\ff^\sigma$ \eqref{flowff} takes the same well-defined form as the convolution expression for the model \eqref{convolution}:
\be 
\label{convolutiongen}
\dot{\ff}^{\sigma}(\vpi,\Lambda) \sim \sum {\rm e}^{\frac{\vpi^2}{4}\Omega_\Lambda} \int^\infty_{-\infty}\frac{d\vpi_1}{2\pi}\,\, \ff^a\!\left(\vpi_1+\frac\vpi2\right) \ff^b\!\left(\vpi_1-\frac\vpi2\right) \, {\rm e}^{-\Omega_\Lambda{\vpi^2_1}}\,,
\ee
where the remaining finite sum is over all such matches to $\sigma$ \eqref{defsigma} independent of loop order $\ell$. 

That this is indeed the correct expression can be verified by 
computing the melonic diagrams for the following particular choice of cutoff function:
\be 
\label{Cexp}
C^\Lambda(p)= C(p^2/\Lambda^2) = \exp(-p^2/\Lambda^2)\,,
\ee
%In preparation for this it is useful to evaluate 
for which the tadpole integral \eqref{Omega}, gives $a\cu=2\sqrt{2}\,\pi$:
\be 
\label{Omegaexp}
\Omega_\Lambda = \frac{\Lambda^2}{(4\pi)^2}\,.
\ee
The standard scalar propagator \eqref{scalarnoreg} in its IR regulated version of  is then
\be 
\label{propIR}
\prop_\Lambda(x) = \frac1{4\pi^2x^2}\,\text{e}^{-\Lambda^2x^2/4}\,.
\ee
This can be derived from $C_\Lambda(p)/p^2$ by differentiating with respect to $\Lambda$, performing the now Gaussian momentum integral, as per conventions \eqref{defs}, and then integrating back up with respect to $\Lambda$. That the integration constant vanishes is confirmed by either the $\Lambda\cu\to\infty$ or $\Lambda\cu\to0$ limits, while integrating $(x^2)^s\prop_\Lambda(x)$ for integer $s\cu\ge0$ over all space, and comparing the result to the corresponding momentum integral, confirms that there is no distributional part. Thus we also have
\be 
\label{propUV}
\prop^{\!\Lambda}(x) = 1 - \prop_\Lambda(x) = 
 \frac1{4\pi^2x^2} \left( 1 - \text{e}^{-\Lambda^2x^2/4} \right)\,.
\ee
Taking into account that propagators appear with a total of $d$ derivatives which up to $\ell$-independent factors,  we can do by inserting $x^{-d}$ as well as the product over propagators \eqref{Fprod} into the integral formula (\ref{Fderiv}) we have, by definition of $c$ \eqref{rcd}, that \notes{[1092.3]}
\beal 
D^m_{rc\ell} &\sim \frac{(-)^{\ell}}{(\ell-1)!} \frac1{(4\pi^2)^\ell}\int\!\!d^4\!x\, (x^2)^{m-c-\ell}\, \text{e}^{-\Lambda^2x^2/2} \left( 1 - \text{e}^{-\Lambda^2x^2/4} \right)^{\ell-1}\,,\nn\\
&\sim \frac{(-)^{\ell}}{(\ell-1)!} \,\Omega^\ell_\Lambda \int^\infty_0\!\!\!\!\!\!dt\ t^{1+m-c-\ell}\, \text{e}^{-2t}\left(1-\text{e}^{-t}\right)^{\ell-1}\,,\nn\\
& = \frac{(-)^{\ell}}{(\ell-1)!} \,\Omega^\ell_\Lambda \left( I^{2+m-c}_{\ell+1} - 2I^{1+m-c}_\ell +I^{m-c}_{\ell-1}\right)\,,
\label{Fell}
\eeal
where we introduced $t\cu=\Lambda^2x^2/4$, used the current value of $\Omega_\Lambda$ \eqref{Omegaexp}, and expressed the result in terms of the numbers
\be 
\label{In}
I^k_{\ell} = \int^\infty_0\!\!\!\!\!\!dt\ t^{k-\ell} \left(1-\text{e}^{-t}\right)^\ell\,,
\ee
which are well defined for (integer) $0\cu\le k\cu< \ell\cu-1$.
%, and $I^0$ ($I^1$) comes from the second (first) term in braces in \eqref{Mnint}.
We show in app. \ref{app:largen} that at large $\ell$,
%At large $\ell$ (see app. \ref{app:largen}), 
\be
\label{largenIn}
I^k_\ell = k! \left({2}/{\ell}\right)^{k+1} +O({\ell^{-k-2}})\,.
%I^0_n = 2/n +O(n^{-2})\,,\qquad I^1_n = 4/{n^2} +O(n^{-3})\,,
\ee
and thus in \eqref{Fell} it is $I^{m-c}_{\ell-1}$ that provides the leading contribution. Since asymptotically, 
\be 
\label{largellsum}
\sum_{\ell}^\infty  \frac{(-)^{\ell}}{(\ell-1)!}\, I^k_\ell u^\ell \sim \,\text{e}^{-u}\,,
\ee
up to $k$-dependent multiplicative powers of $u$ (as can be seen by operating with $(u \,d_u)^k$ on both sides) we see that the na\"\i ve estimate of the sum \eqref{naivel} and thus the general convolution formula for $\ff^\sigma$ \eqref{convolutiongen} are indeed correct asymptotically. Notice that  \eqref{convolutiongen} confirms the assertions below \eqref{coeffgen} in sec. \ref{sec:open}, that the $\vpi$-integral in the stripped coefficient function $\rg{f}^\sigma_\Lambda$ \eqref{ringGexp} fails to converge for $\Lambda\cu>a\Lambda_\p$ due to exponentials with positive exponents, namely \eqref{piexponentials} with $\Lambda'\cu=\Lambda$.

Now we extract from the general convolution formula for $\ff^\sigma$ \eqref{convolutiongen} the general form of the flow of the couplings at large $\Lambda$. 
The monomials in $\Gamma_1$ that yield $\sigma^a$ in the expansion of the RHS one-loop formula  \eqref{twoexpand}, start with dimension $5\cu-\varepsilon_a$ according to whether a factor of $\vp$ has been absorbed in their coefficient function. However $\sigma^a$ will have less fields if some have been eliminated by attaching propagators. This is true equally of $\sigma^b$. Let $n_f$ be the total number of fields eliminated from these monomials in forming $\sigma^a$ and $\sigma^b$. 
%Also recall that $d$ space-time derivatives were eliminated to form \eqref{generalFeynman}. 
The final monomials \eqref{defsigma} arise from applying $(-\Box)^m\partial^r$, gaining $n$ new appearances of $\vp$, as explained below \eqref{kraction}. Recalling that $d$ space-time derivatives were eliminated, and since space-time derivatives and these fields all have dimension one, the monomials $\sigma$ have dimension
\beal 
d_\sigma = [\sigma] &= 10 - \varepsilon_a -\varepsilon_b -n_f - d + 2m + r +n\,,\nn\\
&= 10 + 2(m-c) + n-n_f - \varepsilon_a -\varepsilon_b\,,
\label{dsigma}
\eeal
where in the second line we used the definition of $c$ \eqref{rcd}. Since the propagators make $2(\ell\cu+1)$ attachments, and $\vp$-propagators can also attach to $f^a$ or $f^b$, $n_{a\ell}$ ($n_{b\ell}$) times respectively, we have that
\be 
\label{nanbnf}
n_{a\ell}+n_{b\ell}+n_f = 2(\ell+1)\,.
\ee
The parity of $\rg{f}^\sigma$ is given by the parity of the product in its flow equation \eqref{flowfo}:
\beal
\label{epseqn}
\varepsilon &= \varepsilon_a +\varepsilon_b+n_{a\ell}+n_{b\ell}+n \ \mod 2\,,\nn\\
&= \varepsilon_a +\varepsilon_b+n_f+n \ \mod 2\,,
\eeal
where in the second line we use \eqref{nanbnf}. From dimensions, \eqref{gdimension} and \eqref{dsigma}, its couplings \eqref{coeffgen} have dimension
\be 
\label{gtwodimensions}
[g^\sigma_{2l+\varepsilon}] = 2(l+c-m) +\varepsilon+ \varepsilon_a +\varepsilon_b+n_f-n-5\,.
\ee
Using the parity equation \eqref{epseqn}, we see that their dimension is always odd. There are therefore no marginal couplings at second order. Furthermore since the first order couplings only have even dimensions \eqref{gonedimensions}, it is clear by dimensions that the first-order operators $\sigma\dd\Lambda{n}\cu\in\Gamma_1$ cannot appear on the RHS of the $\rg\Gamma_2$ flow equation \eqref{flowallRHS}, \ie that the first-order couplings $g^a_{2l+\eps_a}$ do not run at second order. 

These results are consequences of the decision to work with coefficient functions of definite parity under $\vp\cu\mapsto-\vp$ \cite{first}, and do not apply without this, evidently so for the second order couplings to have only odd dimension, but also in protecting first-order couplings from running at second order. To see why this is true in the latter case, consider trying to reproduce the first-order operators in the $f^1_\Lambda$ terms  (\ref{Gammaonetwo},\ref{Gammaoneone},\ref{Gammaonezero}), by substituting only these parts into the melonic expansion fig. \ref{fig:melons}, and retaining only lowest order in the derivative expansion. In this case the two first-order monomials contribute three fields each. Therefore in order to reproduce these operators, the propagators have to eliminate three of these fields. This means that the propagators must attach an odd number of times to the $f^1_\Lambda$. If we insist that $f^1_\Lambda$ is even parity, and thus contain only $\dd\Lambda{2n}$, the result must then be overall parity odd and thus contain only  $\dd\Lambda{2n+1}$. One can continue this exercise now including also the $f^{1_1}_\Lambda$ parts and allowing for derivative expansion, and still one always finds that parity excludes the first-order operators from being reproduced at second order. 

However these same considerations demonstrate that at 
%by hand these first-order operators, substituting  (\ref{Gammaonetwo},\ref{Gammaoneone},\ref{Gammaonezero}) into the melonic expansion fig. \ref{fig:melons} and expanding in derivatives as necessary. One finds that if the first order monomial $\sigma$ is reproduced it is always accompanied by a $\dd\Lambda{n}$ of the wrong parity. The same exercise indicates however that at
third order, which will be the generalisation of fig. \ref{fig:melons} to the sum of all one-particle irreducible Feynman diagrams containing exactly three vertices,  it is possible to reproduce the operators in $\Gamma_1$, \ie at third order in perturbation theory the first order couplings will run. Indeed returning again to the purely even parity parts of $\Gamma_1$ and lowest order in the derivative expansion, we now need to eliminate six fields in order to reproduce these operators, which can be done by attaching three propagators, leaving a vertex with $(f^1_\Lambda)^3$ as a factor, and thus now containing the required even parity operators, $\dd\Lambda{2n}$.
 
Taylor expanding the first order conjugate momentum coefficient functions in the general convolution formula for $\ff^\sigma$ \eqref{convolutiongen} we get convergent $\vpi_1$ integrals which yield the $\beta$ functions for $g^\sigma_{2l+\eps}$ in the form we have already seen for the model system \eqref{beta-physical}:
\be 
\label{beta-gen}
\dot{g}^\sigma_{2l+\eps} = \sum_{a,b} \sum_{l_a,l_b=0}^\infty \rg{C}^{\,\sigma,l}_{a,l_a,b,l_b}\, g^{a}_{2l_a+\eps_a} g^{b}_{2l_b+\eps_b}\, \Lambda^{2(l+c-m-l_a-l_b)+\eps-\eps_a-\eps_b+n_f-n-5}\,,
%\dot{g}^\sigma_{2l+\eps}(\Lambda) = \sum_{i_j=1,1_1}\,\sum_{l_1,l_2=0}^\infty \rg{C}^{\,\sigma,l}_{i_1,l_1,i_2,l_2}\, g^{i_1}_{2l_1+\eps_1} g^{i_2}_{2l_2+\eps_2} \Lambda^{2(l+c-m-l_1-l_2)+\eps-\eps_1-\eps_2+n_f-n-5}\,,
\ee
where now the first sum is a sum over the options $a,b=1,1_1$. The second sum is over the first order couplings, the $\rg{C}$ being dimensionless numbers, and the power of $\Lambda$ following by dimensions  \eqref{gonedimensions}. This can be integrated immediately, 
%with integration constant $\mathring{g}^\sigma_{2l+\varepsilon}$, 
giving a closely similar result to the formula for the model system \eqref{gtwofours}: 
\besp 
\label{gtwos}
g^\sigma_{2l+\eps}(\Lambda) = 
\sum_{a,b} \sum_{l_a,l_b=0}^\infty \frac{\rg{C}^{\,\sigma,l}_{a,l_a,b,l_b}\, g^{a}_{2l_a+\eps_a} g^{b}_{2l_b+\eps_b}}{2(l_a\cu+l_b\cu+m\cu-l\cu-c)+5\cu+n\cu+\eps_a\cu+\eps_b\cu-n_f\cu-\eps} \, \Lambda^{2(l+c-m-l_a-l_b)+\eps-\eps_a-\eps_b+n_f-n-5}\\ + \mathring{g}^\sigma_{2l+\varepsilon}\,.
\eesp
In particular by the parity equation \eqref{epseqn} only odd powers of $\Lambda$ appear in \eqref{beta-gen}, as is also clear from the fact that the first order couplings have only even dimensions \eqref{gonedimensions} and the second order couplings only have odd dimensions \eqref{gtwodimensions}. Thus no $\ln\Lambda$ terms are generated by integration.
Using $\Lambda$ to cast everything in dimensionless terms, the result will depend on $\Lambda$ only implicitly through the scaled couplings: 
\be 
\label{dimensionless}
\tg^1_{2n}(\Lambda) = g^1_{2n}/ \Lambda^{2n}\,, \qquad \tg^{1_1}_{2n+1}(\Lambda) = g^{1_1}_{2n+1}/\Lambda^{2n+2}\,.
\ee
In the limit $\Lambda\cu\to\infty$, the sums therefore collapse as before to dependence only on $g^1_0$:
\be 
\label{fpline}
\tilde{g}^\sigma_{2l+\eps}(\Lambda)\to \tilde{g}^\sigma_{2l+\eps\,*} = \frac{\rg{C}^{\,\sigma,l}_{1,0,1,0}}{2(m\cu-c\cu-l)\cu+5\cu+n\cu-n_f\cu-\eps}\, (g^1_0)^2\,, \qquad\text{as}\quad\Lambda\to\infty\,,
\ee
this being the parametrisation of the line of UV fixed points $\tilde{g}^\sigma_{2l+\eps\,*}(g^1_0)$.
Here we have used the fact that the dimensionless version of the integration constant, $\Lambda^{-[g^\sigma_{2l+\varepsilon}]}\,\mathring{g}^\sigma_{2l+\varepsilon}$, vanishes for relevant couplings. These $\mathring{g}^\sigma_{2l+\varepsilon}$ remain as the freely adjustable finite parts. On the other hand $\Lambda^{-[g^\sigma_{2l+\varepsilon}]}\,\mathring{g}^\sigma_{2l+\varepsilon}$ would diverge for irrelevant couplings unless we set their $\mathring{g}^\sigma_{2l+\varepsilon}\cu=0$. The irrelevant couplings are thus solely determined by the sum in \eqref{gtwos}. Therefore, just as we saw in the model system of sec. \ref{sec:model}, the UV regime correctly describes a continuum limit, through its renormalized trajectory, provided that the irrelevant couplings are defined by integrating down from $\Lambda\cu=\infty$, with boundary condition that they vanish there.

As before, the sum in the solution \eqref{gtwos} converges only for $\Lambda\cu>a\Lambda_\p$. However now we notice that the flow equation for $\rg\Gamma_2$ \eqref{flowclosed} can be rewritten as:
\be 
\label{Poexpression}
\partial_t\,\rg{\Gamma}_2 =
 \partial_t \left[\,\half\, (1+\Po_\Lambda)\,\mathrm{e}^{\Po^\Lambda}\, \Gamma_{1\,\text{phys}}\,\Gamma_{1\,\text{phys}}\, \right]\,,
\ee
and can therefore be integrated exactly to give 
\be 
\label{ringGtwo}
\rg{\Gamma}_2 = \half\left[\, (1+\Po_\Lambda)\,\mathrm{e}^{\Po^\Lambda} - (1+\Po)\,\right] \Gamma_{1\,\text{phys}}\,\Gamma_{1\,\text{phys}} +\Gamma_{2\,\text{phys}}\,.
\ee
The additional terms are integration constants and are determined by the requirement that $\rg{\Gamma}_2$ coincide with the physical $\Gamma_{2\,\text{phys}}$ in the limit $\Lambda\cu\to0$, as is clear must be the case from the definition of $\rg\Gamma_2$  \eqref{ringG} since the exponent on the RHS there, and $\Po^\Lambda$, must vanish in this limit, while $\Po=\Po_{\Lambda\to0}$ is the analogous expression to $\Po^\Lambda$ \eqref{Posimple} but without regularisation. The $(1+\Po)$ term can also be seen to be required in order to subtract the disconnected and tree-level part one gets from expanding the first term in the square brackets above. The rest of its expansion gives again the melonic diagrams of fig. \ref{fig:melons}. 

As it stands, expression \eqref{ringGtwo} is however not very useful. Although clearly it is UV regularised, again when $\Lambda\cu>a\Lambda_\p$ it only makes sense in conjugate momentum space  but worse, $\Gamma_{2\,\text{phys}}$ presupposes the answer and by definition has no IR regularisation
%well-defined, \TRM{as we show later,} \notes{or maybe not bother} it is not very useful in this form,
while also
$\mathrm{e}^{\Po^\Lambda}$ appears also without a factor providing IR regularisation. The parts without IR regularisation are defined only at non-exceptional momenta, and have no derivative expansion. We require solutions $\Gamma_2$ (or equivalently through its definition \eqref{ringG}, $\rg\Gamma_{2}$)  that do have a derivative expansion when $\Lambda\cu>0$, both to define the monomials $\sigma$ and thus the RG flow and, as recalled in sec. \ref{sec:prelim}, because it implements locality. The derivative expansion property will only be manifest if we can express the solution in terms of IR regulated propagators. 

At first sight there is an elegant solution using only propagators $\prop_\Lambda$, defining the melonic integrals through appropriate UV subtractions where the UV cutoff has already been removed, much as was done in ref. \cite{\yuji}. Although this defines each Feynman diagram contribution, the resummed solution $\Gamma_{2\Lambda}$ (where we now make explicit the $\Lambda$ dependence) has coefficient functions that become singular for $\Lambda\cu\ge a\Lambda_\p/\sqrt{\textrm{e}-1}$, %sufficiently above the amplitude suppression scale 
after which the flow ceases to exist, as we show in app. \ref{app:fails}. 
A somewhat similar problem appears if we try to compute it from a bare action \cite{\morri}, as we confirmed for the model \eqref{downlimit}, and will see again below. 

Instead we need to define the solution in terms of $\Gamma_{2\mu}$, its value at another finite non-zero scale $\Lambda\cu=\mu\cu>0$, which we can do by subtracting from our solution for $\rg\Gamma_2$ \eqref{ringGtwo} the same expression evaluated at $\mu$: 
\be
\label{ringGtwomu}
\rg{\Gamma}_{2\Lambda} = 
 \half\left[\, (1+\Po_\Lambda)\,\mathrm{e}^{\Po^\Lambda} - (1+\Po_\mu)\,\mathrm{e}^{\Po^\mu}\,\right] \Gamma_{1\,\text{phys}}\,\Gamma_{1\,\text{phys}} +\rg\Gamma_{2\mu}\,.
\ee
Recalling the definitions of UV and IR cutoff propagators, below \eqref{mST}, let us generalise to
\be 
\label{propmuLambda}
\prop^\mu_\Lambda = -\prop^\Lambda_\mu = \prop_\Lambda -\prop_\mu = \prop^\mu - \prop^\Lambda =
\left[ C(p^2/\mu^2) - C(p^2/\Lambda^2) \right] \prop(p)\,.
\ee
This propagator is regularised in both the UV and the IR, with $\mu$ and $\Lambda$ performing these r\^oles as determined by whichever is the larger cutoff. Writing correspondingly $\Po^\mu_\Lambda$ via the obvious change to its definition \eqref{Posimple}, it is evident that the melonic expansion of \eqref{ringGtwomu}  has $\Po^\mu_\Lambda\cu=\Po_\Lambda\cu-\Po_\mu$ as a factor, and thus, recalling the general situation stated above \eqref{allposition}, all melonic Feynman integrals are now IR regulated, by $\Lambda$ and/or $\mu$.

In conjugate momentum space, \eqref{ringGtwomu} is the exact version for which \eqref{ftwofour} is the model, as we will see explicitly in \eqref{ffgeneral}, except that here the $\Lambda'$-integral can be performed exactly (see also the later footnote \ref{foot:exact}). 
%This latter will act as the boundary condition. 
%We will see here also that the solution is well defined for all $\Lambda\cu\ge0$, provided $\mu$ lies in the range \eqref{murange}. 
It is now manifest that additionally $\rg{\Gamma}_{2\Lambda}$ has a derivative expansion provided $\Lambda\cu>0$. Now we show that its derivative expansion leads to well-defined coefficient functions $f^\sigma_\Lambda(\vp)$ \eqref{coeffgen} for all $\Lambda\cu>0$ however provided that, just as we saw in the model, $\mu$ is in the range \eqref{murange}. For this we need the large-$\ell$ asymptotic form which we again compute by specializing to the exponential cutoff \eqref{Cexp}. Although we can now repeat the previous analysis, it is clear that actually the asymptotic behaviour \eqref{naivel} follows from the exponential $\mathrm{e}^{\Po^\Lambda}$ in the flow equation \eqref{flowclosed} independently of the other details. (The detailed changes only affect the values of  $k$ in $I^k_\ell$ \eqref{In}, which only affect the power corrections to the asymptotic value of the sum \eqref{largellsum} as before.) Thus for a given match for $\sigma$ \eqref{defsigma}, we can see that the above solution for $\rg\Gamma_2$ \eqref{ringGtwomu} implies 
\beal
\label{ffgeneral}
\ff^\sigma(\vpi,\Lambda) &= \ff^\sigma(\vpi,\mu)\\ + &\sum_{a,b} \int^\infty_{-\infty}\frac{d\vpi_1}{2\pi}\,\,\ff^a\!\left(\vpi_1\cu+\frac\vpi2\right) \ff^b\!\left(\vpi_1\cu-\frac\vpi2\right) \left\{\, M^{ab}_\Lambda(\vpi_1,\vpi) \,{\rm e}^{\left(\frac{\vpi^2}{4}-\vpi^2_1\right)\Omega_\Lambda}
-M^{ab}_\mu(\vpi_1,\vpi) \,{\rm e}^{\left(\frac{\vpi^2}{4}-\vpi^2_1\right)\Omega_\mu} \right\}\,,\nn
\eeal
where the exponentials capture the asymptotic behaviour of the two parts in square brackets in \eqref{ringGtwomu}, and $M^{ab}$ the multiplicative remainders from the sums analogous to those in the flow equation for $\ff^\sigma$ \eqref{flowff}. Recalling the asymptotic behaviour of the first order coefficient functions for large $\vpi$ \eqref{largevpibar} it is easy to see that 
asymptotically there are no convergence issues in the $\vpi_1$ integral. The second-order coefficient functions \eqref{coeffgen} are constructed by multiplying the above by ${\rm e}^{-\frac{\vpi^2}{2}\Omega_\Lambda+i\vpi\vp}$ and integrating over $\vpi$. There are still no issues with the first term in braces, but by collecting $\vpi^2$ exponents using the first-order formula \eqref{largevpibar}, we see that the second term reproduces the exponential factors we found in the model \eqref{piexponentials}, with $\Lambda'$ replaced by $\mu$. Thus the particular integral part of $f^\sigma_\Lambda(\vp)$ becomes singular before $\Lambda$ reaches zero, unless $\mu$ is in the range \eqref{murange}, just as we saw in the model and with the same implications that were addressed there. 
%This means that $\mu$ is bounded above by $a\Lambda_\p$ unless we arrange the complementary solution $\ff^\sigma(\vpi,\mu)$ to cancel all the leading and subleading poor asymptotic behaviour in the particular integral at all $\Lambda$, just as we saw in the model. 

Now consider the limit $\mu\cu\to0$. In this case of course our subtracted solution for $\rg\Gamma_2$ \eqref{ringGtwomu} returns to the earlier solution \eqref{ringGtwo}. $\rg{\Gamma}_{2\,\Lambda\cu>0}$ still has the derivative expansion property, but it now arises only by delicate cancellation of IR divergences between the $\mu$-dependent terms in the general convolution solution above \eqref{ffgeneral}. In particular $\ff^\sigma(\vpi,\mu)$ is given by the complementary solution \eqref{homoirr}, except again the integral over $\Lambda'$ is exact,\footnote{See the later footnote \ref{foot:exact} for how to construct a model with this property.} fixing the irrelevant couplings uniquely so that they vanish in the $\mu\cu\to\infty$ limit. The opposite limit $\mu\cu\to0$, involves IR divergences for some of these irrelevant couplings which must cancel against $\mu\cu\to0$ divergences in the corresponding coupling in the particular integral in \eqref{ffgeneral}. 
%This observation will prove useful \TRM{later} when we derive the large $\Lambda_\p$ limit. 
It will be useful to extract these, which we do by again casting the particular integral in position space, so that it takes a similar form to the RHS of the flow equation in position space  \eqref{allposition}. At $\ell$-loop order the dominant contribution comes from $\ell\cu+1$ propagators decaying as $1/x^2$ at large $x$ up to the IR cutoff at $x\cu\sim 1/\mu$.  Recalling that these propagators appear $d$ times differentiated, where $d$ is related to $c$ by \eqref{rcd},
we see that the derivative expansion coefficients therefore diverge for $\mu\cu\to0$, to leading order as:
\beal 
\label{smallmupow}
D^m_{rc\ell} &= c^m_{rc\ell}/\mu^{2(1+m-c-\ell)}\,, &\ell<1+m-c\,,\qquad\qquad\\ 
&= c^m_{rc,1+m-c} \ln\mu\,, &\ell=1+m-c\,,\qquad\qquad
\label{smallmuln}
\eeal
for some constants $c^m_{rc\ell}$, where $\ell\cu\ge\ell_\sigma$, the first loop order at which $\sigma$ \eqref{defsigma} appears. (In fact the only cases where $\ell_\sigma\cu>1$, have $\ell_\sigma\cu=2$ and correspond to $\sigma^a\cu=\sigma^b\cu=1$ in the expansion of the flow equation for $\rg\Gamma_2$ \eqref{allexpand}, as is clear from the discussion there. This is still an infinite set however, corresponding to all monomials $\sigma$ generated by the derivative expansion $F_{rc\ell}(i\partial)$ \eqref{Dexp} acting on $f^b$ in the expansion \eqref{allexpand}.) Introducing the index
\be 
\label{isigma}
i_\sigma = 1+m-c-\ell_\sigma\,,
\ee
the IR divergences do not appear if $i_\sigma<0$, otherwise they get progressively weaker for larger $\ell$, and then disappear when $\ell>1\cu+m\cu-c = \ell_\sigma\cu+i_\sigma$. In fact even at fixed $\ell$, there are such subleading divergences,  for example for \eqref{smallmupow} the next-to-leading subdivergences are $\sim\mu^{2(\ell+c-m)}$ for $\ell\cu<m\cu-c$, or $\sim\ln\mu$ for $\ell\cu=m\cu-c$. Overall, we see that the dominant behaviour comes from $\ell\cu=\ell_\sigma$, \ie for $i_\sigma\cu>0$, by the term $\propto1/\mu^{2i_\sigma}$ \eqref{smallmupow}, and for $i_\sigma\cu=0$ by \eqref{smallmuln}.

%%In \eqref{smallmuln} we have recognised that the integral is UV sensitive and regularised by $\Lambda$, although this is not the complete story as we explain below. 
%The IR divergences do not appear if $m\cu< c$,  (In fact even at fixed $\ell$, there are such subleading divergences,  for example for \eqref{smallmupow} the next-to-leading subdivergences are $\sim\mu^{2(\ell+c-m)}$ for $\ell\cu<m\cu-c$, or $\sim\ln\mu$ for $\ell\cu=m\cu-c$.) Therefore the dominant behaviour is determined by $\ell\cu=\ell_\sigma$, the first loop order at which $\sigma$ in \eqref{defsigma} appears. 

The model in sec. \ref{sec:model} thus gives exactly the correct small $\mu$ behaviour of the $m\cu=c\cu=0$ case (at antighost level four), the log divergence appearing at $\ell_\sigma\cu=1$ loop, consistently with \eqref{smallmuln}, since the one-loop contribution is given exactly by the $t$-integral of the RHS of its flow equation  \eqref{twofourRHS} (and which by the Feynman integral \eqref{Dp} after writing $\dot{C}^\Lambda = -\dot{C}_\Lambda$ can be performed exactly). The log divergence appears after taking the lowest order ($D_0$) contribution \eqref{Dzero} in the derivative expansion of the Feynman integral \eqref{D}. Indeed we see from the solution for $\ff^{\sigma_0}$ \eqref{ftwofour} that the particular integral is log divergent as $\mu\cu\to0$. 

Finally, writing $\Gamma_2$ in terms of $\rg\Gamma_2$ \eqref{ringinvert} we cast the subtracted solution \eqref{ringGtwomu} in terms of the \textit{bona fide} coefficient functions and thus in a form that genuinely exists in field-space at all $\Lambda\cu\ge0$.
Substituting the latter into the former, we again split the exponential over the bilinear term  \eqref{WickTwoFactorId}. This gives a factor $\mathrm{e}^{-\Po^\Lambda}$  and converts the two $\Gamma_{1\,\text{phys}}$ to $\Gamma_{1\Lambda}$, as follows from the first-order solution  \eqref{eigenoperatorsol}. Defining 
\be 
\label{complementary}
\Gamma_{2\Lambda}(\mu) = 
\exp\left(-\frac12 {\prop}^{\Lambda\,AB} \frac{\partial^2_l}{\partial\Phi^B\partial\Phi^A}\right)\,
\rg\Gamma_{2\mu} \ =\ \sum_\sigma \left( \sigma f^\sigma_\Lambda(\vp,\mu)+\cdots \right)\,,
\ee
so as to absorb the exponential acting on the last term, we see from expanding $\rg\Gamma_{2\mu}$ using stripped coefficient functions \eqref{ringGexp} with $\Lambda\cu\mapsto\mu$, that the result constructs the complementary solution coefficient functions \eqref{homogeneous}, the top terms being accompanied by tadpole corrections as indicated by ellipses, these being generated with the now-standard prescription \eqref{firstOrder}. 
We have thus shown that the solution of the second-order flow equation \eqref{flowtwo} can be written as
\be 
\label{Gammatwosol}
\Gamma_{2} = \half \left[\, 1+\Po_\Lambda-(1+\Po_\mu)\,\mathrm{e}^{\Po^\mu_\Lambda}\,\right] \Gamma_{1}\, \Gamma_{1} + \Gamma_{2}(\mu)\,.
\ee
Since all parts, $\Gamma_{2\Lambda}$, $\Gamma_{1\Lambda}$ and $\Gamma_{2\Lambda}(\mu)$, are now evaluated at $\Lambda$, we revert to suppressing the $\Lambda$ dependence subscript. 
The complementary solution \eqref{complementary} satisfies just the (linearised) LHS of the second-order flow equation \eqref{flowtwo}, with boundary condition: 
\be 
\label{complementarybc}
\Gamma_2(\mu)  = \Gamma_2 \quad\text{at}\quad\Lambda=\mu \,.
\ee 
Its coefficient functions satisfy the corresponding boundary condition \eqref{equalitymu} as before. Its physical limit $\Gamma_{2\,\text{phys}}(\mu) = \rg{\Gamma}_{2\mu}$  is nothing but the stripped version (compare also the first-order solution (\ref{eigenoperatorsol}) and the complementary solution  (\ref{complementary}) above).
We recognise that the bilinear term in the above solution \eqref{Gammatwosol} is the particular integral corresponding in the model truncation to the second term in \eqref{gensolution}.\footnote{\label{foot:exact}If we alter the last term in the model flow equation \eqref{flowftwofour} to $D_0\,\Omega^2_\Lambda \left(f^{1\prime\prime}_\Lambda\right)^2$, the particular integral in the model can also be done exactly, however such a term does not arise in the truncated flow equation of any coefficient function in the current study. 
In app. \ref{app:alt} we show that the solution \eqref{Gammatwosol} can be derived directly from the second-order flow equation \eqref{flowtwo} without going through $\vpi$-space $\rg\Gamma_2$ expressions.}
%by following similar steps, using:
%\be 
%%\label{integratingfactor}
%\nn
%\frac{\partial}{\partial t}\left\{  \exp\left(-\frac12 {\prop}^{\mu\,AB}_\Lambda \frac{\partial^2_l}{\partial\Phi^B\partial\Phi^A}\right) \Gamma_{2\Lambda} \right\} = -\frac12\exp\left(-\frac12 {\prop}^{\mu\,AB}_\Lambda \frac{\partial^2_l}{\partial\Phi^B\partial\Phi^A}\right)  \text{Str}\, \dot{\prop}_\Lambda \Gamma^{(2)}_{1\Lambda} \propH \Gamma^{(2)}_{1\Lambda} \,.
%\ee}
%then following similar steps to those above.}
Expanding the exponential, the $1+\Po_{\Lambda}$ and $1+\Po_{\mu}$ parts cancel the disconnected and one-particle reducible contribution. The remaining terms form the infinite series of melonic Feynman diagrams, starting at one loop, as in fig. \ref{fig:melons}, all of which are individually finite since they are regularised in both the UV and the IR by $\mu$ and $\Lambda$.

Together with the complementary solution \eqref{complementary}, the solution \eqref{Gammatwosol} can be expanded in a closely similar way to the flow equation for $\rg\Gamma_2$ \eqref{allexpand}: 
\be 
\label{GamtwoExp}
\Gamma_2 %= \sum_\sigma \left( \sigma f^\sigma_\Lambda(\vp)+\cdots \right)
= \sum_\sigma \left( \sigma f^\sigma_\Lambda(\vp,\mu)+\cdots \right)
+  \sum_{rc\ell} \sum_{\sigma^a\sigma^b} \sigma^a\! f^{a(n_{a\ell})}_\Lambda F_{rc\ell}(i\partial)\,\sigma^b\!f^{b(n_{b\ell})}_\Lambda\,,
\ee
%where in the first equality we also recall the expansion over $\sigma$, \eqref{ringinvert}.  
where $F_{rc\ell}$ again takes the form \eqref{Frcl}. 
%\old{Organising the derivative expansion in the same way gives the analogous form to \eqref{flowfo}, where the sum is again over matches \eqref{defsigma}:}
%\be 
%f^\sigma_\Lambda(\vp) + 
Provided $\mu$ is chosen in the range \eqref{murange}, the $\vpi$ integral  \eqref{ringGexp} defining the stripped coefficient functions in $\rg\Gamma_{2\mu}$, converges. Thus it is safe to take the $\Lambda\cu\to0$ limit of the complementary solution \eqref{complementary}.  However the derivative expansion of $F_{rc\ell}(i\partial)$ \eqref{Dexp} ceases to exist in this limit (holding all other quantities fixed). Indeed from the symmetry of the subtracted solution for $\rg\Gamma_2$ \eqref{ringGtwomu}, we get the same behaviour (\ref{smallmupow},\ref{smallmuln}) we previously derived for $\mu$, with the r\^oles $\mu\leftrightarrow\Lambda$ swopped over. But this corresponds to Taylor expanding in $p$ in $F_{rc\ell}(p)$ \eqref{Frcl},  the total external momentum entering through either vertex in the melonic diagrams. Instead in this limit one should treat the external momentum dependence exactly, recovering the usual IR properties expected of a massless theory (see also \eg \cite{\yuji,Morris:2018zgy}). In particular here the results will be finite provided we stay away from the one exceptional momentum case where $p$ vanishes or is null.

By deriving in great generality such a well-defined renormalized trajectory,  we have thus derived in great generality the continuum limit at second order.  Together with the infinite number of couplings at first order, it is parametrised by an infinite collection of new relevant couplings, in fact an infinite number for every new monomial $\sigma$, packaged into the coefficient functions \eqref{coeffgen}. All this will simplify dramatically however when we take the large amplitude suppression scale limit.

\subsection{Large amplitude suppression scale and the mST}
\label{sec:largeASS}

Having formed the continuum limit we are ready now to take the limit of large amplitude suppression scale, leaving everything else finite. All the second order coefficient functions currently have their own `internal' amplitude suppression scales which are set by their relevant couplings and govern only the complementary solution in \eqref{GamtwoExp}. We start with the particular integral, and send the first order amplitude suppression scale $\Lambda_\p\cu\to\infty$, giving the trivialisations  \eqref{fonelimit}. As soon as the $f^{a,b}_\Lambda$ are differentiated more than $\varepsilon_{a,b}$ times, they will vanish as a power of $1/\Lambda_\p$ \eqref{regularity}. Thus in the limit the melonic expansion terminates at two loops, and indeed the particular integral in the solution  \eqref{Gammatwosol} is given by simply sending $\Gamma_1\cu\to\kappa\cG_1$  \eqref{Gammaonelimits}. Now the two-loop term comes exclusively from the level-zero first-order vertex \eqref{Gammaonezero}, and is a pure number (carries no field dependence), so it too can be discarded. Expanding the solution \eqref{Gammatwosol} to $O(\Po^2)$ to pick up the one-loop contribution, and simplifying, we have thus shown that:
\be 
\label{partLASS}
\Gamma_2 %= \sum_\sigma \left( \sigma f^\sigma_\Lambda(\vp)+\cdots \right)
= \sum_\sigma \left( \sigma f^\sigma_\Lambda(\vp,\mu)+\cdots \right)
+\tfrac14\kappa^2\,\text{Str}\!\left[\proph{\mu} \check{\Gamma}^{(2)}_1 \proph{\mu} \check{\Gamma}^{(2)}_1
- \proph{\Lambda} \check{\Gamma}^{(2)}_1 \proph{\Lambda} \check{\Gamma}^{(2)}_1\right]\,.
\ee

%Since there is no tadpole operator \cite{\morrii} on the LHS of \eqref{mSTtwo} (compare  \eqref{flowtwo}), %%This is rubbish: the LASS limit is complicated because its a solution of a flow equation and thus depends on all scales
The large amplitude suppression scale limit of the mST is %much more 
straightforward. Since the last term in the second-order mST \eqref{mSTtwo} is regulated in both the UV and IR by $\Lambda$, 
%and thus in fact vanishes at non-exceptional momenta ($p\cu\ne0$) in the $\Lambda\cu\to0$ limit,
while the first term on its RHS has no momentum integral, we will again be able to neglect any term where an $f^a_\Lambda$ is differentiated more than $\varepsilon_a$ times. Thus here too we can simply make the substitution $\Gamma_1\cu\to\kappa\cG_1$  \eqref{Gammaonelimits}. Note however that the quantum correction in the level-zero first-order vertex \eqref{Gammaonezero}, its last term, will now make a contribution through the antibracket. Nevertheless we see that those parts of the flow equation and the mST in which the first-order vertex makes an explicit appearance, now take a form that is very close to what one would obtain in standard quantisation. This property will be fully explored in ref. \cite{second}.

We are left to decide on the complementary solutions $f^\sigma_\Lambda(\vp,\mu)$ in $\Gamma_2$. Although these are solutions of just the linearised flow equation, \ie LHS of the second-order flow equation \eqref{flowtwo}, they still depend on second-order quantum corrections through the irrelevant couplings in $\ff^\sigma(\vpi,\mu)$ \eqref{homoirr}. 
%therefore to ask whether appropriate complementary solutions $f^\sigma_\Lambda(\vp,\mu)$ in $\Gamma_2$ can be found so that the mST \eqref{mSTtwo} is satisfied. The difficulty is that  
We can however choose their relevant couplings. Therefore we now ask whether it is possible to constrain these so that the mST becomes satisfied in the limit of large amplitude suppression scale.

Consider first the antighost level four contribution that we sketched at the beginning of sec. \ref{sec:second}. Since the only way to attach propagators is to $f^1_\Lambda$, none of its particular integral survives this limit. 
%Indeed by inspection of \eqref{partLASS} using sec. \ref{sec:nontrivialcohomology}, antighost level two is the highest surviving level in the large $\Lambda_\p$ limit of the particular integral.  We still have all the complementary solutions \eqref{complementary} at antighost level three and four to determine however.
 Inspection of the second-order mST \eqref{mSTtwo} shows that it similarly has no RHS in this limit. %above antighost level two. 
Nevertheless we cannot immediately follow the argument %\TRM{\old{above sec. \ref{sec:twofour}}} 
at the beginning of sec. \ref{sec:second} and discard antighost level four, because of the complementary solutions' dependence on second order interactions  \eqref{homoirr}. We therefore have to  seek to constrain their relevant couplings so that the LHS of the mST becomes satisfied on its own.
%in the limit of large amplitude suppression scale. 
To aid in this process we can add further monomials (of antighost number four) with a different structure to the $\sigma$ already generated (\ie different even after integration by parts) such that the antighost level four part is annihilated by $Q_0$  (the first of the descent equations \cite{\morrii,\yuji,first} is satisfied), where $Q_0$ is the only part of the total free quantum BRST operator $\hs_0$ \eqref{mSTone} that does not lower the antighost number. Its only non-vanishing action is \cite{\morrii,first}
 \be 
\label{QH}
Q_0 H_{\mu\nu} = \partial_\mu c_\nu+\partial_\nu c_\mu
\ee
(in gauge invariant basis).  Similarly we must seek to ensure that the lower antighost levels (\ie the rest of the descent equations) are satisfied  in the limit, by adding new vertices with lower antighost number. However all these new vertices must thus also satisfy the linearised flow equations (\ref{flowone}) (and in fact contain only (marginally) relevant couplings). Since the entire complex therefore must be made to satisfy the linearised flow and mST equations, we can apply the quantum BRST cohomology results of sec. 7.2 of ref. \cite{\morrii,first} to prove that the only way to satisfy the mST is if all the coefficient functions trivialise \eqref{flatp}.
% or more generally the Hermite polynomials \eqref{flatp}. 

Despite the fact that the irrelevant couplings in the complementary solutions $f^\sigma_\Lambda(\vp,\mu)$ \eqref{homogeneous} have predetermined dependence on $\Lambda_\p$ (in fact in some cases divergent dependence as we will see), we know that we can place the relevant couplings in a range such that the solutions trivialise in the limit of large amplitude suppression scale, provided that certain convergence conditions are met as determined in sec. \ref{sec:fixcouplings}, namely the vanishing conditions \eqref{vanishesPoalpha} which depend on the dimensionless ratios \eqref{barg}.
Then actually the arguments %\old{above sec. \ref{sec:twofour}} 
at the beginning of sec. \ref{sec:second} now do apply, and we should choose them so that the solutions vanish in this limit, since if the vertices are allowed by the mST they are then anyway just a reparametrisation to an alternative non-trivial BRST cohomology representative \cite{\morrii,\yuji,first}
and if they are not allowed by the mST then we are forced to choose their corresponding coefficient so that $A_\sigma\cu\to0$.

\subsection{Large amplitude suppression scale in the model}
\label{sec:largeASSderiv0}

We gain detailed further insight about how to take the large amplitude suppression scale limit, by analysing  the model in sec. \ref{sec:model}. %Then we will be ready to analyse the full solution. 
First we evaluate its particular integral \eqref{gensolution} in this limit. Since the $\Lambda$ dependence in the exponentials in the convolution expression for $\dot{\ff}^\sigma(\vpi,\Lambda)$ \eqref{convolution}, provides perturbative corrections to the $\Lambda_\p$ dependence in the exponentials provided by $\ff^1$  \eqref{largevpibar}, they can be neglected. Then $\dot{\ff}^{\sigma}(\vpi,\Lambda')\cu\to\dot{\ff}^{\sigma}(\vpi,0)$ becomes independent of $\Lambda'$, so the integral over $\Lambda'$ in the particular integral \eqref{gensolution} is trivial. For the same reasons the exponential prefactor, ${\rm e}^{-\frac{\vpi^2}{2}\Omega_\Lambda}$, in the particular integral can also be neglected. Then the $\vpi$ integral just inverts the Fourier transform taking us back to the inhomogeneous term in the model flow equation \eqref{flowftwofour}, only at $\Lambda\cu=0$. Thus in the large $\Lambda_\p$ limit the model's particular integral collapses to
\be 
\label{modelcollapse}
 - D_0 \left[f^{1\prime\prime}(\vp)\right]^2 \ln\!\left(\frac{\Lambda}{\mu}\right)\,.
\ee
Substituted back into the $\sigma_0$ vertex \eqref{twofour}, this is in fact the correct large $\Lambda_\p$ behaviour of the lowest order term in the derivative expansion \eqref{D} of the antighost level four particular integral, the log dependence having also been identified earlier \eqref{smallmuln} and confirmed correct there. Finally, as already noted for the whole of antighost level four, the particular integral vanishes -- in this case by the detailed convergence relations \eqref{regularity} as $O(1/\Lambda_\p^4)$.

As above we are left to find the complementary solution in the large amplitude suppression scale limit, and in particular its irrelevant couplings.
The irrelevant couplings are computed via the formula for $\ff^\sigma(\vpi,\mu)$ \eqref{homoirr}, in the case of the model, as contained in the Taylor expansion \eqref{firrelevant} and the integrals \eqref{ftwofourirrelevant} evaluated at $\Lambda\cu=\mu$. 
%The limiting behaviour for the irrelevant couplings \eqref{ftwofourirrelevant} is more subtle since t
They are logarithmically IR divergent if $\mu\cu\to0$, but the UV scale appearing in this log will now be set by $\Lambda_\p$. Otherwise their behaviour is set by the $\kappa^2$ extracted from the product of $\ff^{1}$s and by
dimensions, \cf below  \eqref{coefftwofour}, to contain the overall factor of $\kappa^2\Lambda_\p^{2n-3}$, there being no other scales in the problem. Furthermore since the integrand in the explicit expressions \eqref{ftwofourirrelevant} depends on $\Lambda'$ only through its square, we know that corrections will be a power series in $\mu^2/\Lambda_\p^2$. Therefore we have already shown that
\be 
\label{gsigmazeroirrlimit}
g^{\sigma_0}_{2n}(\mu) = \kappa^2\Lambda_\p^{2n-3} \left[ b^{\sigma_0}_{2n}  \ln\!\left(\frac{a\Lambda_\p}{\mu}\right)
+ a^{\sigma_0}_{2n}  + O\!\left(\frac{\mu^2}{\Lambda^2_\p}\right) \right] \,,  \qquad n=0,1,2\,,
\ee
where the factor of $a$  in the log is included for convenience and the $a^{\sigma_0}_{2n}$ and $b^{\sigma_0}_{2n}$ are finite non-universal numbers. We see that  the irrelevant couplings  $g^{\sigma_0}_{0}(\mu)$ and $g^{\sigma_0}_{2}(\mu)$ vanish in the limit $\Lambda_\p\cu\to\infty$, while the irrelevant coupling $g^{\sigma_0}_{4}(\mu)$ actually diverges in this regime.

We now show how to compute these numbers, although we will not actually need them. At the same time this will confirm the large $\Lambda_\p$ form \eqref{gsigmazeroirrlimit} explicitly.
We extract these numbers by splitting the integral \eqref{ftwofourirrelevant} around some point of order $\Lambda_\p$, chosen to be $a\Lambda_\p$ for convenience, after which we can split off the log-divergence. Thus for $g^{\sigma_0}_{0}$ we get:
\besp 
\label{gtwofourzerosplit}
g^{\sigma_0}_{0}(\mu) = D_0\!\int^\infty_{a\Lambda_\p}\! \frac{d\Lambda'}{\Lambda'}\! \int^\infty_{-\infty}\!\frac{d\vpi_1}{2\pi}\,\, \vpi_1^4\, (\ff^1)^2 \, {\rm e}^{-\Omega_{\Lambda'}{\vpi^2_1}} 
+ D_0\!\int^{a\Lambda_\p}_\mu\! \frac{d\Lambda'}{\Lambda'}\! \int^\infty_{-\infty}\!\frac{d\vpi_1}{2\pi}\,\, \vpi_1^4\, (\ff^1)^2 \left( {\rm e}^{-\Omega_{\Lambda'}{\vpi^2_1}} -1\right)\\
+D_0 \ln\left(\frac{a\Lambda_\p}{\mu}\right) \int^\infty_{-\infty}\!\frac{d\vpi_1}{2\pi}\,\, \vpi_1^4\, (\ff^1)^2\,.
\eesp
Substituting as before $\ff^1$ in its general form \eqref{ffsigmaEg}, 
%with $A_\sigma\cu=\kappa$, $\Lambda_\sigma\cu=\Lambda_\p$, and $\bar{\ff}^1(\bar{\vpi}^2)$ having limiting behaviour \eqref{largevpibar}, 
and changing integration variables to the dimensionless
\be 
\label{uvars}
u=\bar{\vpi}_1=\Lambda_\p\vpi_1\qquad\text{and}\qquad v=\Lambda'/(a\Lambda_\p)\,,
\ee 
we release the expected overall $\kappa^2/\Lambda_\p^3$ factor. The first and third integrals then give convergent expressions that are independent of $\Lambda_\p$, while the second integral yields
\be 
\int^1_{\frac{\mu}{a\Lambda_\p}}\! \frac{dv}{v}\! \int^\infty_{-\infty}\!\!\!\!\!\!du\,\, u^4\, (\bar{\ff}^1)^2 \left( {\rm e}^{-u^2v^2/2} -1\right)\,,
\ee
where $\bar{\ff}^1\equiv\bar{\ff}^1(u^2)$.
Thanks to splitting out the $\Lambda'\cu=0$ part, the limit $v=\mu/(a\Lambda_\p)\cu\to0$ is convergent, and indeed we see that corrections to this limit start at $(\mu/a\Lambda_\p)^2$ and form a Taylor series in this small parameter. Thus we have verified the form \eqref{gsigmazeroirrlimit} for $n\cu=0$, and computed its coefficients:
\beal
\label{atwofourzero}
a^{\sigma_0}_0 &= 2\pi D_0\left\{ \int^\infty_1\! \frac{dv}{v}\! \int^\infty_{-\infty}\!\!\!\!\!\!du\,\, u^4\, (\bar{\ff}^1)^2 \, {\rm e}^{-u^2v^2/2} 
+ \int^1_{0}\! \frac{dv}{v}\! \int^\infty_{-\infty}\!\!\!\!\!\!du\,\, u^4\, (\bar{\ff}^1)^2 \left( {\rm e}^{-u^2v^2/2} -1\right) \right\}\,,\\
b^{\sigma_0}_0 &= 2\pi D_0 \int^\infty_{-\infty}\!\!\!\!\!\!du\,\, u^4\, (\bar{\ff}^1)^2\,.
\eeal
%Here $\bar{\ff}^1\equiv\bar{\ff}^1(u^2)$, and in the second integral for $a^{\sigma_0}_{0}$ the lower limit was $v=\Lambda/(a\Lambda_\p)\cu\to0$ was safely taken. 
For example, using the choice \eqref{ffoneEg} for $\bar\ff^1$, and the expression for $D_0$ \eqref{Dzero}, one finds
\be 
a^{\sigma_0}_0 = \frac{1}{8\sqrt{2\pi^3}}\left(1-\frac34\ln2\right)\qquad\text{and}\qquad b^{\sigma_0}_0 = -\frac{3}{32\sqrt{2\pi^3}}\,,
\ee
which indeed are the correct leading terms at large $\Lambda_\p$ followed by a series in $(\mu/a\Lambda_\p)^2$, as one confirms from taking this limit in the previously derived closed-form solution for this example \eqref{gtwofourzeroEg}. As discussed above the small $\mu$ limits \eqref{smallmuln}, the small $\mu$ divergence must cancel between the irrelevant couplings in the complementary solution, and that of the particular integral. The value for $b^{\sigma_0}_0$ verifies this, as can be confirmed by extracting the $g^{\sigma_0}_0$ contribution from this limit of the model's particular integral \eqref{modelcollapse} using the example expression for $f^1$ \eqref{foneEg} and integrating over $\vp$ \eqref{gnfphys}. From the explicit expression for $g^{\sigma_0}_{2}(\mu)$ \eqref{ftwofourirrelevant} we can see that it can be handled in the same way. (In fact the term with the $\Omega_{\Lambda'}$ factor is then IR ($\mu\cu\to0$) safe and so for it the split \eqref{gtwofourzerosplit} is unnecessary.)  Anyway it is immediately clear that we will recover the large $\Lambda_\p$ limiting form \eqref{gsigmazeroirrlimit}, and similarly for $g^{\sigma_0}_{4}(\mu)$.

Now recall that in this case we should constrain the relevant couplings in $\ff^{\sigma_0}_r(\vpi,\mu)$ \eqref{fcdef} so that $f^{\sigma_0}_\Lambda(\vp)$ vanishes in the limit of large amplitude suppression scale.\footnote{Using quantum BRST cohomology arguments, paying attention to further vertices, descendants,  and grading by derivatives \cite{\morrii}, one can 
prove that in fact in this case \emph{only} a vanishing $f^{\sigma_0}_\Lambda(\vp)$ can satisfy the mST.}
From sec. \ref{sec:fixcouplings} we can do this provided the ratios \eqref{barg} involving the irrelevant couplings, diverge slower than $\Lambda_{\sigma_0}^2$. 
%From the characterisation at the end of sec. \ref{sec:newquantisation}, we achieve this if we can find a physical $f^{\sigma_0}(\vp)$ with amplitude suppression scale $\Lambda_{\sigma_0}$, such that it vanishes point-wise in the limit of large $\Lambda_{\sigma_0}$, which must be taken in tandem with $\Lambda_\p\cu\to\infty$ in some way, and such that by \eqref{gnfphys} it also satisfies in this limit
%\be 
%\label{sigma0constraint}
%%\int^\infty_{-\infty}\!\!\!\!\!d\vp\, f^\sigma_{\text{phys}}(\vp) = 0\,,\quad
%%\int^\infty_{-\infty}\!\!\!\!\!d\vp\,\vp^2\, f^\sigma_{\text{phys}}(\vp) = 0\,,\quad
%\frac{1}{4!}\int^\infty_{-\infty}\!\!\!\!\!d\vp\,\vp^4\, f^{\sigma_0}(\vp) \sim
%a^{\sigma_0}_{4} \kappa^2\Lambda_\p + b^{\sigma_0}_{4} \kappa^2\Lambda_\p \ln\!\left(\frac{a\Lambda_\p}{\mu}\right)\,,
%\ee
%together with vanishing $\vp^{n=0,2}$ moments, corresponding to $g^{\sigma_0}_{0}(\mu)\cu=g^{\sigma_0}_{2}(\mu)\cu\to0$.
%
%The key then is to recognise the difference between point-wise and uniform convergence. There are clearly infinitely many solutions. 
Before proceeding further we need to decide on the relation between $\Lambda_{\sigma_0}$ and $\Lambda_\p$ in the large amplitude suppression scale limit. At least at finite order in perturbation theory, infinitely many choices are possible for this also, \eg motivated by the fact that the fourth moment \eqref{gnfphys} is proportional to the diverging $g^{\sigma_0}_{4}(\mu)$, one natural choice here might be to set $\Lambda_{\sigma_0}$ to its divergent part, \viz $a^{\sigma_0}_{4} \Lambda_\p + b^{\sigma_0}_{4} \Lambda_\p \ln\!\left({a\Lambda_\p}/{\mu}\right)$. 
However we find that we can make the simplifying assumption that all amplitude decay scales are identified. Thus we now set  $\Lambda_{\sigma_0}=\Lambda_\p$. Since $f^{\sigma_0}_\Lambda(\vp)$ is even, the simplest solution of the polynomial parametrisation \eqref{ffPoalpha} is to set $\alpha\cu=0$ and $\Po$ to a rank 3 polynomial in $\bar{\vpi}^2$. Setting for example
\be 
\label{Asigma0}
A_{\sigma_0} = \frac{\kappa^2}{\Lambda_\p^4}\,,
\ee
consistent with its dimensions \eqref{dimA}, means that the convergence conditions \eqref{vanishesPoalpha} are met since 
\be 
\label{bargsig0}
\bar{g}^{\sigma_0}_{2n}(\mu) = b^{\sigma_0}_{2n}  \ln\!\left(\frac{a\Lambda_\p}{\mu}\right)
+ a^{\sigma_0}_{2n}  + O\!\left(\frac{\mu^2}{\Lambda^2_\p}\right) \,,  \qquad n=0,1,2\,,
\ee
using the definition of this ratio \eqref{barg}. Then $f^{\sigma_0}_\Lambda(\vp)\cu\to A_{\sigma_0}$, clearly does vanish in the limit $\Lambda_\p\cu\to\infty$, in fact as $O(1/\Lambda_\p^{4})$. (By the discussion on the approach to the limit, \cf above \eqref{regularity}, the log divergence in the above \eqref{bargsig0} appears first at $O\left({\ln\!\Lambda_\p}\,/{\Lambda_\p^{6}}\right)$.) 

Notice that by having the same $\kappa^2$ factor in $A_{\sigma_0}$ as we have in the large $\Lambda_\p$ behaviour for $g^{\sigma_0}_{2n}(\mu)$ \eqref{gsigmazeroirrlimit}, we are guaranteed that $\bar{g}^{\sigma_0}_{2n}(\mu)$ is then independent of $\Lambda_\p$ save for the linear $\ln\!\Lambda_\p$ dependence, since $\bar{g}^{\sigma_0}_{2n}(\mu)$ is dimensionless and depends on $A_{\sigma_0}$ and $g^{\sigma_0}_{2n}(\mu)$ only through their ratio, and all the powers of $\Lambda_\p$ in the ratio \eqref{barg}, the large $\Lambda_\p$ behaviour for $g^{\sigma_0}_{2n}(\mu)$ \eqref{gsigmazeroirrlimit} and the above formula for $A_{\sigma_0}$ \eqref{Asigma0} then follow by dimensions.

We emphasise that we have thus constructed an infinite class of such solutions, since the dimensionless entire function $\bar{\ff}^{\sigma_0}$ appearing in the polynomial parametrisation  \eqref{ffPoalpha} satisfies the asymptotic behaviour \eqref{largevpibar} but is otherwise arbitrary.
There are however infinitely many other ways to achieve the vanishing of $f^{\sigma_0}_\Lambda(\vp)$ in the limit whilst taking into account that its first three couplings are determined and  satisfy the limiting behaviour \eqref{gsigmazeroirrlimit}. The choice above is just one of the simplest. Since we only need $A_{\sigma_0}$ to vanish rather than be fixed to some predetermined finite value, the polynomial parametrisation \eqref{ffPoalpha} is actually over-parametrised. It really only depends on the combination $A_{\sigma_0}\,\bar{g}^{\sigma_0}_{2n}(\mu)$. Then we can make an even simpler solution by taking a rank 2 polynomial $A_{\sigma_0}\,\Po(\bar{\vpi}^2)$ whose coefficients $A_{\sigma_0}\, p_r$ are all fixed by for example \eqref{pr}. In the large $\Lambda_\p$ limit, $f^{\sigma_0}_\Lambda(\vp)$ again vanishes, but this time as $O(\ln\!\Lambda_\p\,/\Lambda_\p^{4})$. 

Since $g^{\sigma_0}_{0}(\mu)$ and $g^{\sigma_0}_{2}(\mu)$ vanish in the limit, a further simplification would be to choose the relevant couplings so as to get vanishing $f^{\sigma_0}_\Lambda(\vp)$ in the case that these $g^{\sigma_0}_{0,2}(\mu)$ vanish identically, \ie base the parametrisation on the general form \eqref{ffforma} with $\alpha\cu=0$ but with $\bar{n}_{\sigma}\cu=2$. However this means that $f^{\sigma_0}_\Lambda(\vp)$ would be something smoothly vanishing plus the remainder $g^{\sigma_0}_{0}(\mu)\,\dd\Lambda0 + g^{\sigma_0}_{2}(\mu)\,\dd\Lambda2$ which, while vanishing for $\Lambda_\p\cu\to\infty$, has a non-smooth physical ($\Lambda\cu\to0$) limit, and thus may be problematic at higher orders. 

\subsection{General form of the large amplitude suppression scale limit}
\label{sec:lassgeneral}

We now work out the behaviour of all the (second-order) complementary solutions $f^\sigma_\Lambda(\vp,\mu)$ in the large-$\Lambda_\p$ limit. For this we need the large-$\Lambda_\p$ behaviour of the irrelevant couplings $g^\sigma_{2l+\varepsilon}(\mu)$ in the $\vpi$-Taylor expansion for all second-order coefficient functions \eqref{coeffgen},
which we can determine by generalising the analysis in the previous section. 
%These couplings are the negative-dimension coefficients in the $\vpi$-Taylor expansion \eqref{coeffgen} and are computed via the $\Lambda'$-integral in \eqref{homoirr}. 
Performing the $\vpi$-Taylor expansion on the convolution formula  for $\dot{\ff}^\sigma(\vpi,\Lambda)$ \eqref{flowff}, we see from its asymptotic form \eqref{convolutiongen} that we are left with convergent $\vpi_1$-integrals. Since the dimensionless couplings tend to the line of fixed points  \eqref{fpline} the result behaves as $\dot{g}^\sigma_{2l+\varepsilon}(\Lambda)\propto \Lambda^{[g^\sigma_{2l+\varepsilon}]}$ for large $\Lambda$, and therefore the $\int^\infty_\mu\! d\Lambda/\Lambda$ integral converges. Recall from the formula for the complementary solution  \eqref{homoirr}, that this integral fixes the irrelevant couplings. Apart from the factor of $\kappa^2$ coming from the two first order coefficient functions, it depends on only two dimensionful parameters, namely $\mu$ and $\Lambda_\p$. Therefore its large-$\Lambda_\p$ dependence is just given by the regime $\mu\ll\Lambda_\p$. The leading behaviour is thus fixed by the minimum power of $\Lambda$ appearing in the flow equation for the stripped coefficient function \eqref{flowfo} as determined by the dimensional relations \eqref{FderivLambda} for all matches for $\sigma$ \eqref{defsigma}, and in particular by the minimum loop $\ell\cu=\ell_\sigma$ where the first match occurs. Performing the $\int^\infty_\mu\! d\Lambda/\Lambda$ integral we see that the leading $\mu$ dependence reproduces the previously derived small-$\mu$ behaviour (\ref{smallmupow},\ref{smallmuln}), and similarly the subleading behaviours,
as required for these small-$\mu$ divergences to cancel between the complementary solution and particular integral. We also see that the $\Lambda_\p$ dependence is then fixed by saturating dimensions, which we can do using the explicit formula for $[g^\sigma_{2l+\eps}]$ \eqref{gtwodimensions}. Thus we have demonstrated that the irrelevant couplings have the following asymptotic behaviour:
\beal
g^\sigma_{2l+\varepsilon}(\mu)\ &\sim\ c^\sigma_{2l+\varepsilon}\, \frac{\kappa^2\ \ }{\mu^{2i_\sigma}}\, \Lambda_\p^{2l+n_f+\varepsilon_a+\varepsilon_b+\varepsilon-n-3}\,, &i_\sigma>0\,,\nn\\
g^\sigma_{2l+\varepsilon}(\mu)\ &\sim\ c^\sigma_{2l+\varepsilon}\, \kappa^2\ln\!\left(\!\frac{\Lambda_\p}{\mu\ }\!\right) \Lambda_\p^{2l+n_f+\varepsilon_a+\varepsilon_b+\varepsilon-n-3}\,, &i_\sigma=0\,,\nn\\
g^\sigma_{2l+\varepsilon}(\mu)\ &\sim\ c^\sigma_{2l+\varepsilon}\, \kappa^2\, \Lambda_\p^{2(l-i_\sigma)+n_f+\varepsilon_a+\varepsilon_b+\varepsilon-n-3}\,, &i_\sigma<0\,,
\label{girrgeneral}
\eeal
where $i_\sigma$ was defined in \eqref{isigma} and the $c^\sigma_{2l+\varepsilon}$ are finite dimensionless non-universal numbers.

In order for the large amplitude suppression scale limit of $\Gamma_2$ \eqref{partLASS} to satisfy the second-order mST \eqref{mSTtwo}, and thus ultimately BRST invariance (in the physical limit $\Lambda\cu\to0$),
we will have to choose the relevant couplings $g^\sigma_{2l+\varepsilon}(\mu)$ in the complementary solutions \eqref{homoirr}, so that these solutions trivialise appropriately \eqref{flatp} \cite{first}. Despite the fact that most of the irrelevant couplings above \eqref{girrgeneral} are diverging, from sec. \ref{sec:fixcouplings} we know we can do this, provided that the dimensionless ratios $\bar{g}^\sigma_{2l+\varepsilon}$ \eqref{barg}, diverge slower than $\Lambda^2_\sigma$. 

%Recall from sec. \ref{sec:prelim} \cite{first}, that 
Now the coefficient appearing in the trivialisation is $A_{\sigma}$, where in general we require a factor of $\vp^\alpha$ to appear in the physical limit \eqref{flatphys} in order to satisfy BRST invariance.
%However if we require a factor of $\vp^\alpha$ to appear in the physical limit in order to satisfy BRST invariance, or for other reasons, the monomial $\sigma$ referenced by $A_\sigma$ is not the one produced by \eqref{defsigma}, but rather the one defined via \eqref{sigmaalpha} where $\sigma$ is actually the monomial produced by \eqref{defsigma}. Equivalently one can just note that by dimensions this factor will appear as $(\vp/\Lambda_\p)^\alpha$, so to make the extra $\Lambda_\p$ dependence explicit we absorb it into $A_\sigma$. Either way we see that the dimension \eqref{dimA} will be corrected to
%\be 
%\label{dimAcorrected}
%[A_\sigma] = 4 - d_\sigma -\alpha\,.
%\ee
%For example we note that s
Since $\alpha$ must have the same parity as $f^\sigma_\Lambda(\vp,\mu)$, it has a minimum value $\alpha=\varepsilon$. 
%Thus if the parity is odd we must take into account that a factor of $\vp$ will in general survive the large amplitude suppression scale limit.
%We have already seen a version of this argument applied in the case of $\sigma'_{m,i}$ to arrive at \eqref{Asigmamip}, and indeed since $\alpha$ must have the same parity as $f^\sigma_\Lambda(\vp,\mu)$ it corresponds to the case where it has its minimum value $\alpha=\varepsilon$. 
Comparing the dimensions of the couplings \eqref{gdimension} and $[A_\sigma]$ \eqref{dimA} gives
\be 
\label{dimgA}
[g^\sigma_{2l+\varepsilon}] = [A_\sigma]+2l+\alpha+\varepsilon+1\,.
\ee
Since all the couplings have odd dimension \eqref{gtwodimensions}, we see that $A_\sigma$ always has even dimension.

%Now we show that for all cases we can do as in sec. \ref{sec:largeASSderiv0}, setting the
Now we show that for all cases we can set the corresponding amplitude suppression scale to a common value $\Lambda_\sigma\cu=\Lambda_\p$, and choose $A_\sigma$ to have the same power-law dependence on $\kappa$ and $\mu$ as the irrelevant couplings 
\eqref{girrgeneral}, with the remaining dependence $\Lambda_\p^{p_\sigma}$ determined by dimensions:
\be 
\label{Adimensional}
{A}_\sigma = a_\sigma\frac{\kappa^2\ \ }{\mu^{2i_\sigma}}\, \Lambda_\p^{n_f+\varepsilon_a+\varepsilon_b-n-\alpha-4}\quad  (i_\sigma\ge0)\,,\quad  a_\sigma\kappa^2\Lambda_\p^{n_f+\varepsilon_a+\varepsilon_b-2i_\sigma-n-\alpha-4}\quad  (i_\sigma\le0)\,,
\ee
where $a_\sigma\cu\ne0$ is a finite  dimensionless number, and note that we have also defined the power  $p_\sigma$.

It is obvious from the dimensionless ratio $\bar{g}^\sigma_{2l+\eps}$ \eqref{barg} and the above relation between dimensions \eqref{dimgA} that this implies $\bar{g}^\sigma_{2l+\varepsilon}\sim (c^\sigma_{2l+\varepsilon}/a_\sigma) \ln(\Lambda_\p/\mu)$ for $i_\sigma=0$, and $\bar{g}^\sigma_{2l+\varepsilon}\to c^\sigma_{2l+\varepsilon}/a_\sigma$ otherwise, and thus that the convergence criteria are satisfied. But such a choice would make no sense if $A_\sigma$ diverges in the large-$\Lambda_\p$ limit.
%for $A_\sigma$ would fail if it meant that it diverged in the large-$\Lambda_\p$ limit. 
At first sight from the above \eqref{Adimensional}, this can easily happen (particularly in the case $i_\sigma<0$). 
However we only require this construction 
%the construction in sec. \ref{sec:fixcouplings} 
if there are irrelevant couplings, in which case 
\be 
[A_\sigma]\le -2 -\varepsilon -\alpha\,,
\ee
as follows from the above relation between dimensions \eqref{dimgA} and the fact that then $[g^\sigma_\varepsilon]\cu\le-1$. Since $\kappa^2$ accounts for the $-2$ if $i_\sigma\cu\le0$, we see that in this case $p_\sigma\cu\le -\varepsilon\cu-\alpha\cu\le0$. On the other hand for $i_\sigma\cu>0$, we see explicitly from \eqref{Adimensional} that $p_\sigma\cu\le2$, since this maximum value is reached only if all parameters take their appropriate extreme values: $n_f\cu=4$, $\varepsilon_a\cu=\varepsilon_b\cu=1$, and $\alpha\cu=n\cu=0$. However this combination of parameters is not possible in practice.
Vertices with $\varepsilon_b\cu=1$ only have two fields in their monomial, both of which are required for propagators since $n_f\cu=4$.\footnote{In fact the contribution would have to come from the second bracket in the level-zero first-order vertex \eqref{Gammaonezero}.} This means that the derivatives $(-\Box)^{m>0}$ must act on $f^b$, in contradiction with the assumed $n\cu=0$. Since $p_\sigma$ is even, we thus learn that actually $p_\sigma\le0$. Altogether we have proved that whenever 
%the construction of sec. \ref{sec:fixcouplings} is needed, 
there are irrelevant couplings, we have $p_\sigma\cu\le0$. 

We have already seen that there are cases where we must be able to choose the coefficient function to vanish in the large-$\Lambda_\p$ limit if we are to satisfy the mST, as discussed above sec. \ref{sec:largeASSderiv0}. For these, we therefore need $p_\sigma\cu<0$.
%In these cases we need $A_\sigma$ to have a negative power of $\Lambda_\p$. 
%For all cases at antighost number four, \cf sec. \ref{sec:largeASS}, this turned out automatically on setting $\alpha\cu=\varepsilon$. It can be shown that this is also true at antighost number three.
In the model, sec. \ref{sec:largeASSderiv0}, this happened automatically and actually this is true in nearly all cases. However as we see in an example below, sometimes $p_\sigma\cu=0$ and thus $A_\sigma$ needs a little help. One way would be to add a linearised solution, \ie a solution to just the LHS of the second-order flow eqaution (\ref{flowtwo}) and thus containing only relevant couplings, which tends to $-A_\sigma$ in the large $\Lambda_\p$ limit. Then we will be left with a combined solution for the coefficient function whose leading behaviour goes as $\sim\vp^\alpha/\Lambda_\p^{\alpha-\varepsilon}$, $\alpha\cu\ge\varepsilon+2$, at large $\Lambda_\p$. Equivalently we can arrange for $\alpha\cu\ge\varepsilon+2$ directly in the complementary solution,  forcing $f^\sigma_\Lambda$ to  the corresponding limit (\ref{flatp},\ref{Hermite}).\footnote{An alternative strategy would be to redefine $A_\sigma$ by multiplying by $\mu/\Lambda_\p$. Then the dimensionless ratio $\bar{g}^\sigma_{2l+\varepsilon}$ \eqref{barg} gains a linear divergence but this would still satisfy the convergence criterion.} 
From the formulae for $A_\sigma$ \eqref{Adimensional} it is always sufficient to choose $\alpha\cu=\varepsilon\cu+2$.

As an example consider the antighost level two contribution appearing in the one-loop formula \eqref{twoexpand} on the RHS of the flow equation, and coming from using $\Gamma^2_1$ as one instance, and the $f^1$ terms of $\Gamma^1_1$ as the other, joining these with a ghost propagator attached to $\partial c$ in $\Gamma^2_1$ \eqref{Gammaonetwo}, and a $\vp$-propagator attached to the two coefficient functions. Thus we have that $\sigma^a = c_\alpha c^*_\beta$, and from  $\Gamma^1_1$ \eqref{Gammaoneone}, $\sigma^b = c_\gamma\partial_\gamma H_{\mu\nu}$ or $h_{\gamma(\mu}\partial_{\nu)}c_\gamma$, and also $n_f\cu=2$, and $\varepsilon_a\cu=\varepsilon_b\cu=0$. In the Feynman diagram expression \eqref{generalFeynman} we have $d\cu=2$. Choosing $c\cu=1$ in $F_{rc}(p)$ \eqref{Frc} (so $r\cu=0$) and $m\cu=0$ (so $n\cu=0$) we have that $f^\sigma_\Lambda$ is even, from the parity equation \eqref{epseqn}. Since the $f^1$s are differentiated, the particular integral does not survive the large-$\Lambda_\p$ limit. Thanks to the absence of contributions at higher antighost number and the similar vanishing of the RHS of the second-order mST \eqref{mSTtwo}, 
this latter would simply be $Q_0\,(\sigma f^\sigma_\Lambda+\cdots)\cu=0$, where the ellipses stand for the rest of $\Gamma^2_2$, while we see by applying the explicit formula for the action of $Q_0$ \eqref{QH}, that $Q_0\,\sigma\cu\ne 0$ since $\sigma\cu=\sigma^a\sigma^b$. The same arguments as above sec. \ref{sec:largeASSderiv0} show again that we should choose $f^\sigma_\Lambda$ to vanish in the limit. From $[g^\sigma_0]\cu=-1$ \eqref{gtwodimensions}, $g^\sigma_0$ is irrelevant and from its large $\Lambda_\p$ behaviour \eqref{girrgeneral} it actually diverges: $g^\sigma_0\cu\sim c^\sigma_0\kappa^2\Lambda_\p$. We therefore require the scheme in sec. \ref{sec:fixcouplings}. However
if we choose $\alpha\cu=0$, then the formula for $A_\sigma$ \eqref{Adimensional} gives a non-vanishing trivialisation $f^\sigma_\Lambda\cu\to A_\sigma\cu=a_\sigma\kappa^2$. We easily fix this problem by choosing instead $\alpha\cu=2$, and thus 
$
f^\sigma_\Lambda\sim a_\sigma\kappa^2(\vp^2+\Omega_\Lambda)/\Lambda_\p^2
$.

Finally, one might worry that there are cases where we need $A_\sigma$ non-vanishing, in order to satisfy the second-order mST \eqref{mSTtwo}, but our prescription for $A_\sigma$ \eqref{Adimensional} would provide $p_\sigma\cu<0$. This would not be an obstruction since we can redefine $A_\sigma$ by multiplying by a sufficiently positive power of $\Lambda_\p/\mu$, making the convergence criteria \eqref{barg} even better satisfied. Actually this issue does not arise, because the highest dimension monomial demanded by the second-order mST is $d_\sigma\cu=6$ as verified for example (schematically) by the vertex $\sigma\cu=h^2(\partial h)^2$, which is also what is generated by expansion of the Einstein-Hilbert action to this order \cite{second}. Then by dimensions the prescription \eqref{Adimensional} must give the non-vanishing $A_\sigma\cu=a_\sigma\kappa^2$. We can confirm this on the above example. The vertex is generated by quantum corrections involving the $f^1_\Lambda$ part of the level-zero first-order vertex \eqref{Gammaonezero}. The scheme of sec. \ref{sec:fixcouplings} is needed since it has an irrelevant coupling $[g^\sigma_0]=-1$ \eqref{gdimension}, that diverges in the large-$\Lambda_\p$ limit as $g^\sigma_0\sim c^\sigma_0 \kappa^2\Lambda_\p$ \eqref{girrgeneral}. We also note that it has vanishing particular integral. The prescription indeed yields $A_\sigma\cu=a_\sigma\kappa^2$,  since the indices work out to be $i_\sigma\cu=-1$, $n_f\cu=2$, and $n\cu=\eps\cu=\eps_a\cu=\eps_b\cu=\alpha\cu=0$. 

We have thus shown that the prescription for $A_\sigma$ \eqref{Adimensional} works in all cases where we require this construction.

\section{Summary}
\label{sec:conclusions}

We finish with a short summary of the main results in this paper. For pure perturbative quantum gravity to second order (but non-perturbatively in $\hbar$) the renormalized trajectory solution  at cutoff scale $\Lambda$, is given by \eqref{Gammatwosol}:
\be 
%\label{Gammatwosol}
\Gamma_{2} = \half \left[\, 1+\Po_\Lambda-(1+\Po_\mu)\,\mathrm{e}^{\Po^\mu_\Lambda}\,\right] \Gamma_{1}\, \Gamma_{1} + \Gamma_{2}(\mu)\,,
\ee
where the $\Po$ operator \eqref{Posimple} attaches a propagator between the two copies of the first-order solution $\Gamma_1$, regularised in the UV(IR) as indicated by the super(sub)script. The first term is the particular integral. Expanding in $\Po$ gives the melonic diagrams of fig. \ref{fig:melons}. For each $\sigma$ term in the derivative expansion \eqref{ringinvert}:
\be
\Gamma_2 = \sum_\sigma \left( \sigma f^\sigma_\Lambda(\vp)+\cdots \right)
\ee
(where the ellipses are given by the formula \eqref{firstOrder}),
the melonic expansion results in convergent coefficient functions $f^\sigma_\Lambda(\vp)$ provided that we choose $\mu$ in the range $0\cu<\mu\cu<a\Lambda_\p$ \eqref{murange}. For the particular integral, if we choose $\mu\cu=0$ the derivative expansion breaks down, while if we choose $\mu\cu>a\Lambda_\p$ the contribution to the coefficient functions will become singular before the physical limit, $\Lambda\cu\to0$, is reached. Although we first see this in a model, we find this also for the full renormalized trajectory in sec. \ref{sec:fullrt}. If we try to define the renormalized particular integral only in terms of $\Lambda$, \ie without introducing the other cutoff scale $\mu$, we again find that it contributes singular coefficient functions, \cf also app. \ref{app:fails}. 

The second term, $\Gamma_2(\mu)$, is the particular solution. It is a solution of the linearised flow equation \eqref{flowone}, which coincides with $\Gamma_2$ at the point $\Lambda\cu=\mu$. If we choose $\mu$ in the range $0\cu<\mu\cu<a\Lambda_\p$, the complementary solution we want also has a derivative expansion for $\Lambda\cu>0$ and non-singular coefficient functions. 
The renormalized trajectory emanates correctly from the Gaussian fixed point, provided that the scaled underlying irrelevant couplings $\tg^\sigma_{2l+\eps}$, in each complementary coefficient function $f^\sigma_\Lambda(\vp,\mu)\in\Gamma_2(\mu)$, vanish as $\Lambda\to\infty$, which determines these irrelevant couplings uniquely in terms of the first-order couplings. Then the limiting behaviour of these irrelevant couplings at large $\Lambda_\p$,  can determined by dimensions up to an overall numerical coefficient, \eqref{girrgeneral}. 
%At worst, they diverge as a power of $\Lambda_\p$ times $\ln(\Lambda_\p)$.

Each $f^\sigma_\Lambda(\vp)$ in $\Gamma_2$, comes with its own amplitude suppression scale, however by choosing the domain of their relevant underlying couplings appropriately we can set all amplitude suppression scales to the first-order common scale, $\Lambda_\p$. The second-order underlying couplings all have odd dimension, so  in particular none of them are marginal. Furthermore at second-order the first order couplings do not run, although they will at third order in perturbation theory. 

In order for the renormalized trajectory to enter the diffeomorphism invariant subspace, as illustrated in fig. \ref{fig:flow}, we must choose the domain of the underlying relevant couplings so that the coefficient functions trivialise: \ie so that as $\Lambda_\p\cu\to\infty$, the physical coefficient functions $f^\sigma(\vp) \cu\to A_\sigma\,\vp^\alpha$, \eqref{flatphys}, for some non-negative integer $\alpha$. Despite the presence in these coefficient functions  of already-determined underlying irrelevant couplings, this can be done provided the reduced irrelevant couplings $\bar{g}^\sigma_{2l+\eps}$, \eqref{barg}, diverge slower than $\Lambda^2_\p$. This in turn can achieved by choosing the coefficient $A_\sigma$ as in \eqref{Adimensional}. We then show that for all $\sigma$ at second order, by choosing $\alpha$ appropriately, this prescription provides the right trivialisations to allow the second order mST to be satisfied. 

This last step, \ie solving for the renormalized trajectory inside the diffeomorphism invariant subspace,  will be treated in ref. \cite{second}. It is already clear however  that dependence on $\Gamma_1$ in the mST becomes essentially that of second order in standard quantisation, and that similarly the particular integral collapses in this limit simply to  standard one-loop self-energy diagrams  \eqref{partLASS}:
\be 
%\label{partLASS}
\Gamma_2 %= \sum_\sigma \left( \sigma f^\sigma_\Lambda(\vp)+\cdots \right)
= \sum_\sigma \left( \sigma f^\sigma_\Lambda(\vp,\mu)+\cdots \right)
+\tfrac14\kappa^2\,\text{Str}\!\left[\proph{\mu} \check{\Gamma}^{(2)}_1 \proph{\mu} \check{\Gamma}^{(2)}_1
- \proph{\Lambda} \check{\Gamma}^{(2)}_1 \proph{\Lambda} \check{\Gamma}^{(2)}_1\right]\,.
\ee

%\newpage
%\mbox{}
%\newpage

\section*{Acknowledgments}
It is a pleasure to thank James Drummond,  \"Omer G\"urdo\u gan, Chrysostomos Kalousios, Jonathan Flynn, Andreas J\"uttner, and  Nick Evans  for discussions. % and to thank David Turton for reminding me of Bank's argument.
I acknowledge support from STFC through Consolidated Grant ST/P000711/1.

%\vfill
%\newpage 

\appendix

\section{Appendix}

\subsection{High loop order dependence of melonic Feynman diagrams}
\label{app:largen}

The $\ell$-dependence of the $\ell$-loop melonic Feynman diagrams follows from the integral $I^k_\ell$ defined in equation \eqref{In}, for $0\cu\le k\cu<\ell\cu-1$. Here we determine the behaviour of $I^k_\ell$ at large $\ell$. It is helpful first to define the following numbers: 
%that will appear in the $(n\cu-1)^\text{th}$ loop diagram:
\be 
\label{zetapn}
\zeta^p_\ell \,=\,\frac{1}{\ell!}\sum^\ell_{m=0} (-)^{m+\ell} \binom\ell{m}\,
m^{\,p+\ell}\,  \,=\, \frac1{\ell!} \left(u\, d_u\right)^{p+\ell}(u-1)^\ell\, \Big|_{u=1}\,. %\, =\, 0 \qquad \text{for}\quad p<\ell\,,
\ee
Clearly they vanish for $-\ell\cu\le p\cu<0$. Evaluating the rightmost derivative  we see that
\be 
\zeta^p_\ell = \frac1{(\ell-1)!}\left(u\, d_u\right)^{p+\ell-1}\left[ u(u-1)^{\ell-1} \right] \, \Big|_{u=1} = \ell\,\zeta^{p-1}_\ell+\zeta^p_{\ell-1}\,.
\ee
From this recurrence relation and directly for small $p$,\footnote{The first three are exactly: $\zeta^0_\ell = 1$,\ $\zeta^1_\ell=\half \ell(\ell\cu+1)$,\ and $\zeta^2_\ell=\ell(\ell\cu+1)(\ell\cu+2)(3\ell\cu+1)/4!$\,.}  one establishes that $\zeta^p_\ell = \alpha_p\, \ell^{2p}$ for large $\ell$ and some $\alpha_p$, and thus substituting back into the recurrence relation we find
\be 
\label{zetaplargen}
\zeta^p_\ell = \frac{\ell^{2p}}{2^pp!}+O(\ell^{2p-1})\,.
\ee
Now we derive from the integral $I^k_\ell$ \eqref{In} a series of alternative exact expressions. We start by noting that
\be 
I^k_{\ell}\, \Omega_\Lambda^{\ell-k-1} = \int^\infty_0\!\!\!\!\!\!dt\ t^{k-\ell} \left(1-\text{e}^{-\Omega_\Lambda t}\right)^\ell\,.
\ee
Thus by expanding the RHS and differentiating $\ell\cu-k\cu-1$ times with respect to $\Omega_\Lambda$, we see that
\be 
(\ell\cu-k\cu-1)!\, I^k_{\ell} = -\lim_{\epsilon\to0} \sum^\ell_{m=0} (-)^{m+\ell+k} \binom{\ell}m\, m^{\ell-k-1} \int^\infty_\epsilon \! \frac{dt}{t} \, \text{e}^{-m\,\Omega_\Lambda t}\,,
\ee
where after expansion, the UV limit $t\cu=\epsilon$ needs to be temporarily introduced. In fact, since
\be 
\int^\infty_\epsilon \! \frac{dt}{t} \, \text{e}^{-m\,\Omega_\Lambda t} =
\int^\infty_{m\,\Omega_\Lambda\epsilon}\! \frac{dt}{t} \, \text{e}^{- t} = -\ln(m) -\ln(\Omega_\Lambda\epsilon) + \int^\infty_0\!\!\!\! dt \, \text{e}^{- t}\ln t\,,
\ee
up to terms that vanish as $\epsilon\cu\to0$, and since $\zeta^{-k-1}_\ell$ \eqref{zetapn}
vanishes for $0\cu\le k\cu< \ell\cu-1$, we see that the limit can be safely taken. Thus we have shown that:\notes{[1093.9]}
\be 
\label{Iclosed}
I^k_{\ell} = \frac{(-)^{\ell+k}}{(\ell\cu-k\cu-1)!} \sum^\ell_{m=0} (-)^m \binom{\ell}m\, m^{\ell-k-1}\ln m\,.
\ee
Thanks to the kind of cancellations we see in $\zeta^p_\ell$ \eqref{zetapn}, the sum is much smaller at large $\ell$ than the terms for $m\cu\approx \ell$ would suggest. To extract the leading behaviour, we relabel $m$ to $\ell\cu-m$, and Taylor expand $L_{\ell-k-1}(x):=x^{\ell-k-1}\ln x$, $x\cu=\ell\cu-m$, about $x\cu=\ell$. Exchanging summations and using \eqref{zetapn}: \notes{[1069.10,1071,1094]}
\be 
\label{Intermediate}
I^k_\ell =  \frac{(-)^k \ell!}{(\ell\cu-k\cu-1)!}\sum_{p=0}^\infty \frac{(-)^p}{(\ell+p)!}\,\zeta^p_\ell\,L^{(\ell+p)}_{\ell-k-1}(\ell)\,.
\ee
Combining $L^{(m)}_m(x) \cu= m! \ln x \,+$ const., and $d^r_x\ln x = (-)^{r+1}(r\cu-1)! /x^r$, for $m\cu=\ell\cu-k\cu-1$ and $r\cu=k\cu+p\cu+1$:
\be L^{(\ell+p)}_{\ell-k-1}(x) = (\ell\cu-k\cu-1)!\, (k\cu+p)!\, (-)^{k+p} x^{-k-p-1}\,. \ee
\notes{1095} Substituting this into \eqref{Intermediate} we still have an exact expression, however now using the asymptotic formula \eqref{zetaplargen} for $\zeta^p_\ell$ and simplifying for large $\ell$ we finally derive the large-$\ell$ behaviour of $I^k_\ell$ already quoted in  \eqref{largenIn}:
\be 
\label{largenInApp}
I^k_\ell = \frac{k!}{\ell^{k+1}}\sum_{p=0}^\infty \binom{p\cu+k}p\frac{1}{2^p} +O({\ell^{-k-2}})  = k! \left({2}/{\ell}\right)^{k+1} +O({\ell^{-k-2}}) \,.
\ee
We have verified numerically that this correctly captures the large $\ell$ behaviour of the integral  $I^k_\ell$.

\subsection{Singular flows in a solution depending only on one cutoff scale}
\label{app:fails}

Although in establishing the range \eqref{murange} we saw that there is  a non-perturbative-in-$\hbar$ obstruction to taking $\mu\cu\to\infty$, we can take this limit formally in our final (subtracted form of) solution \eqref{Gammatwosol} order by order in $\hbar$. 
The UV divergences in the loop integrals are then cancelled by the counterterms in $\Gamma_2(\mu\cu\to\infty)$, leaving behind finite counterterms $\Gamma_2^c(\mu_R)$, where this functional is still defined in the same way as the complementary solution \eqref{complementary} but in terms of $\rg\Gamma^c_2(\mu_R)$, where $\mu_R$ is the usual arbitrary physical mass scale. Thus we now have:
\be 
\label{GammatwoIR}
\Gamma_2 = \half \left( 1+\Po_\Lambda-\mathrm{e}^{\Po_\Lambda}\right) \Gamma_1\, \Gamma_1 + \Gamma^c_2(\mu_R)\,.
\ee
A more elegant derivation is to basically follow app. \ref{app:alt} and start with the $\mu\cu\to\infty$ limit of the integrating factor \eqref{integratingfactor}, which means that we have only IR regulated propagators from the beginning. Then we subtract $\Lambda$-independent UV divergences as part of the definition of the loop integrals \cite{\yuji}. Factoring the exponential into three \eqref{WickTwoFactorId} on the RHS of \eqref{integratingfactor},  will result in the first order solution  \eqref{eigenoperatorsol} being converted into $\Gamma_{1\,\text{phys}}$ plus a series of unregulated massless tadpole integrals. It is natural to define the subtraction to set these to zero, as is the case in dimensional regularisation. The net result is to turn the $\Gamma_{1\Lambda}$ factors into $\Gamma_{1\,\text{phys}}$. Following through with the other steps, we arrive again at the above  \eqref{GammatwoIR}, in particular $\Gamma_2^c(\mu_R) = \exp\left(\frac12 {\prop}^{AB}_\Lambda \frac{\partial^2_l}{\partial\Phi^B\partial\Phi^A}\right) \rg\Gamma^c_2(\mu_R)$ again contains $\Lambda$-independent divergent tadpole integrals which we discard, turning this into the previous definition. 

To see that the result makes sense, order by order in the loop expansion, we note that the above  \eqref{GammatwoIR} 
is a sum over the melonic diagrams fig. \ref{fig:melons} again, that it has a derivative expansion since these are all IR regulated, and that it solves the secon-order flow equation \eqref{flowtwo}. To verify this last assertion, we define the subtracted melonic diagrams iteratively. The differentiated one-loop operator appears as: 
\be 
\label{Mone}
\dot{M}_1=-\half\, \dot{\Po_\Lambda}\Po_\Lambda\,, 
\ee
see the later one-loop expression \eqref{oneloopP} or the one-loop expression derived earlier  \eqref{oneloopPsimple},
and is thus IR and UV regulated. As in ref. \cite{\yuji}, we understand the Feynman integral solution for $M_1$ to be accompanied by appropriate  $\Lambda$-integration constants such as to give a finite answer. %By definition these integration constants are $\Lambda$-independent. 
Evidently these constants should be chosen $\Lambda$-independent, but also 
so as to leave an expression with no dimensionful parameters, or in the case of a logarithmically divergent integral some unavoidable $\ln\mu_R$ dependence ($\mu_R$ being the usual arbitrary finite physical scale), since dimensionful parameters can all be absorbed into the coupling constants in $\Gamma^c_2(\mu_R)$. The higher-loop operators are then defined iteratively:
\be 
\label{recurrence}
\dot{M}_\ell = \dot{\Po_\Lambda} M_{\ell-1}\,.
\ee
Since $M_{\ell-1}$ has already been given as a finite expression, and $\dot{\Po_\Lambda}$ adds another loop which is however clearly UV and IR regulated, a finite solution for $M_\ell$ can also be found by appropriate choice of $\Lambda$-independent integration constants.
Therefore we have expressed our purely IR regulated solution \eqref{GammatwoIR} as the sum:
\be 
\label{GammatwoSum}
\Gamma_2 = \sum_{\ell=1}^\infty M_\ell\, \Gamma_1\, \Gamma_1 + \Gamma^c_2(\mu_R)\,.
\ee
It is straightforward to verify that it indeed solves the second-order flow equation \eqref{flowtwo}:
\beal 
\dot{\Gamma}_2 &= \sum_{\ell=1}^\infty \left(M_\ell\, \dot\Gamma_1\, \Gamma_1 + \dot{M}_\ell\, \Gamma_1\, \Gamma_1 +M_\ell\, \Gamma_1\, \dot\Gamma_1\right) +\dot \Gamma^c_2(\mu_R) \nn\\
&= \sum_{\ell=1}^\infty \left(M_\ell\, \dot\Gamma_1\, \Gamma_1 +\dot{\Po_\Lambda} M_\ell\, \Gamma_1\, \Gamma_1 +M_\ell\, \Gamma_1\, \dot\Gamma_1\right) +\dot \Gamma^c_2(\mu_R) -\half \dot{\Po_\Lambda}\Po_\Lambda\,\Gamma_1\, \Gamma_1\nn\\
&= \half\, \text{Str}\, \dot{\prop}_\Lambda \Gamma^{(2)}_2 
- \half\, \text{Str}\, \dot{\prop}_\Lambda \Gamma^{(2)}_1 \propH \Gamma^{(2)}_1\,,
\eeal
where in the second line we use the recurrence relation \eqref{recurrence} and the one-loop expression \eqref{Mone}, then recognise the action of the tadpole operator \eqref{flowone}, inside the brackets split three ways \eqref{WickTwoFactorId}, and finally reverse the detailed one-loop expression (\ref{oneloopP}/\ref{oneloopPsimple}).

The problem is that the sum \eqref{GammatwoSum}, although well defined term by term, leads to a singular $\Gamma_2$ for sufficiently large $\Lambda$, and thus fails to provide a sensible renormalized trajectory. To show this we again specialise to the exponential regulator \eqref{Cexp} and use position space. Furthermore we demonstrate the problem for $m\cu=0$ and $n_{a\ell} \cu= n_{b\ell}\cu=\ell\cu+1$, \ie  compute $D^0_\ell$, the $O(\partial^0)$ part of $M_\ell$, using only $\vp$-propagators. To define the subtractions we need a UV regulator. We use a simple short distance $x\cu>r_0$ cutoff. Then from the position space %formula for the 
IR cutoff propagator \eqref{propIR},
\be 
\label{quadint}
D^0_1 |_{x>r_0}\ =\ -\frac14\int_{x>r_0}\!\!\!\!\!\!d^4x\ \prop^2_\Lambda(x)\ =\ \frac{\ln(\Lambda^2r^2_0/2)+\gamma_E}{64\pi^2}+O(r^2_0)\,,
\ee
where $\gamma_E$ is Euler's constant, and thus we see that we need to define $M_1$ as the distribution\footnote{Abusing notation, we use $M_\ell$ for the Feynman diagram as well as the functional operator.}
\be 
M_1 = -\frac14\, \prop^2_\Lambda(x)- \frac1{2(4\pi)^2}\ln(\mu_Rr_0)\,\delta(x)\,,
\ee
the latter integrated over all space and the former over $x\cu>r_0$, so that for $D^0_1$ the limit $r_0\cu\to0$ can now be safely taken.
Now using the recurrence relation \eqref{recurrence} (and remembering the sign for the $\vp$-propagator) we have
\be 
\label{dotMtwo}
\dot M_2 = -\dot{\prop}_\Lambda(x)\, M_1 = \frac14\,\dot{\prop}_\Lambda \prop^2_\Lambda (x)
+ \frac1{2(4\pi)^2}\ln(\mu_Rr_0)\, \dot{\prop}_\Lambda(0)\,\delta(x)\,.
\ee
Integrating with respect to $t$, and then over all $x$ to get $D^0_2$, the $\prop^3_\Lambda$ part will need $x\cu>r_0$ regularisation:
\be 
\label{cubeint}
\frac12\frac1{3!}\int_{x>r_0}\!\!\!\!\!\!d^4x\ \prop^3_\Lambda(x)\ =\ \frac1{768\pi^4r^2_0}+\frac{\Lambda^2}{4(4\pi)^4}\left[\ln(3\Lambda^2r^2_0/4)+\gamma_E-1\right]+O(r^2_0)\,.
\ee
The $\Lambda$-dependent $\ln r_0$ divergences cancel when we include the ($t$-integrated) last term from \eqref{dotMtwo} as they must for consistency. However we see that we now need a quadratically divergent counterterm $\propto\delta(x)$. There is now a logarithmic divergence at $m\cu=1$, \ie in the $O(\partial^2)$ part, which we can isolate by integrating $x^2\prop^3_\Lambda$ over all $x\cu>r_0$, and which needs removing with a $\Box\,\delta(x)$ distribution. Thus in total one finds the regularised two-loop contribution
\be 
M_2 = \frac12\frac1{3!}\prop^3_\Lambda(x) -\frac{\Lambda^2}{2(4\pi)^4}\ln(\mu_Rr_0)\,\delta(x)-\frac1{768\pi^4r^2_0}\,\delta(x)+\frac{\ln(\mu_Rr_0)}{3072\pi^4}\,\Box\,\delta(x)\,.
\ee
Proceeding in this way we can define all the higher loop contributions also. A little thought then shows that at $\ell$ loops, $D^0_\ell \propto \Lambda^{2(\ell-1)}\ln\Lambda$ at large $\Lambda$, continuing the pattern established in the one and two loop contributions (\ref{quadint},\ref{cubeint}). By dimensions, this leading term can be extracted by computing the $\Lambda^{2(\ell-1)}\ln r_0$ part of the unregularised integral
\be 
-\frac{(-)^\ell}{2(\ell+1)!}\int_{x>r_0}\!\!\!\!\!\!d^4x\ \prop^{\ell+1}_\Lambda(x)\,.
\ee
But this part is easily extracted by differentiating the above with respect to $r_0$ which gives, up to factors, just $\prop^{\ell+1}_\Lambda(r_0)$. Thus we find \notes{1041}
\be 
D^0_\ell \sim  \frac{\ln\Lambda}{(4\pi)^2} \frac{\left[(\ell+1)\Omega_\Lambda\right]^{\ell-1}}{(\ell+1)!(\ell-1)!}\,,
\ee
where we used the formula  \eqref{Omegaexp} for $\Omega_\Lambda$ with exponential cutoff. Fourier transforming our purely-IR-regulated solution \eqref{GammatwoIR}, we have to compute the sum we saw before \eqref{sumD} with $D^m_{rc\ell}=D^0_\ell$. Using Stirling's approximation, we get asymptotically,
\be 
\sum_\ell D^0_\ell\left(\vpi_1^2-\frac{\vpi^2}{4}\right)^{\!\ell-1} \sim \exp\left\{\rm{e}\,\Omega_\Lambda\left(\vpi^2_1-\frac{\vpi^2}{4}\right)\right\}\,.
\ee
The problem here is the positive exponent for $\vpi_1$ which beats the negative exponents provided asymptotically by the first order coefficient functions  \eqref{largevpibar}, once 
$\Lambda\cu> a\Lambda_\p/\sqrt{\textrm{e}-1}$. Then the $\vpi_1$ integral fails to converge, destroying even the second-order Fourier version, $\ff^\sigma(\vpi,\Lambda)$. 

%of the second order coefficient functions, ceases to exist. 

\subsection{Streamlined derivation of the sum over melonic diagrams}
\label{app:alt}

Using the UV and IR regulated propagator \eqref{propmuLambda} to form the integrating factor, the second-order flow equation \eqref{flowtwo} can be written as
\be 
\label{integratingfactor}
\frac{\partial}{\partial t}\left\{  \exp\left(-\frac12 {\prop}^{\mu\,AB}_\Lambda \frac{\partial^2_l}{\partial\Phi^B\partial\Phi^A}\right) \Gamma_{2\Lambda} \right\} = -\frac12\exp\left(-\frac12 {\prop}^{\mu\,AB}_\Lambda \frac{\partial^2_l}{\partial\Phi^B\partial\Phi^A}\right)  \text{Str}\, \dot{\prop}_\Lambda \Gamma^{(2)}_{1\Lambda} \propH \Gamma^{(2)}_{1\Lambda} \,.
\ee
Instead of the earlier \eqref{oneloopPsimple} we express the one-loop expression as:
\be 
\label{oneloopP}
-\tfrac12\,\text{Str}\, \dot{\prop}^\mu_\Lambda \Gamma^{(2)}_{1\Lambda} \propH \Gamma^{(2)}_{1\Lambda} = -\half\,\dot{\Po}^\mu_\Lambda\,\Po_\Lambda\,\Gamma_{1\Lambda}\,\Gamma_{1\Lambda}\,.
\ee
Factoring the exponential on the RHS into three pieces  \eqref{WickTwoFactorId}, 
those that operate exclusively on their own copy of $\Gamma_{1\Lambda}$, by the first-order solution \eqref{eigenoperatorsol} and the expression for the UV and IR regulated propagator \eqref{propmuLambda}, turn them into $\Gamma_{1\mu}$. Therefore the RHS of \eqref{integratingfactor} is
\be 
%\label{Poexpression}
-\tfrac12 \,\dot{\Po}^\mu_\Lambda \,\Po_\Lambda \,\mathrm{e}^{-\Po^\mu_\Lambda}\, \Gamma_{1\mu}\, \Gamma_{1\mu} = \partial_t \left[\,\half\, (1+\Po_\Lambda)\,\mathrm{e}^{-\Po^\mu_\Lambda}\, \Gamma_{1\mu}\, \Gamma_{1\mu}\, \right]\,,
\ee
and can thus be integrated exactly. Its integration constant
\be 
-\half(1+\Po_\mu)\,\Gamma_{1\mu}\, \Gamma_{1\mu} + \Gamma_{2\mu}\,,
\ee
is determined by the requirement that 
\be 
\label{mubc}
\Gamma_{2\,\Lambda=\mu} = \Gamma_{2\mu}\,.
\ee 
We can now integrate both sides of \eqref{integratingfactor} and multiply through by the inverse of the LHS exponential. The term
\be 
\label{complementaryEquiv}
%\label{complementary}
\Gamma_{2\Lambda}(\mu) = \exp\left(\frac12 {\prop}^{\mu\,AB}_\Lambda \frac{\partial^2_l}{\partial\Phi^B\partial\Phi^A}\right) \Gamma_{2\mu}\,,
\ee
is an equivalent definition of the complementary solution \eqref{complementary}. The rest of the solution involves the exponential acting on the product of the two $\Gamma_{1\mu}$, which we split into three \eqref{WickTwoFactorId}, getting an $\mathrm{e}^{\Po^\mu_\Lambda}$ factor, and factors that act exclusively to restore each $\Gamma_{1\mu}$ back to $\Gamma_{1\Lambda}$. The result is again our final subtracted solution \eqref{Gammatwosol}.

%\vfill
%\newpage 

\bibliographystyle{hunsrt}
\bibliography{references} %%from now on (14/7/15) this is the global references list!

\end{document}